\documentclass{aa}  
\usepackage[table]{xcolor}

\usepackage{graphicx}
\usepackage{txfonts}
\usepackage[colorlinks=true,linkcolor=blue,citecolor=blue]{hyperref}
\usepackage{url}
\usepackage{mathtools}
\usepackage{booktabs}
\usepackage{diagbox}
\usepackage{float}
\restylefloat{table}

\begin{document}

   \title{The slippery slope of dust attenuation curves}

   \subtitle{Correlation of dust attenuation laws with star-to-dust compactness up to $z = 4$}

\author{M. Hamed\inst{1}
          \and
          K. Ma\l{}ek\inst{1,2}
          \and
          V. Buat\inst{2}
          \and
          Junais\inst{1}
          \and
          L. Ciesla\inst{2}
          \and
          D. Donevski\inst{1,3}
          \and
          G. Riccio\inst{1}
          \and
          M. Figueira\inst{1,4}
          }

         \institute{National Centre for Nuclear Research, ul. Pasteura 7, 02-093 Warszawa, Poland
          \and
          Aix Marseille Univ. CNRS, CNES, LAM, 13015 Marseille, France
          \and
           SISSA, Via Bonomea 265, Trieste, Italy
           \and
           Institute of Astronomy, Faculty of Physics, Astronomy and Informatics, Nicolaus Copernicus University, 87-100 Toruń, Poland
          }

   \date{}

 
  \abstract
  {}
  {We investigate dust attenuation of 122 heavily dust-obscured galaxies detected with the Atacama Large Millimeter Array (ALMA) and $Herschel$ in the COSMOS field. We search for \text{correlations} between dust attenuation recipes and the variation of physical parameters, mainly the effective radii of galaxies, their star formation rates, and stellar masses, and aim to understand which of the commonly used laws best describes dust attenuation in dusty star-forming galaxies (DSFGs) at high redshift.}
  {We make use of the extensive photometric coverage of the COSMOS data combined with highly-resolved dust continuum maps from ALMA. We use \texttt{CIGALE} to estimate various physical properties of these dusty objects, mainly their star formation rates (SFR), their stellar masses and their attenuation in the short wavelengths. We infer galaxy effective radii (Re) using GALFIT in the $Y$ band of HSC and ALMA continuum maps. We use these radii to investigate the relative compactness of the dust continuum and the extension of the rest-frame UV/optical Re(y)/Re(ALMA).}
{We find that the physical parameters calculated from our models strongly depend on the assumption of dust attenuation curve. As expected, the most impacted parameter is the stellar mass, which leads to a change in the "starburstiness" of the objects. We find that taking into account the relative compactness of star-to-dust emission prior to SED fitting is crucial, especially when studying dust attenuation of dusty star-forming galaxies. Shallower attenuation curves did not show a clear preference of compactness with attenuation, while the Calzetti attenuation curve preferred comparable spatial extent of unattenuated stellar light and dust emission. The evolution of the R$_e(\mathrm{UV})$/R$_e(\mathrm{ALMA})$ ratio with redshift peeks around the cosmic noon in our sample of DSFGs, showing that this compactness is correlated with the cosmic SFR density of these dusty sources.} 
{}


   \maketitle
%

\section{Introduction}
In its earlier years, the Universe did not only witness more star formation rates (SFRs), but a significant fraction of this SFR was obscured by dust \citep[e.g.,][]{Blain2002,Takeuchi2005,Chapman2005,Bouwens2012,Madau14,Magnelli2014,Bourne2017,Whitaker2017,Gruppioni2020,Khusanova2020}. The heavily attenuated in the ultraviolet (UV) dusty star-forming galaxies \citep[DSFGs, e.g.,][]{Viero2013,Weiss2013,Casey2014,bethermin2015,Strandet2016,Casey2017,Reuter2020} have contributed notably to the cosmic SFR, making them crucial to the general comprehension of galaxy evolution. These dust-rich galaxies have managed to stack up their stellar masses in relatively short timescales, while efficiently depleting their gas reservoirs. Therefore, they might be the direct progenitors to ultramassive passive red galaxies that are often encountered at high redshift \citep[e.g.,][]{Daddi2005,Whitaker2013,Nayyeri2014,Toft2014,Carnall2020}.\smallbreak

Our understanding of the nature of DSFGs have benefited from the plethora of multiwavelength photometry over the last two decades. With a decade worth of observation programs of the Atacama Large Millimeter/submillimeter Array (ALMA), it has become possible to enrich our datasets of infrared (IR) galaxies at higher redshifts, by their multiplicity from their blended, lower resolution detections with \emph{Spitzer} and \emph{Herschel}. This has made modeling dust attenuation at high redshifts a hot topic \citep[e.g.,][]{Popping2017,Wang2017,McLure2018,buat2019,Salim2019,Fudamoto2020}. 

Interstellar dust is highly efficient in absorbing the short wavelength photons, predominantly originating from young UV-bright massive stars, rendering the star-forming cold regions of the DSFGs virtually inaccessible. This attenuated light can be successfully reproduced by assuming a dust attenuation law \citep[e.g.,][]{Burgarella2005,Buat12,Buat14,LoFaro2017,salim2018,Salim2020}. In reality however, a single attenuation law cannot mimic dust extinction in the interstellar medium (ISM) of a large and diverse sample of galaxies \citep[e.g.,][]{Wild2011,Kriek2013,Buat2018,Malek2018,Salim2020}. Different approaches appear to work when modeling dust attenuation in reproducing spectral energy distributions (SEDs) of galaxies. \citet{Calzetti2000} derived a universal effective attenuation law which is used as a screen model, by measuring the extinction in local starburst galaxies. The attenuation curve of \citet{Calzetti2000} succeeds in modeling dust reddening even at high-redshift metal-poor galaxies with bright cold dust component. The double component dust attenuation law of \citet{CharlotFall2000} assumes a more complex, physical mixing of dust and stars. With this approach, newly formed stars are placed in the cold molecular clouds, and experience double attenuation by dust of the molecular clouds and the ISM. Older stars are attenuated by the dust grains of the ISM alone.\\
These attenuation laws, together with their different recipes, are often used in the literature when modeling the SED of galaxies. Such recipes include a steeper curve than that of \citet{Calzetti2000} with a UV bump around 0.217$\mu$m \citep{buat11,Buat12}, and a shallower ISM attenuation than that of \citet{CharlotFall2000} \citep[e.g.,][]{LoFaro2017}. However, when it comes to DSFGs, a proper dust attenuation curve that is able to mimic accurately the absorbed photons is crucial for any SED analysis, especially that in these objects, as dust plays a massive role in their evolution, and is an important agent in their SEDs.\smallbreak

On the other side of the SED, the long-wavelength cutoff of the Rayleigh–Jeans approximated tail is extensively covered by ALMA. Therefore, its cold dust continuum provides a vital element in the far-infrared (FIR) SEDs of DSFGs. However, high-resolution ALMA-detected cold dust emission maps were often found to disagree spatially with detections at shorter wavelengths, such as the UV-emitting star-forming regions or the stellar populations of DSFGs \citep[e.g.,][]{Dunlop2017,Elbaz2018,buat2019}. This disagreement can be either in a form of different dust compactness relative to the higher energy detections, or in some cases a complete physical dissociation of these two components. Such offsets are often found to be more significant than the systematic offsets that arise from instrumental uncertainties or large beam sizes \citep{Faisst2020}. In most cases, radio detections with the Karl G. Jansky Very Large Array (VLA) confirmed the ALMA-detected physical dissociation \citep[e.g.,][]{Rujopakarn16,Dunlop2017,Elbaz2018,Hamed21}. This separation challenges a local energy balance which could be an issue for SED modeling that takes into account a global conservation of energy. This problem was highlighted lately in \citet{buat2019}, since such energetic balance is the core of widely used SED fitting tools \citep[e.g.,][]{daCunha2008,Noll2009,Boquien2019}.\smallbreak

Reverse engineering the spectral distribution of DSFGs does not come without obstacles, despite the advent in current understanding of the physical and chemical processes that such galaxies undergo. The most commonly confronted obstacle in SED fitting is the degeneracy problem. Such degeneracy arises from overlapping different physical contribution in one specific wavelength domain, such as the dust and stellar age degeneracy \citep[e.g.,][]{Hirashita2017}. Although this can be overcome by assuming an energetic balance and using the FIR emission as an additional constrain for dust attenuation, and by choosing an appropriate star formation history (SFH) \citep[see][]{Ciesla2016}, a well-constrained attenuation curve will help limit such degeneracies, leading to better estimated physical parameters.

Dust attenuation curves significantly alter the stellar mass determination of galaxies in general and of DSFGs in particular. As flatter and geometrically complex attenuation curves can dim the light coming from the older stellar populations in the ISM more efficiently than steeper ones, they naturally result in a significant hidden older stellar population in the optical to near-infrared (NIR) range. This, along with other assumptions such as the initial mass function (IMF) and the SFH, lead to large differences in the resulting stellar masses \citep[e.g.,][]{Zeimann2015,Malek2018,buat2019}.

The uncertainty in stellar mass determination of DSFGs \citep[see the stellar mass controversy in][]{Hainline2011,Michalowski2012,Casey2014}, remarkably influences the position of these objects along the commonly named main-sequence (MS) of star-forming galaxies \citep[e.g.,][]{Brinchmann2004,Noeske2007,Daddi2010,Rodighiero2011,Lofaro15,Schreiber15, Hamed21}. Most galaxies seem to follow the tight scatter of the MS independently of the redshift, whose outliers are typically referred to as starbursts. However, the strong dependence of the stellar mass estimation from SED fitting techniques on the assumed attenuation curve, varies massively this scatter and affects the "starburstiness" of already active DSFGs. It is therefore crucial to choose suitable attenuation laws in order to limit biases on the stellar mass determination.

Recently, \citet[]{Donevski2020} investigated the dust and gas contents of a large sample of DSFGs, linking dust abundance to other physical characteristics such as their SFRs. Despite the growing understanding of these objects, we still lack a complete picture of how dust attenuates their stellar radiation. Properly quantifying dust attenuation of DSFGs is crucial for quantifying the cosmic SFR and get a better grip on galaxy evolution. To achieve this goal, we study closely the dust attenuation curves in DSFGs and investigate the possible physical properties that they might depend on. With the ever-growing understanding of the nature of these dusty galaxies, in recent years many works studied the relation between dust attenuation and other physical properties \citep[e.g.,][]{Fudamoto2020,Lin2021,Lower2021,Boquien2022}. In these studies, links between dust attenuation properties and other physical observables were studied, such as the dust grain sizes, star-dust geometry, and the star formation activity in these galaxies. The most widely used attenuation laws are that of \citet{Calzetti2000} and \citet{CharlotFall2000}. Both of these laws are used interchangebly when modeling galaxies' photometry at high redshifts. However, we still lack a complete knowledge of the preference of attenuation laws in DSFGs at different redshift range.\\
In this paper, we aim at answering the question of which of the commonly used laws best describes dust attenuation in DSFGs at high redshift. Measuring the slopes of attenuation is beyond the scope of this paper. We make use of a large statistical sample with available dust continuum maps and their UV/optical counterparts. We study the effect of UV/optical to dust continuum compactness on the preferred attenuation law for our sample.\\
\\
This paper is structured as follows: in Section \ref{sample_selection} we describe the data analyzed in this work, both the photometry and the images. In Section \ref{galfit} we detail the method we use to estimate the circularized effective radii of our sources, and Section \ref{sed} provides a description of the SED fitting procedures we used to achieve the physical properties of our sources.\\
The results and their respective discussions are presented in Section \ref{results}, and the summary is concluding this paper in Section \ref{summary}. Throughout this paper, we adopt the stellar IMF of \citet{Chabrier2003} and a $\Lambda$CDM cosmology parameters (WMAP7, \citealp{Komatsu2011}): H$_0$ = 70.4 km s$^{-1}$ Mpc$^{-1}$, $\Omega_{M}$ = 0.272, and $\Omega_{\Lambda}$ = 0.728.

\section{Sample selection}\label{sample_selection}
The large two square degree COSMOS field, centered on R.A., Dec. = (10h00m27.9171s, +02d12m35.0315s) \citep{Scoville2007,Ilbert2013,Laigle2016}, has been observed with an unmatched commitment from different instruments, covering a wide range of wavelength observation of galaxies up to redshift of six. This unique survey design offers rich data sets of millions of identified galaxies, allowing deep investigation of galaxy evolution at various redshifts.\\

The choice of COSMOS field galaxies in this work is motivated by the abundance of multiwavelength data spanning across a wide range of redshifts, and  the significant number of ALMA detections that this field enjoys. This set of optical, infrared and submillimeter detections allows us to build a statistical sample of DSFGs. This sample is ideal to investigate the evolution of DSFGs properties and the crucial role that dust attenuation plays in their evolution.

\begin{table}[]
\tiny
 \caption{Summary of available photometric data in each band with its centered wavelength, the mean of S:N, and the number of detections in our sample. The detections of different ALMA bands (6 and 7) concern different galaxies.}
  \begin{center}
  \begin{tabular}{|c | c c c c|}
         \hline
     \  Telescope/  \  &\ Band\  & \ $\lambda$    \   & Median  \     &  \   N\textsuperscript{\underline{o}} of \  \\
 \  Instrument \    &  \ \ &  \ $(\mu m)$  \    & \ S:N  \    &  \  detections\  \\

    \hline     \hline

      CFHT/MegaCam             & $u$   & 0.38 & 25.13 &  107\\
                & $g$   & 0.49 & 32.56 &  117\\
                    & $r$  & 0.62 & 66.10 & 122\\
                  & $i$   & 0.75 & 58.59 &  122\\
                     & $z$   & 0.89 & 38.01 & 122\\
     \hline
     Subaru/Suprime-Cam            & B   & 0.437 & 28.00 &  122\\
             & V   & 0.544 & 28.38 &  117\\
                      & Y   & 0.98 & 25.61  &  122\\
    \hline
    VISTA                  & J   & 1.25 & 81.01 &  122\\
                     & H   & 1.64 & 93.16 &  122\\
                      & Ks  & 2.15 & 100.82 &  122\\
    \hline
   Spitzer/IRAC             & ch1 & 3.56 & 146.06 &  122\\
                  & ch2 & 4.50 & 157.26 &  122\\
                     & ch3 & 5.74 & 12.97  &  122\\
                      & ch4 & 7.93 & 10.12   &  122\\
    \hline
    Spitzer/MIPS        & MIPS1 & 23.84 & 47.76 &  122\\
     \hline
    Herschel/PACS           & 100 $\mu m$ & 102.61 & 5.41 &  122\\
                    & 160 $\mu m$ & 167.13 & 2.68 &  122\\
    \hline
    Herschel/SPIRE           & 250 $\mu m$ & 251.50 & 12.62 &  122\\
                   & 350 $\mu m$ & 352.83 & 7.19  &  122\\
                       & 500 $\mu m$ & 511.60 & 3.63 &  122\\
    \hline

                    & 7   & 947     & 12.47  & 85\\
      ALMA                 & 6   & 1255    & 12.89 & 37\\
                       &    &     &    & total = 122\\
     \hline
  JVLA            & S (3\ GHz)   & 1.3$\times 10^5$  & 14.10 &122\\
    \hline

     \end{tabular}
    
     \end{center}
     \label{tab:Table1}
\end{table}

\subsection{ALMA data}
Since the main science objective in this study is to quantify the effect of the distribution of dust emission relative to the stellar continuum, at different redshift ranges, the core of our sample was built around ALMA detections. For that we used ALMA fluxes and continuum maps from the A$^3$COSMOS automated ALMA data mining in the COSMOS field \citep{Liu2019}. This data set assembles hundreds of identified galaxies from the ALMA archive into a single catalogue. In our work we use the primary-beam-corrected ALMA maps. 

 The main advantage in our work is having access to dust continuum morphology relative to the spatial distribution of the star formation region and the stellar population in our sample. For this reason, we carefully select a sample of galaxies characterized by good quality detections.
 
 For submillimeter images, we use the A$^3$COSMOS generated maps \citep{Liu2019}. These images were deconvolved using a robust cleaning with a Brigg's parameter of 2, i.e. a natural weighting of visibilities. This results in significantly better signal-to-noise (S:N) of the innermost regions of a source in the uv plane, thus defining its outermost "borders" in the image plane (see section 2.1 in \citealt{Liu2019} for a more detailed description of produced ALMA continuum images). To study dust attenuation through cosmic time, we select ALMA-detected galaxies with S:N higher than 5. Our preliminary sample is composed of 1,335 ALMA detected individual galaxies. \smallbreak
 
\subsection{Ancillary maps}
For the shorter wavelength (rest-frame UV and optical) images, we used the third data release of the deep field continuum maps detected with the $y$ band of the Hyper Suprime-Cam (HSC) of \emph{Subaru} \citep{Miyazaki2018,Aihara22}. These images have high angular resolution (0.64$\arcsec$) which allows a physical comparison with their ALMA counterparts.

\begin{figure}[h!]
    \centering
        \includegraphics[width=0.5\textwidth]{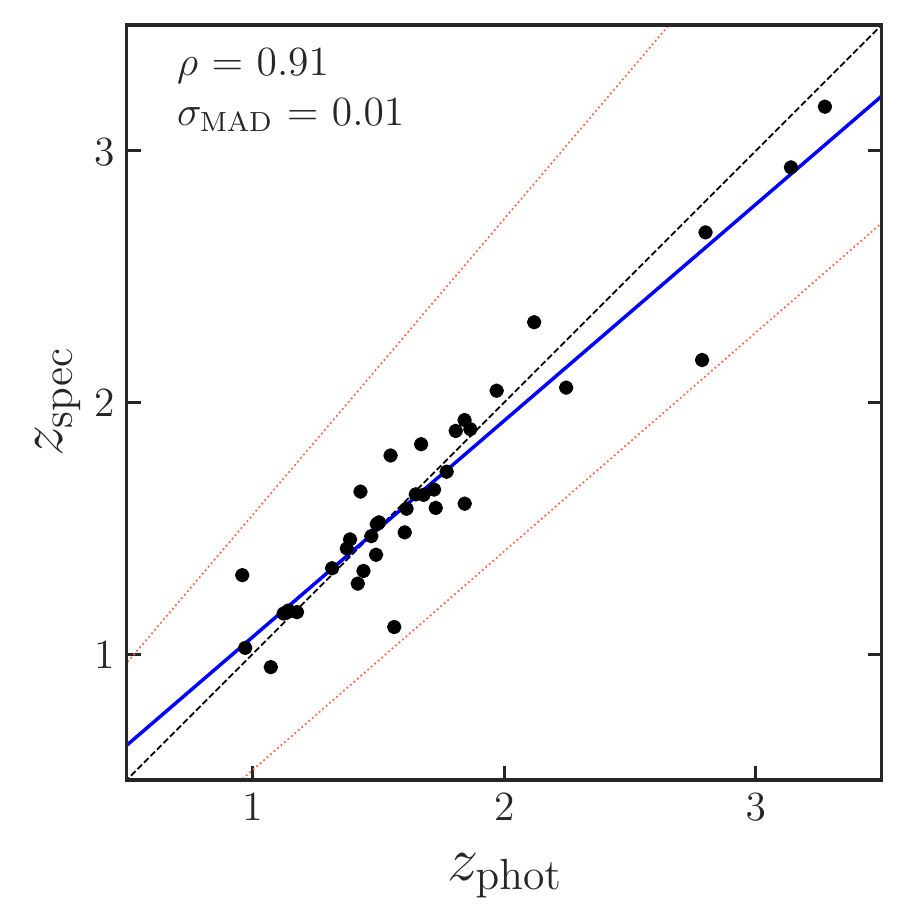}
        \caption{Comparison between the  photometric redshifts of 43 galaxies from the final sample, for which spectroscopic redshifts are available from the literature \citep[][]{Liu2019}. Spearman's rank correlation coefficient is shown as $\rho$, with the photometric redshift accuracy given by the $\sigma_{MAD}$ \citep[][]{Ilbert2009}. Red dotted lines correspond to $z_{p} = z_{s}\pm0.15(1 + z_{s})$ \citep{Ilbert2009}.}
        \label{fig:zspec}
\end{figure}

\subsection{Auxiliary photometric data}

We used the photometric data from the \emph{Herschel} Extragalactic Legacy Project (HELP) panchromatic catalogue \citep{Shirley2019}. This catalogue was achieved taking into account visible to mid-infrared (MIR) range surveys as a prior, homogenizing them and extracting fluxes from \emph{Herschel} maps whose angular resolution is significantly lower than their short wavelength counterparts.\\
To build the HELP catalogue, \emph{Herschel} fluxes were extracted using the probabilistic deblender XID+ \citep{Hurley2017}, which was used on SPIRE maps, taking into into account the positions of sources observed with the high-resolution detections from \emph{Spitzer} at~24$\mu m$. This technique was shown to increase the accuracy of photometric redshift estimations \citep{Duncan2018}.\\

We perform a positional cross-match between ALMA catalogue and the HELP catalogue with a rather conservative 1$\arcsec$ search radius. Although galaxies whose dust continuum or molecular gas emission is significantly dissociated from the shorter wavelength continua are not uncommon \citep[e.g.,][]{Elbaz2018,Hamed21}, due to various factors such as astrometry problems, positional errors as a result of the beam size, or otherwise physical factors, this timid search radius avoids false matches \citep[e.g.,][]{buat2019,Liu2019}. This cross-matching procedure resulted in 383 individual galaxies ranging from z$=$0.3 to z$=$5.5.

To better constrain the physical properties of our statistical sample through SED modeling, we discarded galaxies that have less than three detections in the UV-NIR wave range. This requirement rejected 30$\%$ of the sources. Additionally, we required minimum six detections in the MIR-FIR bands (8--1000 $\mu m$) out of which at least five detections having a S:N$>$3. As a result of these selection criteria, the finally-selected sources have at least 10 detections in the UV to NIR (0.3--8 $\mu$m) range with S:N$>$5.  For objects that had detections in similar bandpass filters, we used the detections coming from the deeper survey. This has a high importance in SED fitting procedures especially because measurements at similar wavelengths may be order of magnitudes different from each other, which will negatively affect the quality of their fitted spectrum. Moreover, dense coverage of a very short part of the SED could add too much weight during the SED fitting process and bias the final fit \citep{Malek2018}. Our UV--NIR photometric data as well as the FIR counterparts have overall high S:N (mean of 60 for the former and 13.23 for the latter), while in the IR side of the spectrum, MIPS measurements at 24~$\mu m$, as well as SPIRE detections at 250 and 350~$\mu$m, and evidently all ALMA detections have high S:N (mean of 12.68 for all bands). All \emph{Herschel}'s SPIRE fluxes (at 250, 350 and 500~$\mu m$) are essential since they cover the thermal part of the total SED up to $z \sim 4$, which contains the Rayleigh-Jeans dust emission tail. To even better constrain the IR--submilimiter part of the SED fits, we appended our sources with VLA detections at 3 GHz from \citet[][]{Smolcic2017}. 

\subsection{Final sample}
The above described selection yields in the final sample of 122 galaxies with panchromatically high S:N, covering a redshift range of $1 < z < 4$. Forty-three galaxies of our sample have spectroscopic redshifts from \citet[][]{Liu2019}, and for the rest of the sample we use the reliable photometric redshifts provided by the HELP catalogue. Figure \ref{fig:zspec} shows a comparison between the spectroscopic and photometric redshifts of the galaxies in our sample that possess both measurements. We calculate the photometric redshift accuracy \citep{Ilbert2009} as $\sigma_{MAD} = 1.48 \times \mathrm{median}(|z_{p} - z_{s}|/(1 + z_{s}))$ where MAD is the mean absolute deviation. This redshift accuracy resulted in reliable photometric redshifts of our sample with $\sigma_{MAD} = 0.014$.\\

Table \ref{tab:Table1} shows the photometric bands and the associated S:N for the final sample of 122 DSFGs. Almost half of the final sample, 60 galaxies, have a S:N of $Y$ band detection higher than 5, and all of the sources had a VLA detection. We want to stress that only 15 DSFGs from our sample (12$\%$) do not have $u$ band detection, and we miss five detections in $g$ and V bands. With the exception of those galaxies, the rest of the sample have full set of 23 photometric bands, assuring excellent spectral coverage, essential for a  detailed  SED fitting.


\section{Methods}\label{section3}
\subsection{Size measurements}\label{galfit}
To study the spatial extent of dust emission and that of the stellar populations and the star-forming regions in our sample, we derived homogeneous effective radii (R$_{e}$) of the dust continuum maps and their short (UV-optical) wavelength counterparts. To achieve this, we used \texttt{GALFIT} \citep{Peng2002}, parametrically fitting two-dimensional S\'ersic profiles in the primary-beam-corrected images of our sample. With the S\'ersic index ($n^{\mathrm{S\acute{e}rsic}}$), obtained from the fitting procedure, it is possible to quantify the concentration of light in a galaxy, which can provide important information about its morphology. Moreover, \texttt{GALFIT} provides the values of the ratio between the minor and major axis, and this allows for calculating the effective circularized radii (hereafter simply effective radii or R$_{e}$). With spectroscopic and reliable photometric redshift information, we can convert the real sizes of our objects. In this work, we analyze the evolution of IR radii between redshift four and one, but also simultaneously changes in the rest-frame UV-optical.\\

For our analysis we used the $Y$ band of HSC, and ALMA (bands 6 and 7) images to compute physical sizes of our final sample of DSFGs. The mean angular resolution of the ALMA detections in our sample is 0.75$\arcsec$, and equivalently 0.64$\arcsec$ for the HSC $Y$ band detection. This renders valid a comparison between sizes estimated from each of these bands directly, without the need to degrade them on resolution.\\ 

\begin{figure*}[h]
    \centering
        \includegraphics[width=1.0\textwidth]{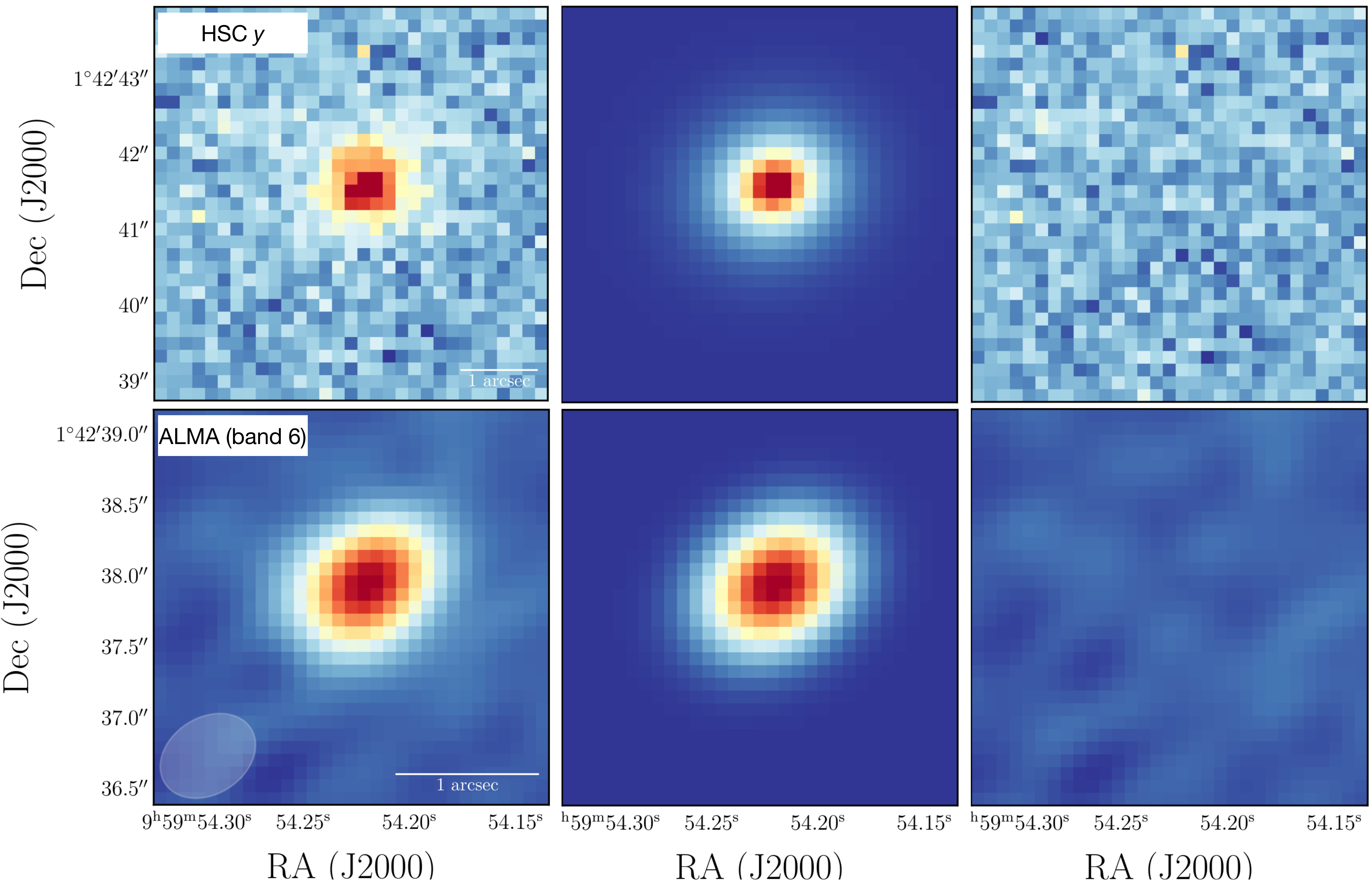}
        \caption{Example of \texttt{GALFIT} fitting of a galaxy from our sample \mbox{(HELP-J095953.305+014250.922)} with its modeled light profiles (second column) and its subsequent residuals (third  column), at two different wavelengths from top to bottom as follows \textit{Top:} HSC-$Y$ band detection. \textit{Middle:} ALMA band 6 detection. The left side of each row shows the scientific primary-beam-corrected image, with the beam size and the corresponding 1 arcsecond scale.}

        \label{fig:GALFIT_mosaique}
\end{figure*}
In computing the S\'ersic profiles, we adopted a similar approach than \citet{Elbaz2018}, by leaving ($n^{\mathrm{S\acute{e}rsic}}$) free. For the sake of comparison, we also fixed $n^{\mathrm{S\acute{e}rsic}}=1$ \citep[e.g.,][]{Hodge2016,Elbaz2018}. While 5$\%$ of galaxies in our sample did not reach convergence when fixing $n^{\mathrm{S\acute{e}rsic}}$, $R_{e}$ was found to be in good agreement in both cases (on average 7$\%$ larger when fixing $n^\mathrm{S\acute{e}rsic}$ for ALMA detections), and we found an agreement within 23$\%$ in resulted $n^{\mathrm{S\acute{e}rsic}}$ with a fixed and a free $n^{\mathrm{S\acute{e}rsic}}$.\\

In our profile fittings, we initially used an automated computation with \texttt{GALFIT}, and carefully checked the resulted models and their residuals. This was performed to get a general understanding of our sample's range of effective radii in each of the two used wavelength domains. This allowed us to test different input values of $n^{\mathrm{S\acute{e}rsic}}$, including a Gaussian model. This is an important step when using this tool, since it permits validating the initial parameters needed for \texttt{GALFIT}. We found that varying $n^{\mathrm{S\acute{e}rsic}}$  between an exponential disk profile ($n^{\mathrm{S\acute{e}rsic}} = 1$) and a Gaussian profile ($n^{\mathrm{S\acute{e}rsic}} = 0.5$) leads to a slight change in the models and their consequent effective radii, with R$_{e}$ $_{ALMA}^{n=0.5}$ being $26\%$ smaller than R$_{e}$ $_{ALMA}^{n=1}$.\\

After the aforementioned tests, we individually fitted S\'ersic profiles to each of our sources in the two bands (UV/optical and IR), while monitoring the resulted models, the residuals and the profile parameters. This individual fitting was especially required for galaxies that did not fit in the automated computation ($\sim$ 5$\%$ of the total sample), in which case a simple and slight parameter adaptation managed to fit these sources. The distribution of the computed $n^{\mathrm{S\acute{e}rsic}}$ was rather narrow, ranging between 0.4 and 1.6 for ALMA detections, and between 0.5 and 1.4 for the $Y$ band.\\

Our technique in fitting S\'ersic profiles in two bands of our sample was performed homogeneously in the same methodical approach. This is important to accurately quantify ratios of different R$_{e}$ at varying wavelengths. Our primary effort was to calculate our effective circularized radii in a homogenized method - with the same tool and approach, which reduces possible biases for the final physical interpretation. To test the reliability of our size measurements, we also computed the minimum possible size that can be accurately measured using the formula by
\citet{visal12} and \citet{Gomez}:
\[\theta_{min}=0.88\times \frac{\theta_{beam}}{\sqrt{S/N}}\]
where the minimum size for each source ($\theta_{min}$) is in units of the synthesized beam FWHM ($\theta_{beam}$), depending on the S:N of the source. All size measurements of our sample were above that limit.\\

Figure \ref{fig:GALFIT_mosaique} shows an example of a galaxy \texttt{HELP-J095953.305+014250.922} at redshift z=2.15 seen in the HSC $Y$ and ALMA band 6.   Figure \ref{fig:GALFIT_mosaique} shows the original image of that galaxy at two different wavelengths, with their light profiles fitted with \texttt{GALFIT}, and their subsequent residuals. The median radii and their errors (the median absolute deviation) at different redshift ranges are presented in Table \ref{tab:radii}, and the redshift evolution of these derived sizes along with the evolution of their star-to-dust compactness will be shown in Figure \ref{fig:radii} (left panel). This Figure shows the change in the radii sizes (also related to the sample selection of DSFGs) of ALMA and HSC $y$ as well as a comparison with similar work done by \citealt{buat2019} based on galaxies at z$\sim$2. The evolution of these derived radii with the dust luminosity and stellar mass of our sample are shown in Appendix \ref{appendix:a2}.

\begin{table}[]
 \caption{Summary of the derived effective radii of our sample from the available detections at two different wavelengths.}
  \begin{center}
  \begin{tabular}{|c c c|}
     \hline
     
   \  \ \ \ \ redshift  \ \ \ \ \ &\ \ \ \ \ $R_{e}^{ALMA}$ [kpc]\ \ \ \ \ &\ \ \ \ \ $R_{e}^{UV-opt}$ [kpc] \ \ \ \ \   \\
    \hline \hline
     $1<z<2$ & 2.59$\pm$0.25 & 5.75$\pm$0.18 \\
     \hline
    $2<z<3$ & 2.12$\pm$0.32 & 5.03$\pm$0.22 \\
      \hline
    $3<z<4$ & 2.54$\pm$0.15 & 3.54$\pm$0.26 \\

    \hline

     \end{tabular}
    
     \end{center}
     \label{tab:radii}
\end{table}

\subsection{SED fitting method}\label{sed}

\begin{figure*}[h]
    \centering
    \minipage{0.5\textwidth}
    \includegraphics[width=\linewidth]{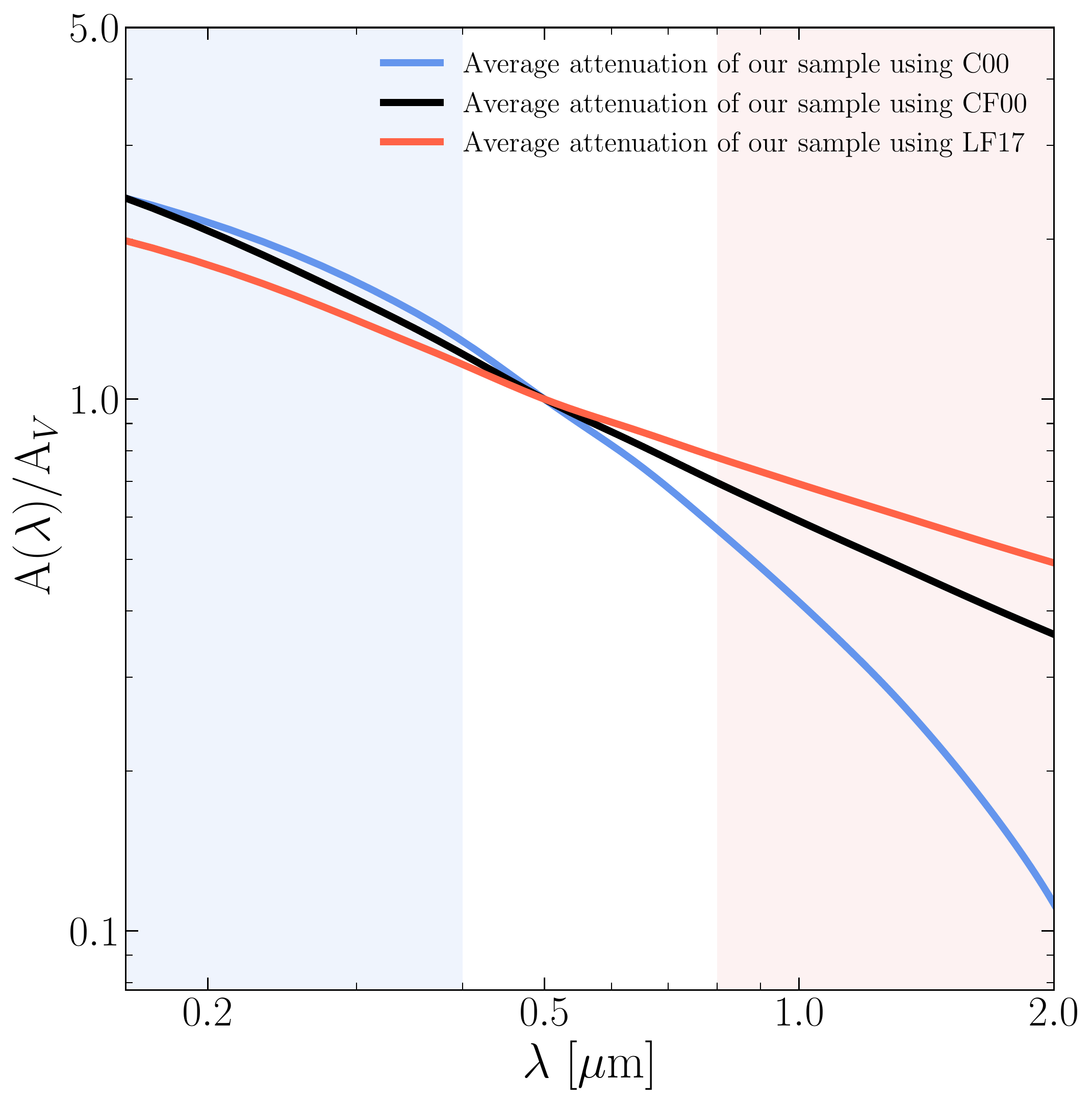}
    \endminipage
    \minipage{0.5\textwidth}
    \includegraphics[width=9cm]{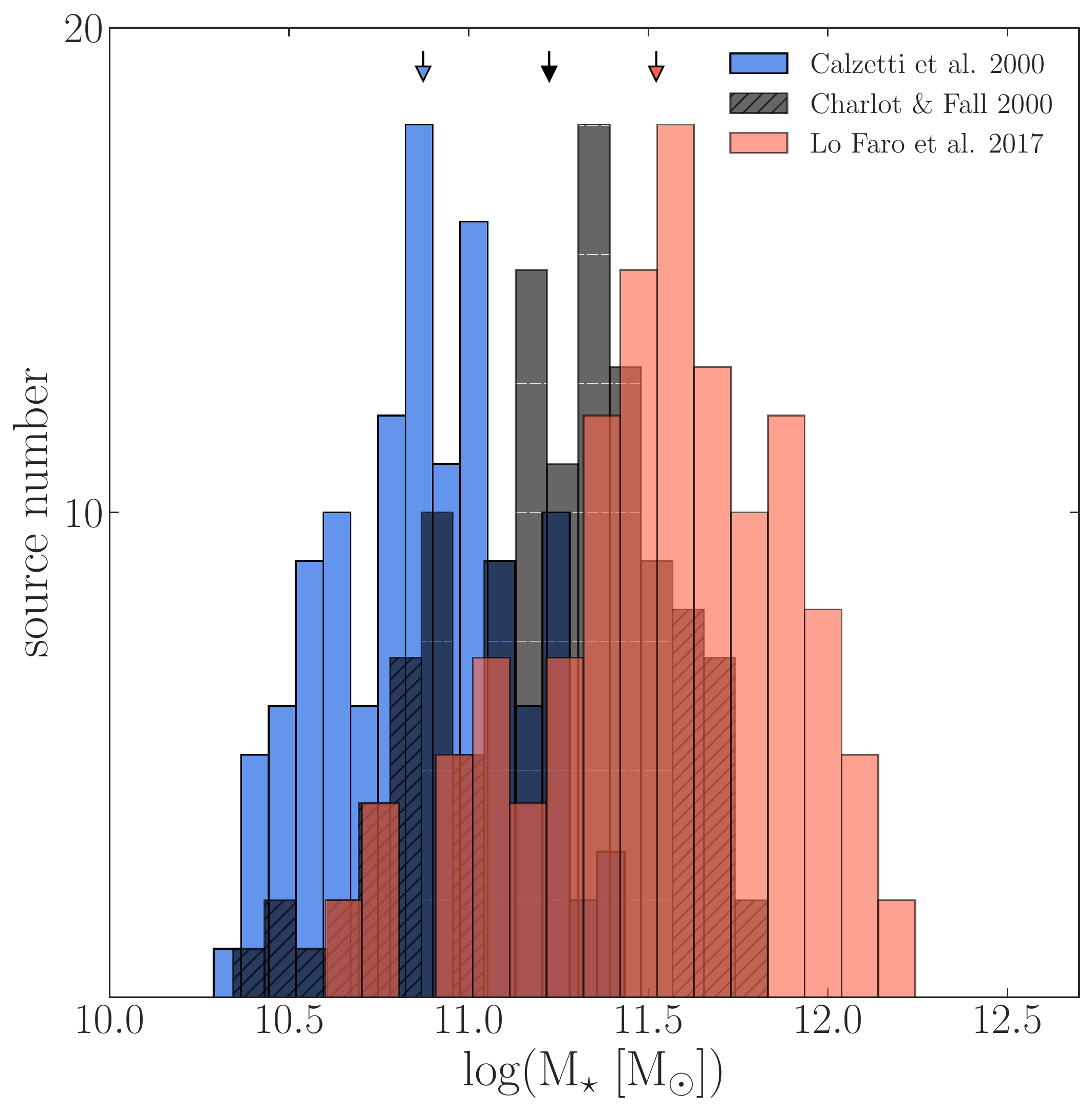}
    \endminipage
    \caption{\emph{Left panel:}The average attenuation curves of our sample using three different attenuation laws/recipes \citep[][]{Calzetti2000,CharlotFall2000,LoFaro2017}. In the case of CF00 and LF17 curves, we are showing the combined curve of the ISM and the BC. \emph{Right panel:} Stellar masses of our sample derived using the three attenuation curves as in the left panel. The arrows show the mean of the specific distribution.}
    \label{fig:attenuation_mass}
\end{figure*}

To derive the physical properties of our well-constrained multiwavelength sample, we use the Code Investigating GALaxy Emission \texttt{CIGALE}\footnote{https://cigale.lam.fr/}, an energy balance technique of the SED fitting \citep{Noll2009,Boquien2019}. This technique of SED fitting takes into account the balance between the energy absorbed in the rest-frame UV-NIR part of the total galaxy emission, and its rest-frame IR emission. The mediator agent in this energy balance is the dust, since it will absorb a significant part of the short wavelength photons emitted by the stars, and emanate in the form of thermal emission in the FIR.

Reverse-engineering the total spectrum of a galaxy is not an easy quest. Some physical processes are completely unrelated, such as the synchrotron emission of accelerated electrons, which dominates the radio part of the SED, and the UV photons whose origin is traced directly to the young stars. However, some physical processes release photons of the same frequency range, such as the MIR range which can have different contributors like Active Galactic Nuclei (AGN) and polycyclic aromatic hydrocarbons (PAHs), resulting in degeneracies. Therefore, carefully choosing physically-motivated templates and parameters is crucial in order to deduce key physical properties that galaxies are experiencing, since these parameters depend on the assumptions made \citep[e.g.,][]{Ciesla15,Leja2018, Carnall2018}.

\begin{table*}
\caption{Input parameters used to fit the SEDs of our sample.}

  \begin{center}

    \begin{tabular*}{\textwidth}{l@{\extracolsep{\fill}}  r}
\toprule\hline
     Parameter & Values \\\toprule
\hline
    \multicolumn{2}{|c|}{Star formation history}   \\
     \hline    \multicolumn{2}{|c|}{delayed with a recent burst}   \\\hline

     Stellar age \tablefootmark{(a)}  &   redshift-dependent (0.5, 1, 2, 3, 5, 6, 8\,Gyr)  \\
     e-folding time &  1, 3, 6\,Gyr \\
     Age of recent burst &   5, 10, 50, 100, 200, 300\,Myr \\
     Strength of the burst  & 0.001,0.005,0.01,0.2,0.3 \\
\hline\hline
     \multicolumn{2}{|c|}{Dust attenuation laws}   \\
\hline
    \multicolumn{2}{|c|}{\citep{Calzetti2000}}   \\
\hline
    Colour excess of young stars  E(B-V) & 0.1 - 1 by a bin of 0.1 \\
    f$_{att}$\tablefootmark{(b)}  & 0.3, 0.5, 0.8, 1.0\\
\hline
    \multicolumn{2}{|c|}{\citep{CharlotFall2000}, \citep{LoFaro2017}}    \\
\hline
    V-band attenuation in the ISM  A$_V^{ISM}$ &  0.3 - 6 by a bin of 0.1 \\
    Av$_{V}^{ISM}$ / (A$_V^{BC}+A_V^{ISM})$ &  0.3, 0.5, 0.8, 1 \\
    Power law slope of the ISM &-0.7, -0.48\tablefootmark{(c)} \\
    Power law slope of the BC &   -0.7 \\
\hline\hline
 \multicolumn{2}{|c|}{Dust emission}   \\
\hline
    \multicolumn{2}{|c|}{\citep{dl2014}}   \\
    \hline
    Mass fraction of PAH &  1.77, 2.50, 3.19 \\
    Minimum radiation field U$_{min}$ & 10, 25, 30, 40 \\
    Power law slope $\alpha$ & 2 \\
\hline
\hline
    \multicolumn{2}{|c|}{Synchrotron emission}   \\
    \hline
    FIR/radio correlation coefficient &   2.2, 2.4. 2.6\\
    Power law slope slope & 0.3, 0.6, 0.9  \\ \hline
\bottomrule
\tablefoottext{a}{The age of the main stellar population.}\\
\tablefoottext{b}{Color excess of old stars.}\\ 
\tablefoottext{c}{Power law slope of LF17.}\\ 
    \end{tabular*}
    \label{tab:Table2}
   \end{center}

    \end{table*}
In the following subsections, we describe the different aspects of our SED fitting strategy, and motivate our choice of certain laws and parameters. The SED modules description used in our work are presented below with the stellar part and the dust part.
    
\subsubsection{Stellar SED}
\begin{figure*}[h]
    \centering
    \minipage{0.33\textwidth}
    \includegraphics[width=\linewidth]{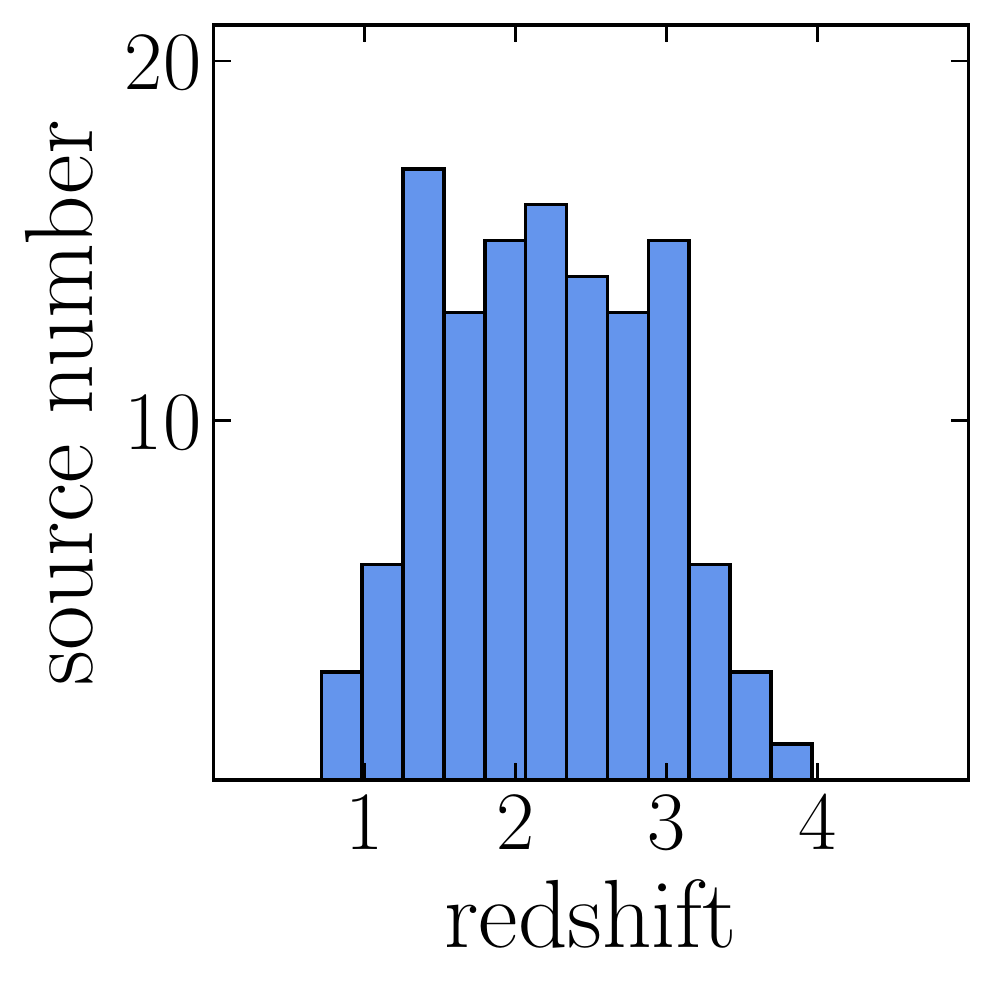}
    \endminipage
    \minipage{0.33\textwidth}
    \includegraphics[width=\linewidth]{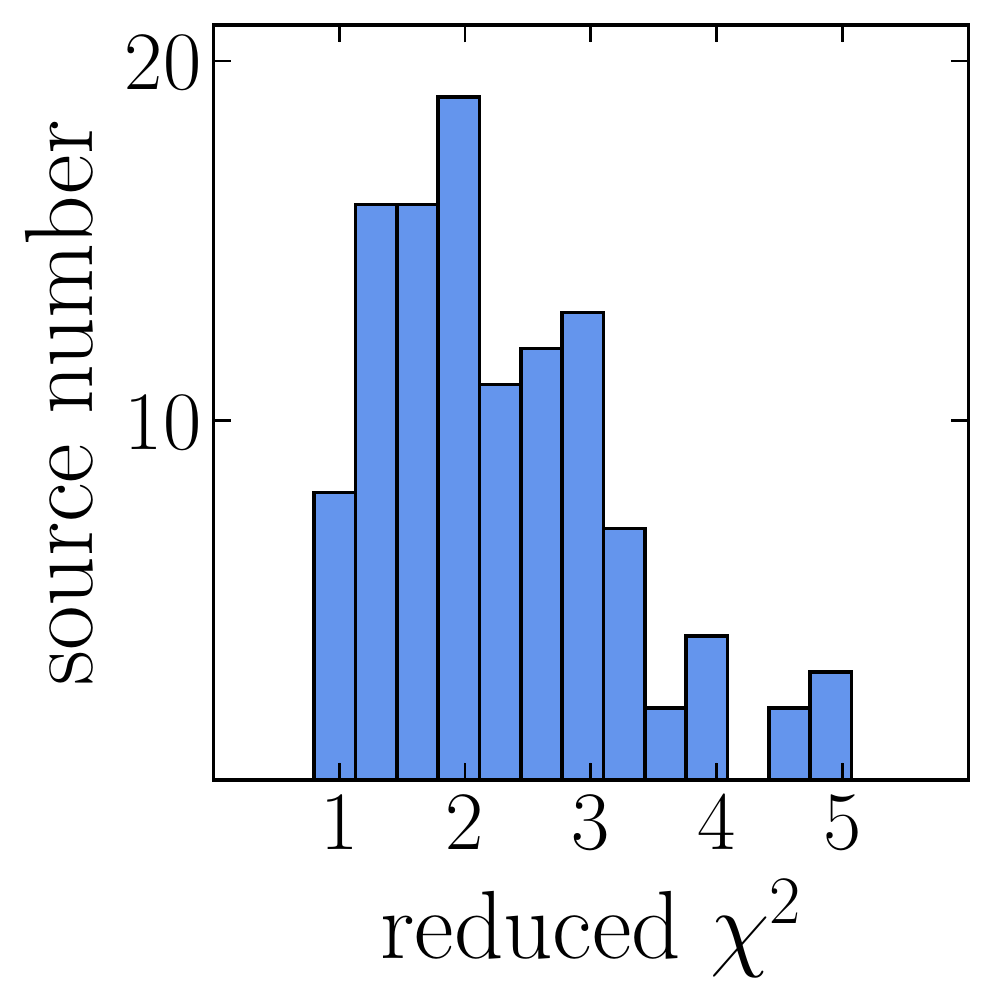}
    \endminipage
    \minipage{0.33\textwidth}
    \includegraphics[width=\linewidth]{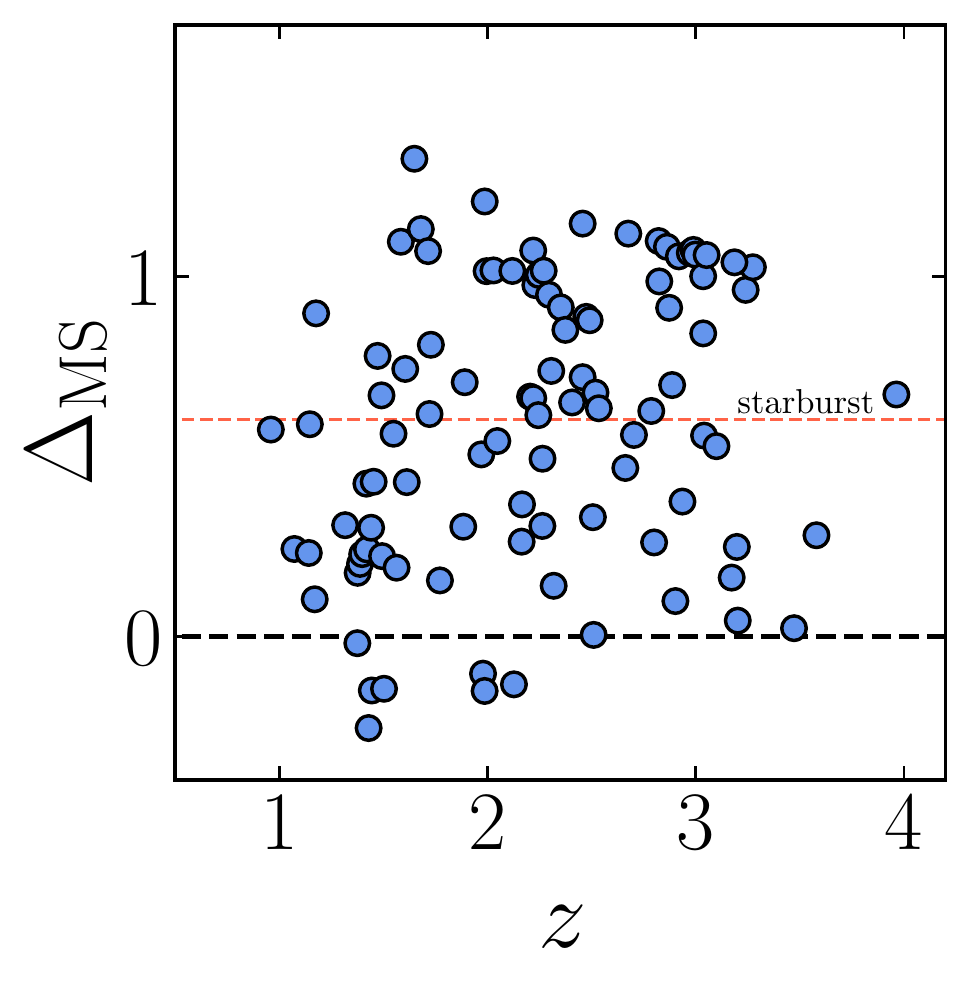}
    \endminipage
    \caption{Left panel: Redshift distribution of our sample of SFGs. Middle panel: Reduced $\chi^2$ of the best fits of the galaxies in our sample using \texttt{CIGALE}. Right panel: Variation of $\Delta$MS (distance to the main sequence of SFGs of \citealt{speagle2014}) of the sample with redshift. The red dotted line shows the area above which a galaxy is considered a starburst.}
    \label{fig:selection}
\end{figure*}
To forge the SED of a galaxy, we first assume a stellar population that is behind its direct and indirect emission. This means that we should take into account a stellar population library and its spectral evolution, stars with different ages and a certain metallicity. In this work, we use the stellar population library of \cite{BC03}, a solar metallicity and an initial mass function (IMF) of \cite{Chabrier2003}, which takes into account a single star IMF as well as a binary star systems.\\

The stellar population models are then convoluted with an assumed star formation history (SFH). SFHs are sensitive to many complex factors including galaxy interaction, merging, gas accumulation and its depletion \citep[e.g.][]{Elbaz2011,Ciesla2018,Schreiber2018,Pearson2019}. The SFHs have a significant effect on fitting the UV part of the SED, and consequently affecting the derived physical parameters such as the stellar masses and the SFRs. \citet{Ciesla2016, Ciesla2017} showed that simple SFH models (such as a delayed model) are not enough to reproduce a precise fit of the UV data, especially for galaxies that are undergoing a starburst or quenching activity. \smallbreak

    To model the SEDs of our IR-bright sample, we use a delayed SFH with a recent exponential burst \citep[e.g.][]{Malek2018,Buat2018,Donevski2020}. This recent burst is motivated by the ALMA detection which makes very likely a numerous populations of young stars manifesting their presence through dust. And in such scenario, a galaxy will build the majority of its stellar population in its earlier evolutionary phase, then the star formation activity slowly decreases over time. This is followed by a recent burst of SFR. The SFR evolution over time is hereby modeled with:
\begin{equation}
    SFR(t)\ \propto\ \frac{t}{\tau^2}\ e^{-t/\tau}\ +\ e^{-t^{'}/\tau^{'}},
\end{equation}
    where the first term translates into a delayed SFH slowed by the factor of $\tau^2$, which is the e-folding time of the main stellar population, and extended over the large part of the age of the galaxy. The second term is the exponential decrease of recent SFR, where $t^{'}$ and $\tau^{'}$ are the age of the burst and the e-folding time of the burst episode respectively. This SFH provided better fits compared to the simpler delayed SFH, especially for non-quiescent galaxies. We vary $\tau$ as shown in Table \ref{tab:Table2}, to give a comprehensive flexibility of the delayed formation of the main stellar population. We discuss the choice of this SFH over a truncated version in Appendix \ref{appendix:b}.

\subsubsection{Dust SED}
\begin{figure}[htb!]
    \centering
    \minipage{0.5\textwidth}
    \includegraphics[width=\linewidth]{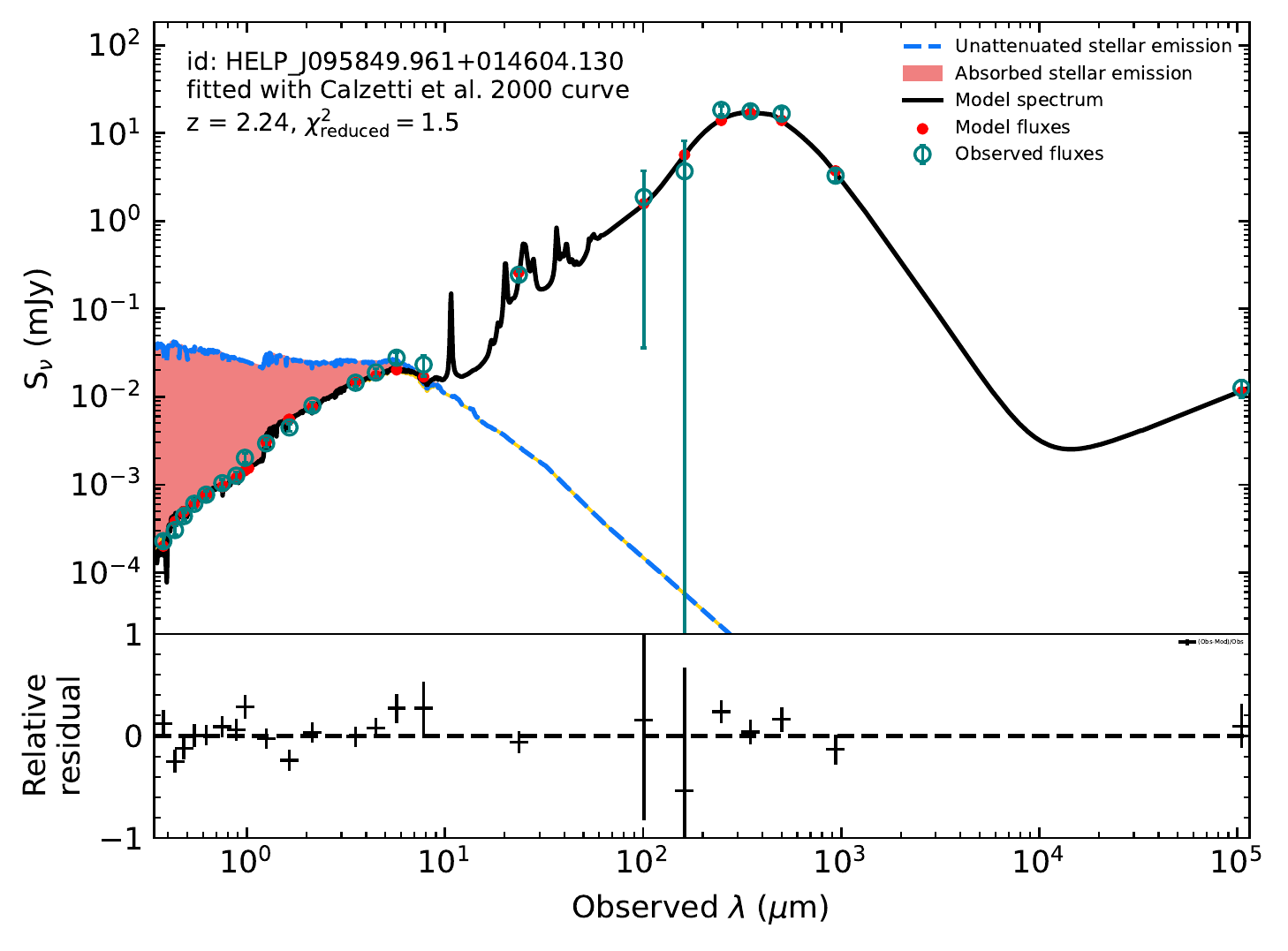}
    \endminipage\\
    \minipage{0.5\textwidth}
    \includegraphics[width=\linewidth]{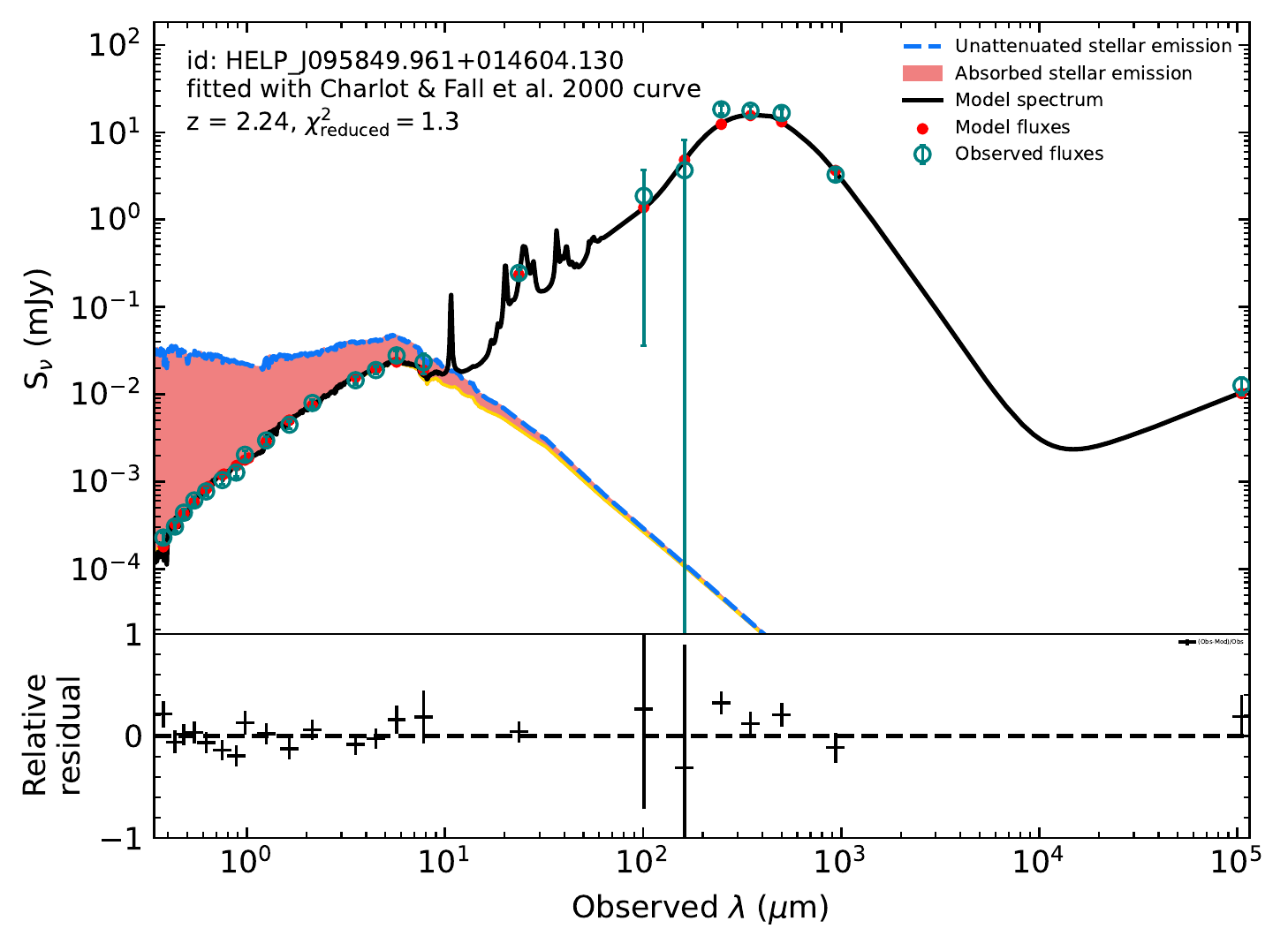}
    \endminipage\\
    \minipage{0.5\textwidth}
    \includegraphics[width=\linewidth]{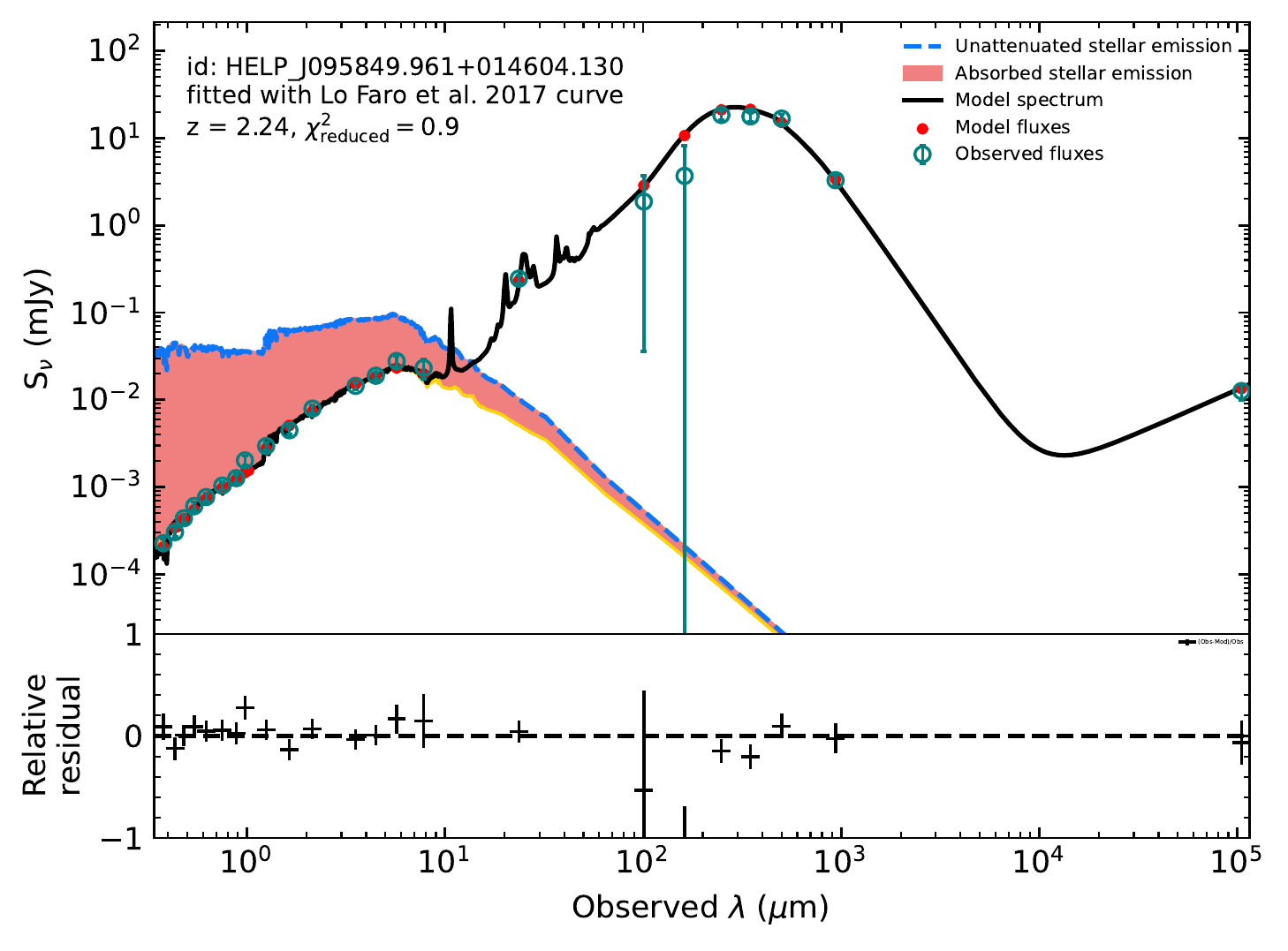}
    \endminipage\\
    \caption{Computed SEDs of a galaxy from our sample $\mathrm{(HELP-J095849.961+014604.130)}$, fitted with three different attenuation laws, from left to right: \citet[][]{Calzetti2000}, \citet[][]{CharlotFall2000}, and \citet[]{LoFaro2017}. The $\chi^2$ of every fit is shown on the figures. Red filled area represents the attenuated stellar light.}
    \label{fig:SED}
\end{figure}
The dust content of our sample of DSFGs is presumed to be the driving component of the shape of their SEDs. This makes the modeling of dust attenuation important to extract accurate physical properties.\smallbreak
To model the effect of dust, we use two different attenuation laws for our SED fitting: the approach of \citet[][henceforth C00]{Calzetti2000} and that of \citet[][henceforth CF00]{CharlotFall2000}. While these two attenuation laws are relatively simple, they differ on how to attenuate a given stellar population.

The attenuation curve of C00 was tuned to fit a sample of starbursts in the local Universe. This curve attenuates a stellar population assuming a screen model: 
\begin{equation}
k(\lambda) = \frac{A(\lambda)}{E(B-V)},
\end{equation}
where k($\lambda$) is the attenuation curve at a given wavelength $\lambda$, A($\lambda$) is the extinction curve, and E(B-V) is the color excess, which is the difference between the observed B-V color index and the intrinsic value for a given population of stars.

Despite its simplicity, this attenuation curve, with its modifications, is widely used in the literature \citep[e.g.][]{Burgarella2005,Buat12,Malek2014,Malek2017,Pearson2017,Elbaz2018,Buat2018,Ciesla2020}. However, it does not always succeed in reproducing the UV extinction of galaxies at higher redshifts \citep{Noll2009,LoFaro2017,buat2019}.\smallbreak
Another approach is to also consider dust present in birth clouds. This is the core of the attenuation curve of CF00. In this approach, dust is considered to attenuate the dense and cooler molecular clouds (hereafter MCs) differently than ambient diffuse interstellar media (ISM). This configuration is expressed by the following analytical expression:
\begin{equation}
A(\lambda)_{\text{ISM}} \propto \left(\dfrac{\lambda}{\lambda_{V}}\right)^{\delta_{ISM}} \text{and}\ \ A(\lambda)_{\text{MC}} \propto \left(\dfrac{\lambda}{\lambda_{V}}\right)^{\delta_{MC}},
\end{equation}
where $\delta_{ISM}$ and $\delta_{MC}$ are the slopes of attenuation in the ISM and the MCs respectively. 
Young stars that are in the MCs will therefore be attenuated twice: by the surrounding dust and additionally by the dust in the diffuse ISM. A ratio of Av$_{V}^{ISM}$ / (A$_V^{BC}+A_V^{ISM})$ is also considered to account for the attenuation of young stars residing in the birth clouds, and the older stars residing in the ISM.\\
CF00 found that $\delta_{ISM}$ = $\delta_{MC}$ = -0.7 satisfied dust attenuation in nearby galaxies, however, this curve is frequently used at higher redshifts \citep[e.g.][]{Malek2018,Buat2018,Pearson2018,Salim2020}. By attenuating at higher wavelengths (until the NIR) more efficiently than C00, this approach considers a more attenuated older stellar population.\smallbreak
\citet[][henceforth LF17]{LoFaro2017} have found that a shallower attenuation curve reproduces the attenuation in ultra-luminous and luminous IR galaxies (ULIRGs and LIRGs) at z$\sim$2. For their sample, LF17 have found $\delta_{ISM}$ = -0.48. This curve was used in \citet{Hamed21} for a heavily dust-obscured ALMA-detected galaxy at z$\sim$2, and provided an overall better fit that other steeper attenuation laws.\smallbreak
To model dust attenuation of the galaxies of our sample, we use the aforementioned laws, with the parameters presented in Table \ref{tab:Table2}. The mean normalized attenuation of our sample as a result of the three attenuation curves are shown in Fig.~\ref{fig:attenuation_mass} left panel. To obtain the left panel of Fig. \ref{fig:attenuation_mass}, we computed the attenuation values in the UV-NIR bands. Then we averaged the attenuation in each band for the whole sample. The curve of C00 is steeper than the double-component power-laws of CF00 and LF17, especially in the NIR domain.

\subsubsection{Hot and cold dust components}

Dust grains heated by AGN, along with the vibrational modes of polycyclic aromatic hydrocarbons (PAHs), dominate the MIR part of the SED of a galaxy. Thus, it is important to include AGN modeling in our SED fitting procedures, as well as taking into account PAH contribution to the overall dust emission. Our initial analysis of the IRAC photometry of \emph{Spitzer} did not suggest AGN candidates. We also included AGN-heated dust templates of \citet{Fritz2006} in our SED fitting procedure, but found no AGN contribution in our sample.\smallbreak

To model the IR emission in our SED models, we use the templates of \citealt{dl2014}. These templates take into consideration different sizes of grains of carbon and silicate, hence, allowing different temperatures of dust grains. These templates rely on observations and are widely used in the literature to fit FIR SEDs.

\subsection{SED quality, model assessment}

\begin{figure}
              \centering
              \includegraphics[width=0.5\textwidth]{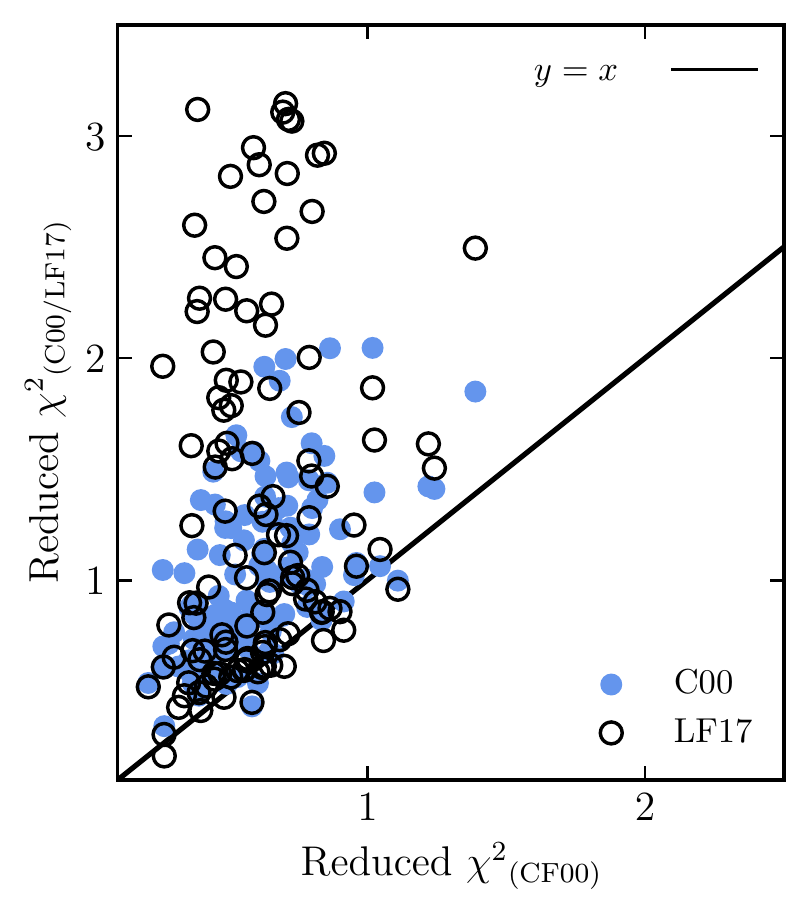}%
              \caption{\footnotesize Comparison between reduced $\chi^2$ of the mock sample obtained with CF00, C00, and LF17 attenuation laws.}
             \label{referee_test}
             \end{figure}

\begin{figure}
    \centering
       \includegraphics[width=0.5\textwidth]{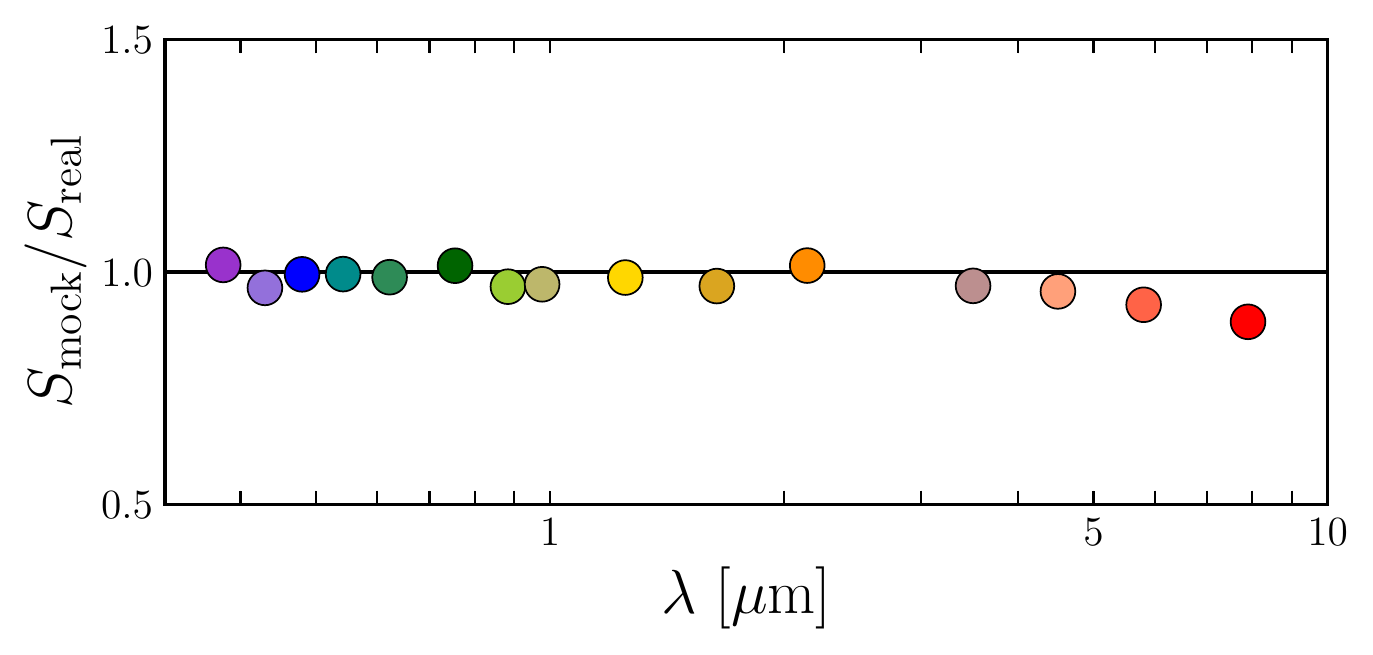}
       \caption{\footnotesize Ratio of mock fluxes to real fluxes of our sample. The mock fluxes were obtained with perturbations to the initial fluxes of our sample by employing a Gaussian distribution with the standard deviation equal to the errors associated with each flux of every source. This process was repeated ten times to generate ten mock galaxy samples. This figure shows the ratio of the average perturbed fluxes to the fluxes of the real sample.}
       \label{fig:mock_fluxes}
   \end{figure}

In assessing which SED provides the best fit for modeling the galaxies of our sample, we adopt a similar methodology as in \citet{buat2019}. We compare the reduced $\chi^2$ of the resulted fits with the three attenuation recipes used, and in the case of different attenuation laws providing a good reduced $\chi^2$, we checked the Bayesian inference criterion (BIC) defined as $BIC=\chi^2+k\times ln(N)$, where $k$ is the number of free parameters and $N$ is the number of data points. The mean reduced $\chi^2$ was found to be 2.4 for our sample (\ref{sample_selection} left panel, where we show the reduced $\chi^2$ for the best fits of our sample).\\ \smallbreak
To test the robustness of the method we use to select the best attenuation law for each galaxy, we perform the following test:
\begin{itemize}
  \item We fit all galaxies of our sample with the attenuation law of CF00. We then take the best fit values for each filter of each galaxy.
  \item We perturb these “best” fluxes (obtained from the fit) using the initial photometry errors to obtain a mock catalog.
  \item This mock catalog was then fitted with: Calzetti law, CF00 law and LF17 law, in the same way as the initial real photometry was treated.
\end{itemize}
The reduced $\chi^2$ obtained from those fits are then compared for each source. We show this comparison in Fig. \ref{referee_test}. The obtained reduced $\chi^2$ for CF00 fits of the mock sample are consistently smaller than that obtained using the two other attenuation recipes.  Precisely, 93$\%$ of mock galaxies preferred CF00 attenuation law using this test. The other 7$\%$ had a reduced $\chi^2$ of $\sim$0.1 lower using the other attenuation laws.\\ \smallbreak
The choice of the best attenuation law that better describes the observed fluxes is crucial for our study. Therefore we used an additional method in reliably attributing the best attenuation law for each galaxy. We introduced perturbations to the fluxes by applying a Gaussian distribution with a standard deviation that corresponds to the uncertainties of each band. To make the calculation time faster, and since the main task was to check the reliability of the best attenuation law for each galaxy, we generated mock fluxes until the mid-IR bands of IRAC. We created ten mock catalogues with this method, and applied the same approach of SED fitting for our original sample to the mock samples. Fig. \ref{fig:mock_fluxes} shows the ratio of mock fluxes to real fluxes of our sample.
             \smallbreak
The results of these tests are presented in Tab. \ref{tab:mock_tab}, where the vast majority of our mock galaxies ($>94\%$) preferred the same attenuation law of the initial real galaxies. This shows that the usage of the reduced $\chi^2$ in assessing the best attenuation law in our case is valid. This is directly linked with the good S:N ratios of our photometric data, that were shown in Tab \ref{tab:Table1}.
\begin{table}
\caption{Summary of the results of the best fits using the same SED method on the fluctuated photometry preferring the same attenuation law (average of ten realizations) and the best attenuation laws for the real galaxies.}
\begin{center}
\begin{tabular}{|c||c c c|}
\toprule\hline
\backslashbox{Real}{Mock} &\ \ \ \ \ C00 \ \ \ \ \ &\ \ \ \ \ CF00\ \ \ \ \ &\ \ \ \ \ LF17\ \ \ \ \ \\\hline
\hline
 C00 & 94$\%$ & 5$\%$ & 1$\%$\\
 CF00 & 0$\%$ & 97$\%$ & 3$\%$\\
 LF17 & 0$\%$ & 3$\%$&97$\%$\\
 \hline
    \bottomrule
     \end{tabular}
     \end{center}
     \label{tab:mock_tab}
\end{table}
\smallbreak
An example of a computed SED of our sample is given in Fig.~\ref{fig:SED}, where we attenuated the same galaxy in three different ways. In this example, the shallower attenuation of \citealp[][]{LoFaro2017} was preferred since it provided a significantly better fit.\\ 
To test the reliability of our SED models, with \texttt{CIGALE} we generated a mock galaxy sample and fitted SEDs with the same methods applied to our sample. We show the comparison between the real physical properties that we derived for our sample and its mock equivalent in Fig.~\ref{fig:mock}.    

\section{Results $\&$ Discussions}\label{results}

\subsection{Galaxy properties and dust attenuation}

We applied a bouquet of attenuation slopes in fitting our sample of DSFGs. The attenuation curve of C00 results in lower stellar masses compared to the ones obtained with the shallower double component attenuation laws of CF00 and LF17 (with a mean stellar mass of $10^{10.87}$ M$_{\odot}$ for C00,  $10^{11.22}$ M$_{\odot}$ and  $10^{11.52}$ M$_{\odot}$ for CF00 and LF17 respectively). The distribution of the obtained stellar masses was shown in Fig.~\ref{fig:attenuation_mass} right panel with the mean M$_{*}$ for the whole sample portrayed for every attenuation law used. Stellar masses computed using shallower attenuation slopes are higher than the one produced with steeper curves.\\

Star formation rates computed from the panchromatic SED fitting using the three aforementioned attenuation laws do not change, similar to the results found in \citet[][]{Malek2018}. The mean values of the log$_{10}$(SFR) of the sample fitted with C00, CF00, and LF17 are 2.75, 2.63, and 2.60 M$_{\odot}$ yr$^{-1}$ respectively, which is of a similar range of galaxies studied in \citealp[][]{buat2019}.\\ The dust masses computed with the three attenuation laws are invariant (with a mean of 1.80$\times$10$^9$ M$_{\odot}$ for the whole sample). This is mainly due to the strong constrain of the FIR part of the SED provided by the ALMA detections as well as the good fitting of the spectrum.\\

The significant difference in produced stellar masses using the different attenuation law slopes result in a clear distinction in the ``starburstiness'' of galaxies, and also affects the quiescent systems \citep{Lofaro15}. In our sample, the number of starburst galaxies decrease with a shallower attenuation curve (60$\%$ for C00, 25$\%$ for CF00, and 14$\%$ for LF17).\\
Despite its simplicity, the attenuation law given by \citet[][]{Calzetti2000} provided good fits in building the SEDs and was favored over the shallower curves of \citet[][]{CharlotFall2000} and \citet[][]{LoFaro2017} in 49$\%$ of the whole sample, that is 61 sources (by comparing the resulted reduced $\chi^2$). This was mainly noticed below redshift of $z=2$. The attenuation curve adapted by \citet[][]{LoFaro2017} provided better fits for 38 galaxies in total, but 79$\%$ of these galaxies fell in the redshift range of $1.5<z<2.5$, which supports the initial tuning of the ISM attenuation at -0.48 \citep[][]{LoFaro2017}.\\

\begin{figure}[h]
\centering
  \includegraphics[width=0.5\textwidth]{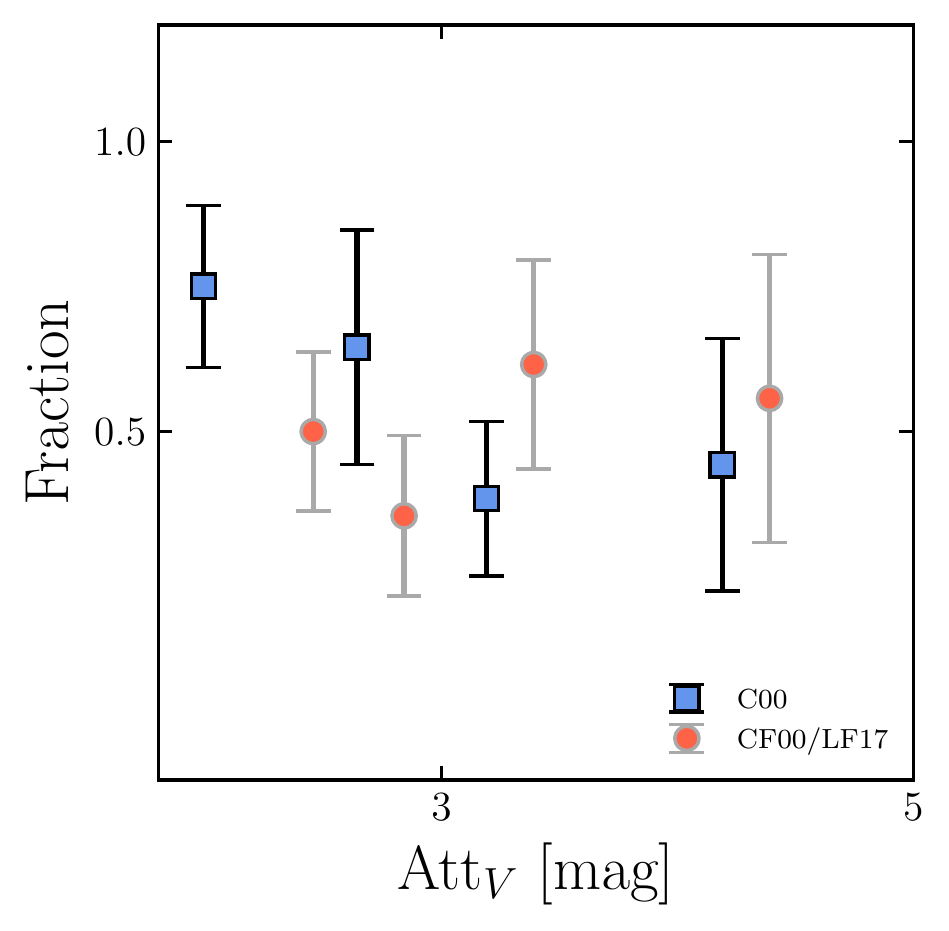}%
        \caption{\footnotesize Preference of attenuation laws of our sample according to the attenuation in the V band. To facilitate the reading of this plot, we shifted the bins of CF00/LF17 slightly to the right (+0.1).}
        \label{fig:pref_mstar}
    \end{figure}

\begin{figure}
\centering
  \includegraphics[width=0.5\textwidth]{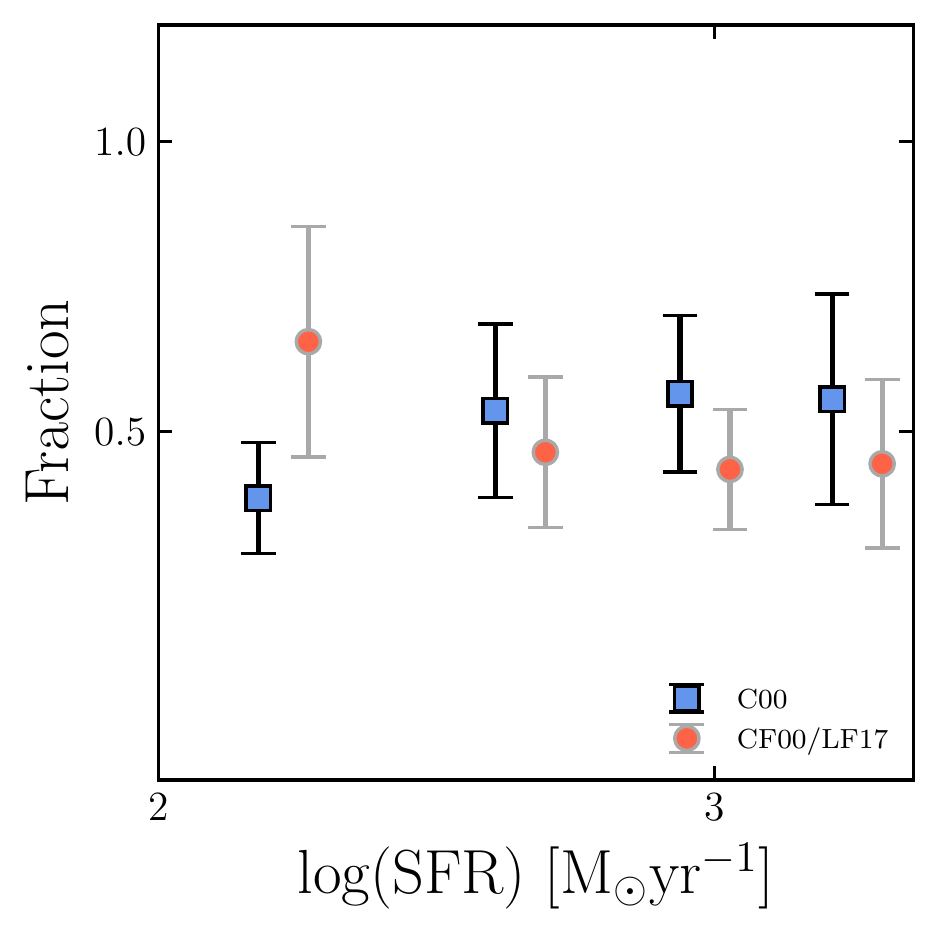}%
        \caption{\footnotesize Preference of attenuation laws of our sample according to the SFR. To facilitate the reading of this plot, we shifted the bins of CF00/LF17 slightly to the right (+0.1).}
        \label{fig:pref_sfr}
\end{figure}

We show the preference of the attenuation laws for our sample based on the V band attenuation and the SFR in Figs. \ref{fig:pref_mstar} and \ref{fig:pref_sfr} respectively. We find no clear correlation between the attenuation in the V band and the preference of the attenuation laws in our sample. We also checked the correlation with the stellar masses. But since these masses are directly a byproduct of the attenuation law used, as it was shown in Figs. \ref{fig:attenuation_mass}, we cannot tell if this correlation is physical. Galaxies that are preferring CF00 double-component attenuation law and its shallower version LF17, by construction, result in a significantly higher older stellar population, therefore increasing the stellar mass \citep[e.g.,][]{Malek2018, buat2019, Hamed21, Figueira22}.\\
We also checked the preference of attenuation laws used with SFR for our galaxies. This is shown in Fig. \ref{fig:pref_sfr}. We found that towards higher SFRs, there are no preferences in attenuation laws for our sample. However, in the lower limit of SFR, a double-component attenuation was slightly preferred, but still within the error bars. In the sample we had only 18 galaxies with log(SFR)< 2.4 and $38\%$ of them were fitted with C00 while the rest preferred CF00/LF17. The small statistics at this lower end of SFR for our sample does not allow us to make a strong statement about the correlation of the attenuation laws for low-SFR galaxies.

\begin{figure}
\centering
  \includegraphics[width=0.5\textwidth]{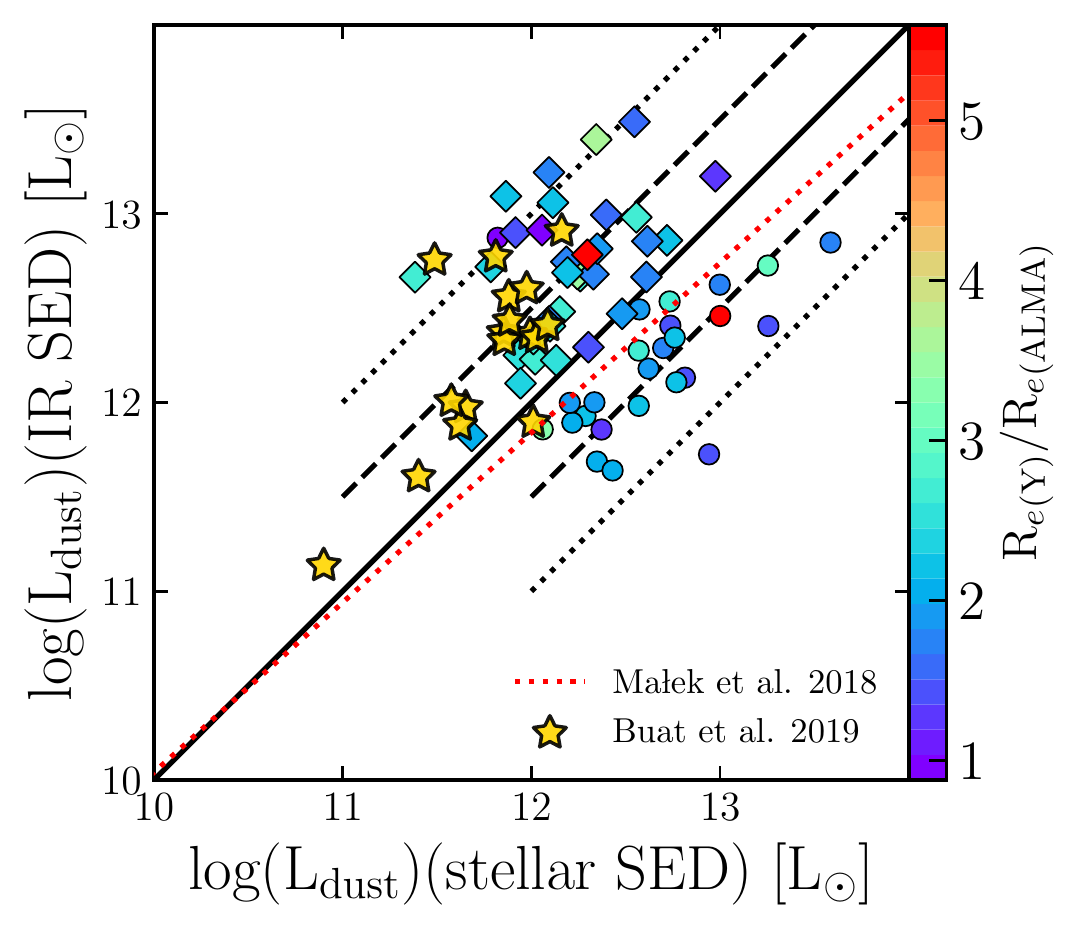}%
%

        \caption{\footnotesize Infrared luminosities of our sample produced from the IR data alone using the \citet{dl2014} dust emission templates ($y$-axis) with respect to the IR luminosity evoked indirectly through the energy balance based on the short wavelength data only.}
        \label{fig:dust_stellar}
    \end{figure}
    
\subsection{Stellar vs. dust components}\label{dust_stars}
To analyze the energy balance of our sample of DSFGs, especially with the available dust and stars emission and images, we follow the method introduced in \citet[][]{Malek2018} and \citet[][]{buat2019} by dissecting the stellar continuum of our sample of galaxies without taking into account the FIR detections. Equivalently, we also fit the FIR continua of our SFGs.\\
To model the stellar continuum, we use the photometric bands of CFHT, \emph{Subaru}, \emph{VISTA} and available IRAC bands. This is done in order to ensure the attenuation curve requirements without adding the energy balance constrain to the global SED fitting method. As shown in \citealp[e.g.,][]{buat2019} and \citealp[][]{Hamed21}, the dichotomy between the stellar SED and its dust counterpart is important in testing the validity of the energy balance concept that is the basis of most of panchromatic SED fitting tools. Moreover, this method is critical in cases where the dust continuum maps are not centered on their short wavelength counterparts.\\
When fitting the short wavelength part of the SEDs, we model the stellar light taking into account the delayed SFH boosted by a recent burst, a stellar population library of \citet[]{BC03}, and dust attenuation laws that are discussed in \ref{sed}. We also modeled the IR emission using \citet[][]{dl2014} dust emission templates, but without taking into account the IR photometry, allowing the energy balance to dictate dust luminosities and masses based on the amount of the stellar light that is attenuated. \\
A total of 61$\%$ of our sample (75 sources) provided better fits with the simple power law of C00. While a shallower attenuation was needed to reproduce the spectra with the least $\chi^2$ for the rest (14$\%$ with CF00 and 25$\%$ with an even grayer LF17 slope).\\

The steep law of C00 provided better fits for more galaxies when taking into account the stellar emission only. This result was also found in \citet[][]{buat2019} for a smaller sample. This tendency will be confirmed in future studies based on the new generation of IR datasets from JWST and well constrained short wavelength counterparts from LSST.\\

Equivalently, we estimated dust luminosities based on the rest frame UV to NIR photometry of our sample, assuming an energy balance between dust absorption and dust emission. We compare these IR luminosities with the ones calculated from the IR photometric points with \citet[]{dl2014} templates. The results are shown in Fig.~\ref{fig:dust_stellar}.\\
We confirm the scatter initially found in \citet[][]{buat2019} and \citet{Hamed2022} with dust-rich galaxies which significantly differs from the normal SFGs found in \citet[][]{Malek2018}. We find that the dust emission evoked from a pure energy balance based on the short wavelength does not always explain the one calculated from the FIR photometry. Galaxies that are fitted with C00 attenuation law are found to inhibit more attenuation and they produce higher L$_{\mathrm{IR}}$ from their dust content rather than the stellar parts. The star-to-dust compactness did not seem to play a role in this trend. \\
This shows that the dust luminosity concluded from the direct and attenuated UV photons based on the energy balance is not enough to reproduce the dust luminosity observed from the actual IR photometry.\\


\subsection{Dust attenuation and sizes}\label{dust_morph}
\begin{figure*}[h]
    \centering
    \minipage{0.5\textwidth}
      \includegraphics[width=1\textwidth]{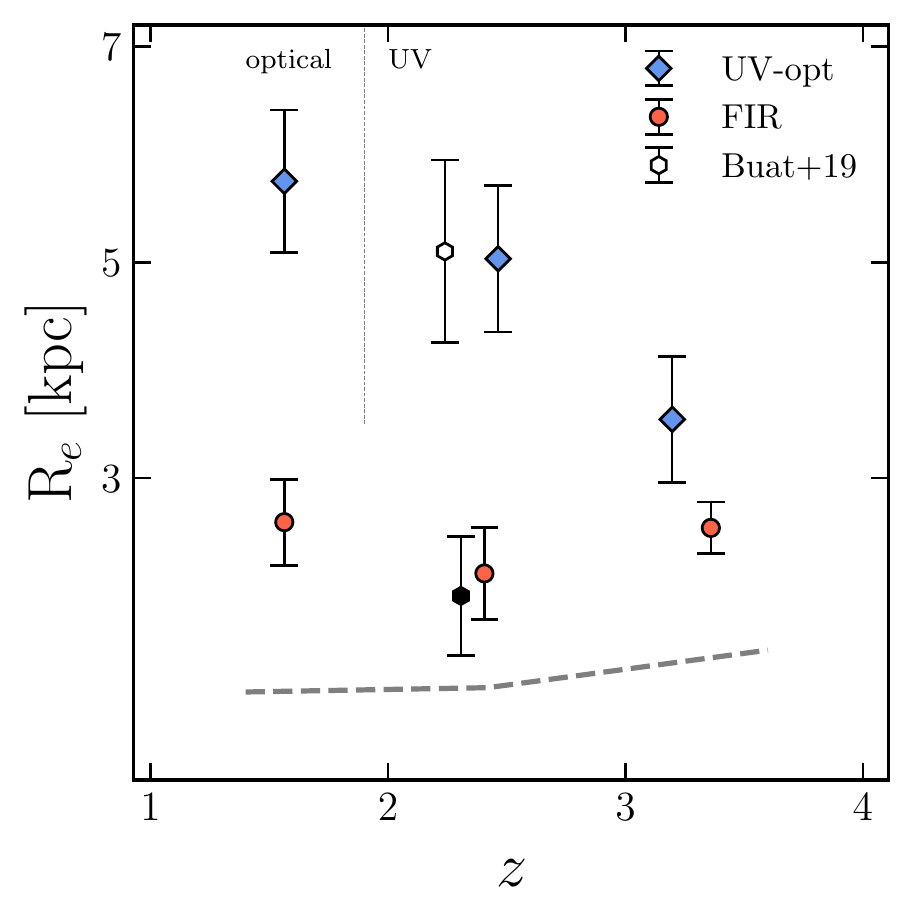}
    \endminipage
    \minipage{0.5\textwidth}
     \includegraphics[width=1\textwidth]{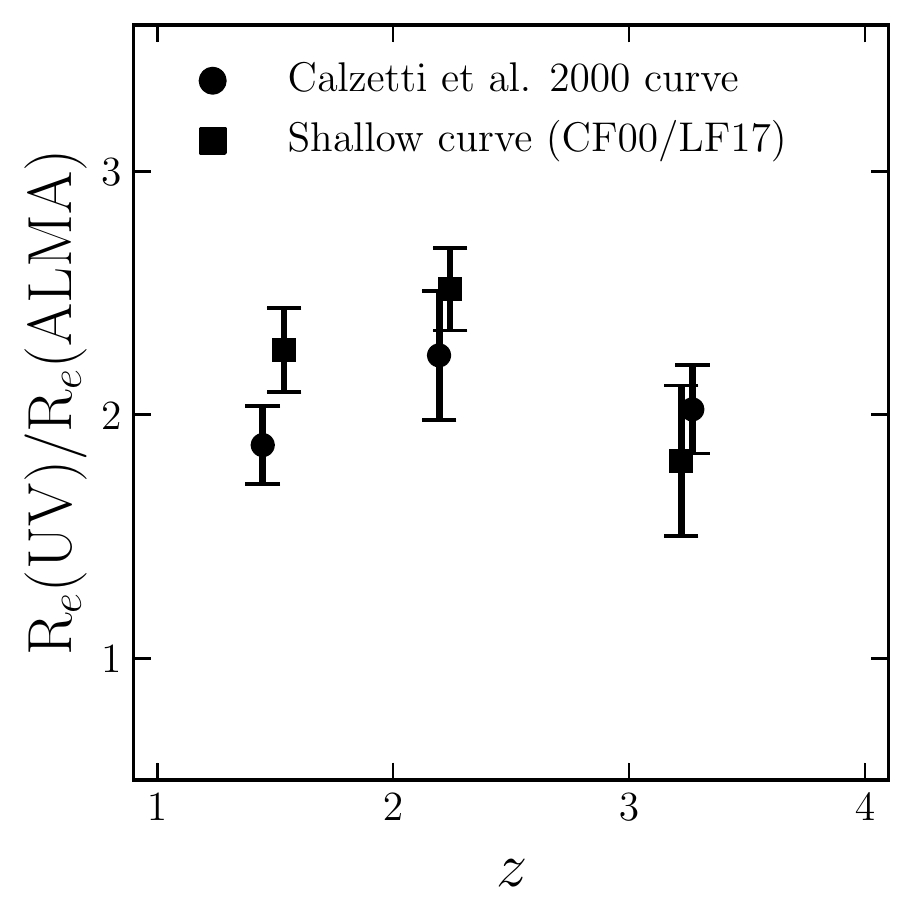}
     \endminipage
        \caption{\footnotesize (Left panel): Redshift evolution of the derived effective radii at two different wavelengths of our sample. The dashed line represents the redshift at which the $Y$ band starts to probe the restframe UV. Hexagons show the binned sample of 17 galaxies from \citet{buat2019} around $z\sim2$, taken from \citet{Elbaz2018} and \citet{Dunlop2017}. Empty hexagons show the rest-frame UV radii while the filled ones show the ALMA detection. The dashed line shows the minimal size that can be measured with ALMA \citep{Gomez}. (Right panel): The evolution of $\mathrm{R}_e(\mathrm{UV})/\mathrm{R}_e(\mathrm{ALMA})$ ratio with redshift in our sample of DSFGs.}
        \label{fig:radii}
\end{figure*}

We study the effects of star-to-dust compactness of the unobscured star-forming regions and the stellar population emission to the extent of dust emission detected by ALMA. We define the ratio of the short wavelength radii to their FIR counterparts as $\mathrm{R}_e(\mathrm{UV})/\mathrm{R}_e(\mathrm{ALMA})$ where $\mathrm{R}_e(\mathrm{UV})$ is the circularized effective radii measured from the HSC $Y$ bands of our sources. The UV radii of our sample decrease with redshift, while the ALMA counterparts are rather constant accross the studied redshift range in this work. This is shown in Fig.~\ref{fig:radii} and is explained by the bright IR ALMA-selected sources, and their active star-forming/stellar population regions.\\
Despite the fact that the attenuation curve of \citet[][]{Calzetti2000} managed to fit the largest sub-sample of our DSFGs around the cosmic noon and at lower redshift ranges, the distribution of these galaxies for which a steeper attenuation curve provided the best fit is relatively compact. A relative star-to-dust compactness was found to have an average of 1.7, while for the shallower attenuation curve recipes were found to not follow a specific preference, and they are rather scattered accross all the studied redshift bins. We show the redshift distribution of the preferred attenuation laws in  Appendix \ref{appendix:a2}. The range of the resulted ($n^{\mathrm{S\acute{e}rsic}}$) was too small (0.4 to 1.6), and in that narrow range we did not find any correlation with other observables.\\
\begin{figure}[h!]
    \centering
        \includegraphics[width=0.5\textwidth]{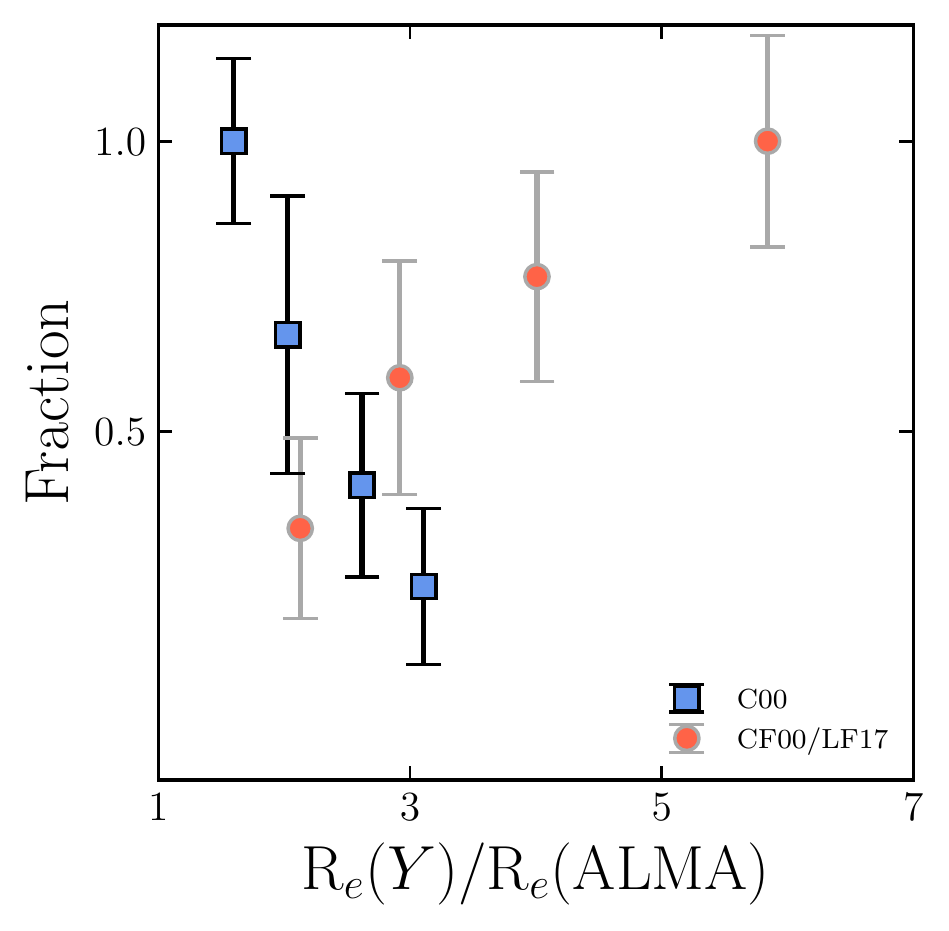}
        \caption{Preference of attenuation laws of our sample according to the star-to-dust compactness. To facilitate the reading of this plot, we shifted the bins of CF00/LF17 slightly to the right (+0.1).}
        \label{fig:pref_compact}
\end{figure}
We find that the ratio of $\mathrm{R}_e(\mathrm{UV})$ to $\mathrm{R}_e(\mathrm{ALMA})$ changes accross the redshift in our sample of ALMA-detected DSFGs. This distribution is found to peak at $z=2$ around the cosmic noon, as shown in the right panel of Fig.~\ref{fig:radii}. This change is more prominent at higher redshift with the decrease of the star-to-dust compactness is directly connected to the rapidly decreasing rest-frame UV sizes of these DSFGs at higher redshift. This might be explained by the more intense star formation around that cosmic epoch, especially that DSFGs contributed significantly to the total star formation activity in the Universe. Moreover, this peak is found to be stronger for galaxies that require a shallower attenuation curve. This might correlate with the higher need of cold star forming regions in these galaxies to explain the higher SFRs. This result shows that dust attenuation and the star-to-dust compactness of dust in DSFGs might be correlated with the cosmic SFR density.\\

These findings partly agree with the smaller sample size studied in \citet[][]{buat2019}. However, our statistically larger sample size allows for an extrapolation of this correlation not only at different redshift ranges, but also showed that shallower attenuation curves do not favor higher star-to-dust compactness, but rather a scattered trend was found, unlike the steeper curves which clearly preferred relatively smaller $\mathrm{R}_e(\mathrm{UV})/\mathrm{R}_e(\mathrm{ALMA})$ ($\sim2$) at different redshift ranges. One possible explanation for this can be the fact that for galaxies with relatively smaller sizes of their unobscured star-forming regions and their dust content, might be translated by a screen model of attenuation, due to a non-effective mixing of stars and dust. On the other hand, a very compact dust emission requires a more complex mixing of dust and stars which is translated in a shallower attenuation curve.\\
A correlation is visible between the fraction of our sample that is fitted with a specific attenuation law and the relative compactness. We show this in Fig. \ref{fig:pref_compact}. \\
\smallbreak
We find that the galaxies with smaller opt/UV sizes relative to dust radii largely prefer the C00 attenuation law in our sample. For larger R$_e$(Y)/R${_e}$(ALMA) ratio (>3) only CF00 and its shallower modification LF17 fitted the galaxies better. This shows that taking relative compactness into account is highly important when performing SED fitting. One can infer the most likely attenuation law for a given galaxy by taking into account its sizes in the unattenuated SF region and the dust continuum. This in turn enables a reliable procedure for fitting its spectral energy distribution (SED).
\begin{figure*}[h]
    \centering
    \minipage{0.5\textwidth}
      \includegraphics[width=1\textwidth]{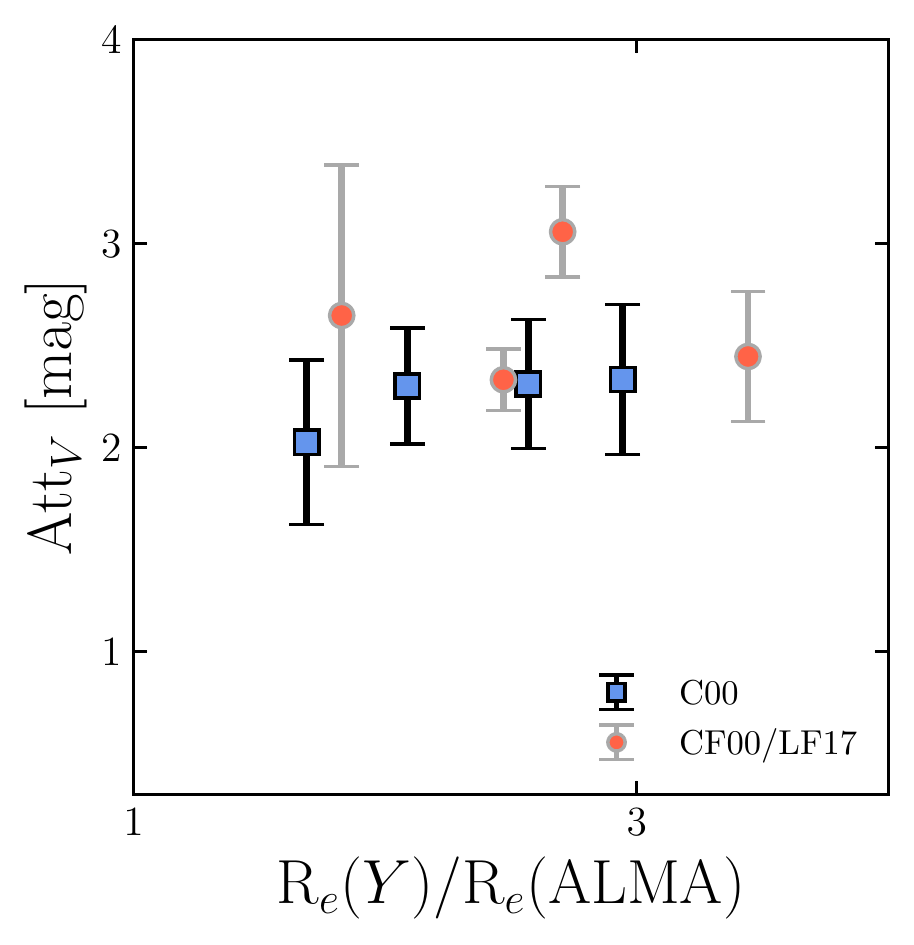}
    \endminipage
    \minipage{0.5\textwidth}
     \includegraphics[width=1\textwidth]{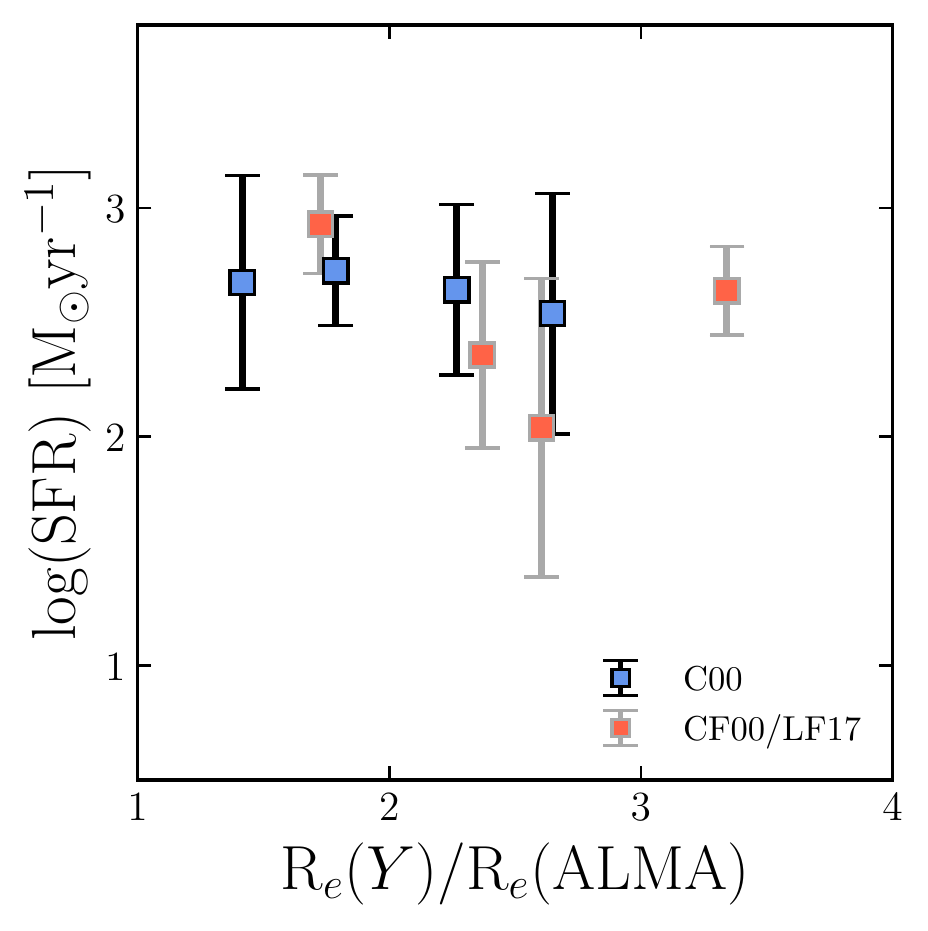}
     \endminipage
        \caption{\footnotesize (Left panel): Variation of the V band attenuation with the star-to-dust compactness. (Right panel): Variation of SFRs with the star-to-dust compactness.}
        \label{fig:removing_bias}
\end{figure*}

\section{Summary}\label{summary}
In this paper, we studied a statistical sample of 122 DSFGs not hosting AGN accross a wide range of redshift ($1<z<4$). We derived their circularized effective radii in two different bands, HSC's $Y$ band when available, and equivalently their dust components' radii with ALMA detections. On the other hand, we carefully analyzed their SEDs, modeling them with varied dust attenuation laws, particularly that of \citet[][]{Calzetti2000} and the shallower curves of \citet[][]{CharlotFall2000} and their recipes.\\

We also dissected the stellar SEDs alone and their IR counterparts as done in \citep[e.g.,][]{buat2019, Hamed21} to investigate the validity of the energy balance when taking into account the ALMA detections of our sources. We found that even if most of our sources seem to produce the same dust emission when relying on an energetic balance from the short wavelengths, and when fitting the IR photometry apart, some galaxies expressed dimmer star formation when the \citet[][]{Calzetti2000} attenuation law was favored. This was translated into an under-supply of dust emission from the stellar population alone.\\

We found that knowing the information of the sizes of DSFGs especially their stellar and dust contents in the analysis, constrains the used attenuation curves to fit the photometry of these galaxies. We found that a starburst curve of \citet[][]{Calzetti2000} was favored in the reproduction of the SEDs of DSFGs with comparable star-to-dust radii ratios, precisely at $1.2<\mathrm{R}_e(\mathrm{UV})/\mathrm{R}_e(\mathrm{ALMA})<2.5$. However, despite the seemingly irrelevant star-to-dust ratios with the attenuation curves of \citet[][]{CharlotFall2000} and the shallower version \citep[][]{LoFaro2017}, we found that compact dust emission and extended stellar radii needed shallow curves and double exponential attenuation laws to account for the missing photons absorbed by dust. This shows that when fitting SEDs using broad band photometry, a careful analysis of the radii of different components should be taken into account, before using a unique attenuation law which will result in wrongly-estimated stellar masses.\\

We stress that recent ALMA studies of smaller samples of z$\sim$2 galaxies suggested that compact IR sizes could be connected to rapid growth of supermassive black holes during the SMG star-formation phase \citep{Ikarashi2017}. Semi-analytical models (i.e. \citealt{Lapi2018}) predict that such sources would experience forthcoming/ongoing AGN feedback, which is thought to trigger the morphological transition from star-forming discs to early-type galaxies. However, our galaxies do not show AGN activity, and are suggestive of SF galaxies caught in the compaction phase characterised by clump/gas migration toward the galaxy center, where the highly intense dust production takes place and most of the stellar mass is accumulated (\citealt{Pantoni20212} and \citealt{Pantoni2021}). Interestingly, dust compaction phase is suggested to play role in metallicity enrichment efficiency, which further affects dust growth in ISM \citep{Pantoni2019,Donevski2020}. We expect that our dusty galaxies have relatively wide range of gas metallicities and dust growth efficiencies, which would reflect on different attenuation slopes, rather than favoring the single one. This expectation is also in line with results from recent cosmological simulations that found dependence of attenuation on dust compactness and/or geometry \citep[e.g.,][]{schulz2020} and the ratio between the small and large dust grains \citep[e.g.,][]{Hou2019}.\\

In our sample, we have observed that the C00 attenuation law is mostly favored by galaxies with opt/UV sizes two times larger than the dust radii. However, for galaxies with R$_{e}$(Y)/R$_{e}$(ALMA) >3 the CF00 and its shallower modification LF17 were found to fit better. These findings suggest significant importance of considering the relative compactness when conducting SED fitting. By taking into account the sizes of a galaxy in the unattenuated SF region and the dust continuum, one can deduce the most probable attenuation law for that galaxy, thus providing a dependable approach for fitting its SED.\\

We conducted a test to ensure that the trend observed between the preferred attenuation law and relative compactness is not influenced by other physical properties. Specifically, we investigated the potential relationship between the relative compactness and two other properties - attenuation in the V band and SFR (Fig. \ref{fig:pref_mstar} and Fig. \ref{fig:pref_sfr}). Fig. \ref{fig:removing_bias} presents the results of this analysis. We find no significant correlation between these properties and relative compactness. This finding supports the conclusion that the observed trend between relative compactness and preferred attenuation law can be robust and not influenced by these physical properties.\\

We find that the star-to-dust compactness of the unobscured star-forming regions/stellar population regions to dust emission of these DSFGs peaks around the cosmic noon (z$\sim$2). This is notable at higher redshift. A possible correlation might be with the cosmic SFR density for which the DSFGs were a major contributor in the early Universe.\\

These results are promising in the era of highly resolved deep-field detections with the LSST and JWST, where dust attenuation and size measurements are getting more precise. Combining these detections with the FIR information especially with ALMA, is unparalleled to deal with the dust attenuation curve problem at different redshift ranges.


    

\begin{acknowledgements}
M.H. acknowledges support from the National Science Centre (UMO-2022/45/N/ST9/01336). K.M and M.H. have been supported by the National Science Centre (UMO-2018/30/E/ST9/00082). D.D. acknowledges support from the National Science Centre (grant SONATA-16, UMO-2020/39/D/ST9/00720). M.F. has been supported by the First TEAM grant of the Foundation for Polish Science No. POIR.04.04.00-00-5D21/18-00. We acknowledge and thank the referee for a thorough and constructive report, which helped improving this work. This paper makes use of ALMA data. ALMA is a partnership of ESO (representing its member states), NSF (USA) and NINS (Japan), together with NRC
(Canada), MOST and ASIAA (Taiwan), and KASI (Republic of Korea), in cooperation with the Republic of Chile. The Joint ALMA Observatory is operated by ESO, AUI/NRAO and NAOJ. This research is based [in part] on data collected at the Subaru Telescope, which is operated by the National Astronomical Observatory of Japan. We are honored and grateful for the opportunity of observing the Universe from Maunakea, which has the cultural, historical, and natural significance in Hawaii. We used observations obtained with MegaPrime/MegaCam, a joint project of CFHT and CEA/DAPNIA, at the Canada-France-Hawaii Telescope (CFHT) which is operated by the National Research Council (NRC) of Canada, the Institut National des Science de l'Univers of the Centre National de la Recherche Scientifique (CNRS) of France, and the University of Hawaii. The observations at the Canada-France-Hawaii Telescope were performed with care and respect from the summit of Maunakea which is a significant cultural and historic site. We also used VLA data, operated under The National Radio Astronomy Observatory.
\end{acknowledgements}

\bibliographystyle{aa}
\bibliography{aanda.bib}

\begin{thebibliography}{110}
\expandafter\ifx\csname natexlab\endcsname\relax\def\natexlab#1{#1}\fi

\bibitem[{{Aihara} {et~al.}(2022){Aihara}, {AlSayyad}, {Ando}, {Armstrong},
  {Bosch}, {Egami}, {Furusawa}, {Furusawa}, {Harasawa}, {Harikane}, {Hsieh},
  {Ikeda}, {Ito}, {Iwata}, {Kodama}, {Koike}, {Kokubo}, {Komiyama}, {Li},
  {Liang}, {Lin}, {Lupton}, {Lust}, {MacArthur}, {Mawatari}, {Mineo},
  {Miyatake}, {Miyazaki}, {More}, {Morishima}, {Murayama}, {Nakajima},
  {Nakata}, {Nishizawa}, {Oguri}, {Okabe}, {Okura}, {Ono}, {Osato}, {Ouchi},
  {Pan}, {Plazas Malag{\'o}n}, {Price}, {Reed}, {Rykoff}, {Shibuya},
  {Simunovic}, {Strauss}, {Sugimori}, {Suto}, {Suzuki}, {Takada}, {Takagi},
  {Takata}, {Takita}, {Tanaka}, {Tang}, {Taranu}, {Terai}, {Toba}, {Turner},
  {Uchiyama}, {Vijarnwannaluk}, {Waters}, {Yamada}, {Yamamoto}, \&
  {Yamashita}}]{Aihara22}
{Aihara}, H., {AlSayyad}, Y., {Ando}, M., {et~al.} 2022, \pasj, 74, 247

\bibitem[{B{\'e}thermin {et~al.}(2015)B{\'e}thermin, Daddi, Magdis, Lagos,
  Sargent, Albrecht, Aussel, Bertoldi, Buat, Galametz,
  {et~al.}}]{bethermin2015}
B{\'e}thermin, M., Daddi, E., Magdis, G., {et~al.} 2015, A\&A, 573, A113

\bibitem[{{Blain} {et~al.}(2002){Blain}, {Smail}, {Ivison}, {Kneib}, \&
  {Frayer}}]{Blain2002}
{Blain}, A.~W., {Smail}, I., {Ivison}, R.~J., {Kneib}, J.~P., \& {Frayer},
  D.~T. 2002, \physrep, 369, 111

\bibitem[{{Boquien} {et~al.}(2022){Boquien}, {Buat}, {Burgarella}, {Bardelli},
  {B{\'e}thermin}, {Faisst}, {Ginolfi}, {Hathi}, {Jones}, {Koekemoer},
  {Lemaux}, {Narayanan}, {Romano}, {Schaerer}, {Vergani}, {Zamorani}, \&
  {Zucca}}]{Boquien2022}
{Boquien}, M., {Buat}, V., {Burgarella}, D., {et~al.} 2022, \aap, 663, A50

\bibitem[{{Boquien} {et~al.}(2019){Boquien}, {Burgarella}, {Roehlly}, {Buat},
  {Ciesla}, {Corre}, {Inoue}, \& {Salas}}]{Boquien2019}
{Boquien}, M., {Burgarella}, D., {Roehlly}, Y., {et~al.} 2019, \aap, 622, A103

\bibitem[{{Bourne} {et~al.}(2017){Bourne}, {Dunlop}, {Merlin}, {Parsa},
  {Schreiber}, {Castellano}, {Conselice}, {Coppin}, {Farrah}, {Fontana},
  {Geach}, {Halpern}, {Knudsen}, {Micha{\l}owski}, {Mortlock}, {Santini},
  {Scott}, {Shu}, {Simpson}, {Simpson}, {Smith}, \& {van der
  Werf}}]{Bourne2017}
{Bourne}, N., {Dunlop}, J.~S., {Merlin}, E., {et~al.} 2017, \mnras, 467, 1360

\bibitem[{{Bouwens} {et~al.}(2012){Bouwens}, {Illingworth}, {Oesch}, {Franx},
  {Labb{\'e}}, {Trenti}, {van Dokkum}, {Carollo}, {Gonz{\'a}lez}, {Smit}, \&
  {Magee}}]{Bouwens2012}
{Bouwens}, R.~J., {Illingworth}, G.~D., {Oesch}, P.~A., {et~al.} 2012, \apj,
  754, 83

\bibitem[{{Brinchmann} {et~al.}(2004){Brinchmann}, {Charlot}, {White},
  {Tremonti}, {Kauffmann}, {Heckman}, \& {Brinkmann}}]{Brinchmann2004}
{Brinchmann}, J., {Charlot}, S., {White}, S.~D.~M., {et~al.} 2004, \mnras, 351,
  1151

\bibitem[{{Bruzual} \& {Charlot}(2003)}]{BC03}
{Bruzual}, G. \& {Charlot}, S. 2003, \mnras, 344, 1000

\bibitem[{{Buat} {et~al.}(2018){Buat}, {Boquien}, {Ma{\l}ek}, {Corre}, {Salas},
  {Roehlly}, {Shirley}, \& {Efstathiou}}]{Buat2018}
{Buat}, V., {Boquien}, M., {Ma{\l}ek}, K., {et~al.} 2018, \aap, 619, A135

\bibitem[{{Buat} {et~al.}(2019){Buat}, {Ciesla}, {Boquien}, {Ma{\l}ek}, \&
  {Burgarella}}]{buat2019}
{Buat}, V., {Ciesla}, L., {Boquien}, M., {Ma{\l}ek}, K., \& {Burgarella}, D.
  2019, \aap, 632, A79

\bibitem[{{Buat} {et~al.}(2011){Buat}, {Giovannoli}, {Heinis}, {Charmandaris},
  {Coia}, {Daddi}, {Dickinson}, {Elbaz}, {Hwang}, {Morrison}, {Dasyra},
  {Aussel}, {Altieri}, {Dannerbauer}, {Kartaltepe}, {Leiton}, {Magdis},
  {Magnelli}, \& {Popesso}}]{buat11}
{Buat}, V., {Giovannoli}, E., {Heinis}, S., {et~al.} 2011, \aap, 533, A93

\bibitem[{{Buat} {et~al.}(2014){Buat}, {Heinis}, {Boquien}, {Burgarella},
  {Charmandaris}, {Boissier}, {Boselli}, {Le Borgne}, \& {Morrison}}]{Buat14}
{Buat}, V., {Heinis}, S., {Boquien}, M., {et~al.} 2014, \aap, 561, A39

\bibitem[{{Buat} {et~al.}(2012){Buat}, {Noll}, {Burgarella}, {Giovannoli},
  {Charmandaris}, {Pannella}, {Hwang}, {Elbaz}, {Dickinson}, {Magdis}, {Reddy},
  \& {Murphy}}]{Buat12}
{Buat}, V., {Noll}, S., {Burgarella}, D., {et~al.} 2012, \aap, 545, A141

\bibitem[{{Burgarella} {et~al.}(2005){Burgarella}, {Buat}, \&
  {Iglesias-P{\'a}ramo}}]{Burgarella2005}
{Burgarella}, D., {Buat}, V., \& {Iglesias-P{\'a}ramo}, J. 2005, \mnras, 360,
  1413

\bibitem[{{Calzetti} {et~al.}(2000){Calzetti}, {Armus}, {Bohlin}, {Kinney},
  {Koornneef}, \& {Storchi-Bergmann}}]{Calzetti2000}
{Calzetti}, D., {Armus}, L., {Bohlin}, R.~C., {et~al.} 2000, \apj, 533, 682

\bibitem[{{Carnall} {et~al.}(2018){Carnall}, {McLure}, {Dunlop}, \&
  {Dav{\'e}}}]{Carnall2018}
{Carnall}, A.~C., {McLure}, R.~J., {Dunlop}, J.~S., \& {Dav{\'e}}, R. 2018,
  \mnras, 480, 4379

\bibitem[{{Carnall} {et~al.}(2020){Carnall}, {Walker}, {McLure}, {Dunlop},
  {McLeod}, {Cullen}, {Wild}, {Amorin}, {Bolzonella}, {Castellano}, {Cimatti},
  {Cucciati}, {Fontana}, {Gargiulo}, {Garilli}, {Jarvis}, {Pentericci},
  {Pozzetti}, {Zamorani}, {Calabro}, {Hathi}, \& {Koekemoer}}]{Carnall2020}
{Carnall}, A.~C., {Walker}, S., {McLure}, R.~J., {et~al.} 2020, \mnras, 496,
  695

\bibitem[{{Casey} {et~al.}(2017){Casey}, {Cooray}, {Killi}, {Capak}, {Chen},
  {Hung}, {Kartaltepe}, {Sanders}, \& {Scoville}}]{Casey2017}
{Casey}, C.~M., {Cooray}, A., {Killi}, M., {et~al.} 2017, \apj, 840, 101

\bibitem[{{Casey} {et~al.}(2014){Casey}, {Scoville}, {Sanders}, {Lee},
  {Cooray}, {Finkelstein}, {Capak}, {Conley}, {De Zotti}, {Farrah}, {Fu}, {Le
  Floc'h}, {Ilbert}, {Ivison}, \& {Takeuchi}}]{Casey2014}
{Casey}, C.~M., {Scoville}, N.~Z., {Sanders}, D.~B., {et~al.} 2014, \apj, 796,
  95

\bibitem[{{Chabrier}(2003)}]{Chabrier2003}
{Chabrier}, G. 2003, \pasp, 115, 763

\bibitem[{{Chapman} {et~al.}(2005){Chapman}, {Blain}, {Smail}, \&
  {Ivison}}]{Chapman2005}
{Chapman}, S.~C., {Blain}, A.~W., {Smail}, I., \& {Ivison}, R.~J. 2005, \apj,
  622, 772

\bibitem[{{Charlot} \& {Fall}(2000)}]{CharlotFall2000}
{Charlot}, S. \& {Fall}, S.~M. 2000, \apj, 539, 718

\bibitem[{{Ciesla} {et~al.}(2020){Ciesla}, {B{\'e}thermin}, {Daddi}, {Richard},
  {Diaz-Santos}, {Sargent}, {Elbaz}, {Boquien}, {Wang}, {Schreiber}, {Yang},
  {Zabl}, {Fraser}, {Aravena}, {Assef}, {Baker}, {Beelen}, {Boselli},
  {Bournaud}, {Burgarella}, {Charmandaris}, {C{\^o}t{\'e}}, {Epinat},
  {Ferrarese}, {Gobat}, \& {Ilbert}}]{Ciesla2020}
{Ciesla}, L., {B{\'e}thermin}, M., {Daddi}, E., {et~al.} 2020, \aap, 635, A27

\bibitem[{{Ciesla} {et~al.}(2016){Ciesla}, {Boselli}, {Elbaz}, {Boissier},
  {Buat}, {Charmandaris}, {Schreiber}, {B{\'e}thermin}, {Baes}, {Boquien}, {De
  Looze}, {Fern{\'a}ndez-Ontiveros}, {Pappalardo}, {Spinoglio}, \&
  {Viaene}}]{Ciesla2016}
{Ciesla}, L., {Boselli}, A., {Elbaz}, D., {et~al.} 2016, \aap, 585, A43

\bibitem[{{Ciesla} {et~al.}(2015){Ciesla}, {Charmandaris}, {Georgakakis},
  {Bernhard}, {Mitchell}, {Buat}, {Elbaz}, {LeFloc'h}, {Lacey}, {Magdis}, \&
  {Xilouris}}]{Ciesla15}
{Ciesla}, L., {Charmandaris}, V., {Georgakakis}, A., {et~al.} 2015, \aap, 576,
  A10

\bibitem[{{Ciesla} {et~al.}(2017){Ciesla}, {Elbaz}, \& {Fensch}}]{Ciesla2017}
{Ciesla}, L., {Elbaz}, D., \& {Fensch}, J. 2017, \aap, 608, A41

\bibitem[{{Ciesla} {et~al.}(2018){Ciesla}, {Elbaz}, {Schreiber}, {Daddi}, \&
  {Wang}}]{Ciesla2018}
{Ciesla}, L., {Elbaz}, D., {Schreiber}, C., {Daddi}, E., \& {Wang}, T. 2018,
  \aap, 615, A61

\bibitem[{{da Cunha} {et~al.}(2008){da Cunha}, {Charlot}, \&
  {Elbaz}}]{daCunha2008}
{da Cunha}, E., {Charlot}, S., \& {Elbaz}, D. 2008, \mnras, 388, 1595

\bibitem[{{Daddi} {et~al.}(2010){Daddi}, {Elbaz}, {Walter}, {Bournaud},
  {Salmi}, {Carilli}, {Dannerbauer}, {Dickinson}, {Monaco}, \&
  {Riechers}}]{Daddi2010}
{Daddi}, E., {Elbaz}, D., {Walter}, F., {et~al.} 2010, \apjl, 714, L118

\bibitem[{{Daddi} {et~al.}(2005){Daddi}, {Renzini}, {Pirzkal}, {Cimatti},
  {Malhotra}, {Stiavelli}, {Xu}, {Pasquali}, {Rhoads}, {Brusa}, {di Serego
  Alighieri}, {Ferguson}, {Koekemoer}, {Moustakas}, {Panagia}, \&
  {Windhorst}}]{Daddi2005}
{Daddi}, E., {Renzini}, A., {Pirzkal}, N., {et~al.} 2005, \apj, 626, 680

\bibitem[{{Donevski} {et~al.}(2020){Donevski}, {Lapi}, {Ma{\l}ek}, {Liu},
  {G{\'o}mez-Guijarro}, {Dav{\'e}}, {Kraljic}, {Pantoni}, {Man}, {Fujimoto},
  {Feltre}, {Pearson}, {Li}, \& {Narayanan}}]{Donevski2020}
{Donevski}, D., {Lapi}, A., {Ma{\l}ek}, K., {et~al.} 2020, \aap, 644, A144

\bibitem[{{Draine} {et~al.}(2014){Draine}, {Aniano}, {Krause}, {Groves},
  {Sandstrom}, {Braun}, {Leroy}, {Klaas}, {Linz}, {Rix}, {Schinnerer},
  {Schmiedeke}, \& {Walter}}]{dl2014}
{Draine}, B.~T., {Aniano}, G., {Krause}, O., {et~al.} 2014, \apj, 780, 172

\bibitem[{{Duncan} {et~al.}(2018){Duncan}, {Brown}, {Williams}, {Best}, {Buat},
  {Burgarella}, {Jarvis}, {Ma{\l}ek}, {Oliver}, {R{\"o}ttgering}, \&
  {Smith}}]{Duncan2018}
{Duncan}, K.~J., {Brown}, M. J.~I., {Williams}, W.~L., {et~al.} 2018, \mnras,
  473, 2655

\bibitem[{{Dunlop} {et~al.}(2017){Dunlop}, {McLure}, {Biggs}, {Geach},
  {Micha{\l}owski}, {Ivison}, {Rujopakarn}, {van Kampen}, {Kirkpatrick},
  {Pope}, {Scott}, {Swinbank}, {Targett}, {Aretxaga}, {Austermann}, {Best},
  {Bruce}, {Chapin}, {Charlot}, {Cirasuolo}, {Coppin}, {Ellis}, {Finkelstein},
  {Hayward}, {Hughes}, {Ibar}, {Jagannathan}, {Khochfar}, {Koprowski},
  {Narayanan}, {Nyland}, {Papovich}, {Peacock}, {Rieke}, {Robertson},
  {Vernstrom}, {Werf}, {Wilson}, \& {Yun}}]{Dunlop2017}
{Dunlop}, J.~S., {McLure}, R.~J., {Biggs}, A.~D., {et~al.} 2017, \mnras, 466,
  861

\bibitem[{{Elbaz} {et~al.}(2011){Elbaz}, {Dickinson}, {Hwang},
  {D{\'\i}az-Santos}, {Magdis}, {Magnelli}, {Le Borgne}, {Galliano},
  {Pannella}, {Chanial}, {Armus}, {Charmandaris}, {Daddi}, {Aussel}, {Popesso},
  {Kartaltepe}, {Altieri}, {Valtchanov}, {Coia}, {Dannerbauer}, {Dasyra},
  {Leiton}, {Mazzarella}, {Alexander}, {Buat}, {Burgarella}, {Chary}, {Gilli},
  {Ivison}, {Juneau}, {Le Floc'h}, {Lutz}, {Morrison}, {Mullaney}, {Murphy},
  {Pope}, {Scott}, {Brodwin}, {Calzetti}, {Cesarsky}, {Charlot}, {Dole},
  {Eisenhardt}, {Ferguson}, {F{\"o}rster Schreiber}, {Frayer}, {Giavalisco},
  {Huynh}, {Koekemoer}, {Papovich}, {Reddy}, {Surace}, {Teplitz}, {Yun}, \&
  {Wilson}}]{Elbaz2011}
{Elbaz}, D., {Dickinson}, M., {Hwang}, H.~S., {et~al.} 2011, \aap, 533, A119

\bibitem[{{Elbaz} {et~al.}(2018){Elbaz}, {Leiton}, {Nagar}, {Okumura},
  {Franco}, {Schreiber}, {Pannella}, {Wang}, {Dickinson}, {D{\'\i}az-Santos},
  {Ciesla}, {Daddi}, {Bournaud}, {Magdis}, {Zhou}, \& {Rujopakarn}}]{Elbaz2018}
{Elbaz}, D., {Leiton}, R., {Nagar}, N., {et~al.} 2018, \aap, 616, A110

\bibitem[{{Faisst} {et~al.}(2020){Faisst}, {Schaerer}, {Lemaux}, {Oesch},
  {Fudamoto}, {Cassata}, {B{\'e}thermin}, {Capak}, {Le F{\`e}vre}, {Silverman},
  {Yan}, {Ginolfi}, {Koekemoer}, {Morselli}, {Amor{\'\i}n}, {Bardelli},
  {Boquien}, {Brammer}, {Cimatti}, {Dessauges-Zavadsky}, {Fujimoto},
  {Gruppioni}, {Hathi}, {Hemmati}, {Ibar}, {Jones}, {Khusanova}, {Loiacono},
  {Pozzi}, {Talia}, {Tasca}, {Riechers}, {Rodighiero}, {Romano}, {Scoville},
  {Toft}, {Vallini}, {Vergani}, {Zamorani}, \& {Zucca}}]{Faisst2020}
{Faisst}, A.~L., {Schaerer}, D., {Lemaux}, B.~C., {et~al.} 2020, \apjs, 247, 61

\bibitem[{{Falkendal} {et~al.}(2019){Falkendal}, {De Breuck}, {Lehnert},
  {Drouart}, {Vernet}, {Emonts}, {Lee}, {Nesvadba}, {Seymour}, {B{\'e}thermin},
  {Kolwa}, {Gullberg}, \& {Wylezalek}}]{Falkendal2019}
{Falkendal}, T., {De Breuck}, C., {Lehnert}, M.~D., {et~al.} 2019, \aap, 621,
  A27

\bibitem[{{Figueira} {et~al.}(2022){Figueira}, {Pollo}, {Ma{\l}ek}, {Buat},
  {Boquien}, {Pistis}, {Cassar{\`a}}, {Vergani}, {Hamed}, \&
  {Salim}}]{Figueira22}
{Figueira}, M., {Pollo}, A., {Ma{\l}ek}, K., {et~al.} 2022, \aap, 667, A29

\bibitem[{{Fritz} {et~al.}(2006){Fritz}, {Franceschini}, \&
  {Hatziminaoglou}}]{Fritz2006}
{Fritz}, J., {Franceschini}, A., \& {Hatziminaoglou}, E. 2006, \mnras, 366, 767

\bibitem[{{Fudamoto} {et~al.}(2020){Fudamoto}, {Oesch}, {Faisst},
  {B{\'e}thermin}, {Ginolfi}, {Khusanova}, {Loiacono}, {Le F{\`e}vre}, {Capak},
  {Schaerer}, {Silverman}, {Cassata}, {Yan}, {Amorin}, {Bardelli}, {Boquien},
  {Cimatti}, {Dessauges-Zavadsky}, {Fujimoto}, {Gruppioni}, {Hathi}, {Ibar},
  {Jones}, {Koekemoer}, {Lagache}, {Lemaux}, {Maiolino}, {Narayanan}, {Pozzi},
  {Riechers}, {Rodighiero}, {Talia}, {Toft}, {Vallini}, {Vergani}, {Zamorani},
  \& {Zucca}}]{Fudamoto2020}
{Fudamoto}, Y., {Oesch}, P.~A., {Faisst}, A., {et~al.} 2020, \aap, 643, A4

\bibitem[{{Fujimoto} {et~al.}(2017){Fujimoto}, {Ouchi}, {Shibuya}, \&
  {Nagai}}]{Fujimoto2017}
{Fujimoto}, S., {Ouchi}, M., {Shibuya}, T., \& {Nagai}, H. 2017, \apj, 850, 83

\bibitem[{{G{\'o}mez-Guijarro} {et~al.}(2022){G{\'o}mez-Guijarro}, {Elbaz},
  {Xiao}, {B{\'e}thermin}, {Franco}, {Magnelli}, {Daddi}, {Dickinson},
  {Demarco}, {Inami}, {Rujopakarn}, {Magdis}, {Shu}, {Chary}, {Zhou},
  {Alexander}, {Bournaud}, {Ciesla}, {Ferguson}, {Finkelstein}, {Giavalisco},
  {Iono}, {Juneau}, {Kartaltepe}, {Lagache}, {Le Floc'h}, {Leiton}, {Lin},
  {Motohara}, {Mullaney}, {Okumura}, {Pannella}, {Papovich}, {Pope}, {Sargent},
  {Silverman}, {Treister}, \& {Wang}}]{Gomez}
{G{\'o}mez-Guijarro}, C., {Elbaz}, D., {Xiao}, M., {et~al.} 2022, \aap, 658,
  A43

\bibitem[{{Gruppioni} {et~al.}(2020){Gruppioni}, {B{\'e}thermin}, {Loiacono},
  {Le F{\`e}vre}, {Capak}, {Cassata}, {Faisst}, {Schaerer}, {Silverman}, {Yan},
  {Bardelli}, {Boquien}, {Carraro}, {Cimatti}, {Dessauges-Zavadsky}, {Ginolfi},
  {Fujimoto}, {Hathi}, {Jones}, {Khusanova}, {Koekemoer}, {Lagache}, {Lemaux},
  {Oesch}, {Pozzi}, {Riechers}, {Rodighiero}, {Romano}, {Talia}, {Vallini},
  {Vergani}, {Zamorani}, \& {Zucca}}]{Gruppioni2020}
{Gruppioni}, C., {B{\'e}thermin}, M., {Loiacono}, F., {et~al.} 2020, \aap, 643,
  A8

\bibitem[{{Hainline} {et~al.}(2011){Hainline}, {Shapley}, {Greene}, \&
  {Steidel}}]{Hainline2011}
{Hainline}, K.~N., {Shapley}, A.~E., {Greene}, J.~E., \& {Steidel}, C.~C. 2011,
  \apj, 733, 31

\bibitem[{{Hamed} {et~al.}(2021){Hamed}, {Ciesla}, {B{\'e}thermin}, {Ma{\l}ek},
  {Daddi}, {Sargent}, \& {Gobat}}]{Hamed21}
{Hamed}, M., {Ciesla}, L., {B{\'e}thermin}, M., {et~al.} 2021, \aap, 646, A127

\bibitem[{{Hamed} \& {Ma{\l}ek}(2022)}]{Hamed2022}
{Hamed}, M. \& {Ma{\l}ek}, K. 2022, in XL Polish Astronomical Society Meeting,
  ed. E.~{Szuszkiewicz}, A.~{Majczyna}, K.~{Ma{\l}ek}, M.~{Ratajczak},
  E.~{Niemczura}, U.~{B{\k{a}}k-St{\k{e}}{\'s}licka}, R.~{Poleski},
  M.~{Bilicki}, \& {\L}.~{Wyrzykowski}, Vol.~12, 38--41

\bibitem[{{Hirashita} {et~al.}(2017){Hirashita}, {Burgarella}, \&
  {Bouwens}}]{Hirashita2017}
{Hirashita}, H., {Burgarella}, D., \& {Bouwens}, R.~J. 2017, \mnras, 472, 4587

\bibitem[{{Hodge} {et~al.}(2016){Hodge}, {Swinbank}, {Simpson}, {Smail},
  {Walter}, {Alexander}, {Bertoldi}, {Biggs}, {Brandt}, {Chapman}, {Chen},
  {Coppin}, {Cox}, {Dannerbauer}, {Edge}, {Greve}, {Ivison}, {Karim},
  {Knudsen}, {Menten}, {Rix}, {Schinnerer}, {Wardlow}, {Weiss}, \& {van der
  Werf}}]{Hodge2016}
{Hodge}, J.~A., {Swinbank}, A.~M., {Simpson}, J.~M., {et~al.} 2016, \apj, 833,
  103

\bibitem[{{Hou} \& {Gao}(2019)}]{Hou2019}
{Hou}, L.~G. \& {Gao}, X.~Y. 2019, \mnras, 489, 4862

\bibitem[{{Hurley} {et~al.}(2017){Hurley}, {Oliver}, {Betancourt}, {Clarke},
  {Cowley}, {Duivenvoorden}, {Farrah}, {Griffin}, {Lacey}, {Le Floc'h},
  {Papadopoulos}, {Sargent}, {Scudder}, {Vaccari}, {Valtchanov}, \&
  {Wang}}]{Hurley2017}
{Hurley}, P.~D., {Oliver}, S., {Betancourt}, M., {et~al.} 2017, \mnras, 464,
  885

\bibitem[{{Ichikawa} {et~al.}(2012){Ichikawa}, {Kajisawa}, \&
  {Akhlaghi}}]{Ichikawa12}
{Ichikawa}, T., {Kajisawa}, M., \& {Akhlaghi}, M. 2012, \mnras, 422, 1014

\bibitem[{{Ikarashi} {et~al.}(2017){Ikarashi}, {Caputi}, {Ohta}, {Ivison},
  {Lagos}, {Bisigello}, {Hatsukade}, {Aretxaga}, {Dunlop}, {Hughes}, {Iono},
  {Izumi}, {Kashikawa}, {Koyama}, {Kawabe}, {Kohno}, {Motohara}, {Nakanishi},
  {Tamura}, {Umehata}, {Wilson}, {Yabe}, \& {Yun}}]{Ikarashi2017}
{Ikarashi}, S., {Caputi}, K.~I., {Ohta}, K., {et~al.} 2017, \apjl, 849, L36

\bibitem[{{Ilbert} {et~al.}(2009){Ilbert}, {Capak}, {Salvato}, {Aussel},
  {McCracken}, {Sanders}, {Scoville}, {Kartaltepe}, {Arnouts}, {Le Floc'h},
  {Mobasher}, {Taniguchi}, {Lamareille}, {Leauthaud}, {Sasaki}, {Thompson},
  {Zamojski}, {Zamorani}, {Bardelli}, {Bolzonella}, {Bongiorno}, {Brusa},
  {Caputi}, {Carollo}, {Contini}, {Cook}, {Coppa}, {Cucciati}, {de la Torre},
  {de Ravel}, {Franzetti}, {Garilli}, {Hasinger}, {Iovino}, {Kampczyk},
  {Kneib}, {Knobel}, {Kovac}, {Le Borgne}, {Le Brun}, {Le F{\`e}vre}, {Lilly},
  {Looper}, {Maier}, {Mainieri}, {Mellier}, {Mignoli}, {Murayama}, {Pell{\`o}},
  {Peng}, {P{\'e}rez-Montero}, {Renzini}, {Ricciardelli}, {Schiminovich},
  {Scodeggio}, {Shioya}, {Silverman}, {Surace}, {Tanaka}, {Tasca}, {Tresse},
  {Vergani}, \& {Zucca}}]{Ilbert2009}
{Ilbert}, O., {Capak}, P., {Salvato}, M., {et~al.} 2009, \apj, 690, 1236

\bibitem[{{Ilbert} {et~al.}(2013){Ilbert}, {McCracken}, {Le F{\`e}vre},
  {Capak}, {Dunlop}, {Karim}, {Renzini}, {Caputi}, {Boissier}, {Arnouts},
  {Aussel}, {Comparat}, {Guo}, {Hudelot}, {Kartaltepe}, {Kneib}, {Krogager},
  {Le Floc'h}, {Lilly}, {Mellier}, {Milvang-Jensen}, {Moutard}, {Onodera},
  {Richard}, {Salvato}, {Sanders}, {Scoville}, {Silverman}, {Taniguchi},
  {Tasca}, {Thomas}, {Toft}, {Tresse}, {Vergani}, {Wolk}, \&
  {Zirm}}]{Ilbert2013}
{Ilbert}, O., {McCracken}, H.~J., {Le F{\`e}vre}, O., {et~al.} 2013, \aap, 556,
  A55

\bibitem[{{Khusanova} {et~al.}(2020){Khusanova}, {B{\'e}thermin}, {Le
  F{\`e}vre}, {Capak}, {Faisst}, {Schaerer}, {Silverman}, {Cassata}, {Yan},
  {Ginolfi}, {Fudamoto}, {Loiacono}, {Amorin}, {Bardelli}, {Boquien},
  {Cimatti}, {Dessauges-Zavadsky}, {Gruppioni}, {Hathi}, {Jones}, {Koekemoer},
  {Lagache}, {Maiolino}, {Lemaux}, {Oesch}, {Pozzi}, {Riechers}, {Romano},
  {Talia}, {Toft}, {Vergani}, {Zamorani}, \& {Zucca}}]{Khusanova2020}
{Khusanova}, Y., {B{\'e}thermin}, M., {Le F{\`e}vre}, O., {et~al.} 2020, arXiv
  e-prints, arXiv:2007.08384

\bibitem[{{Komatsu} {et~al.}(2011){Komatsu}, {Smith}, {Dunkley}, {Bennett},
  {Gold}, {Hinshaw}, {Jarosik}, {Larson}, {Nolta}, {Page}, {Spergel},
  {Halpern}, {Hill}, {Kogut}, {Limon}, {Meyer}, {Odegard}, {Tucker}, {Weiland},
  {Wollack}, \& {Wright}}]{Komatsu2011}
{Komatsu}, E., {Smith}, K.~M., {Dunkley}, J., {et~al.} 2011, \apjs, 192, 18

\bibitem[{{Kriek} \& {Conroy}(2013)}]{Kriek2013}
{Kriek}, M. \& {Conroy}, C. 2013, \apjl, 775, L16

\bibitem[{{Laigle} {et~al.}(2016){Laigle}, {McCracken}, {Ilbert}, {Hsieh},
  {Davidzon}, {Capak}, {Hasinger}, {Silverman}, {Pichon}, {Coupon}, {Aussel},
  {Le Borgne}, {Caputi}, {Cassata}, {Chang}, {Civano}, {Dunlop}, {Fynbo},
  {Kartaltepe}, {Koekemoer}, {Le F{\`e}vre}, {Le Floc'h}, {Leauthaud}, {Lilly},
  {Lin}, {Marchesi}, {Milvang-Jensen}, {Salvato}, {Sanders}, {Scoville},
  {Smolcic}, {Stockmann}, {Taniguchi}, {Tasca}, {Toft}, {Vaccari}, \&
  {Zabl}}]{Laigle2016}
{Laigle}, C., {McCracken}, H.~J., {Ilbert}, O., {et~al.} 2016, \apjs, 224, 24

\bibitem[{{Lapi} {et~al.}(2018){Lapi}, {Pantoni}, {Zanisi}, {Shi}, {Mancuso},
  {Massardi}, {Shankar}, {Bressan}, \& {Danese}}]{Lapi2018}
{Lapi}, A., {Pantoni}, L., {Zanisi}, L., {et~al.} 2018, \apj, 857, 22

\bibitem[{{Leja} {et~al.}(2018){Leja}, {Johnson}, {Conroy}, \& {van
  Dokkum}}]{Leja2018}
{Leja}, J., {Johnson}, B.~D., {Conroy}, C., \& {van Dokkum}, P. 2018, \apj,
  854, 62

\bibitem[{{Lin} {et~al.}(2021){Lin}, {Hirashita}, {Camps}, \& {Baes}}]{Lin2021}
{Lin}, Y.-H., {Hirashita}, H., {Camps}, P., \& {Baes}, M. 2021, \mnras, 507,
  2755

\bibitem[{Liu {et~al.}(2019)Liu, Lang, Magnelli, Schinnerer, Leslie, Fudamoto,
  Bondi, Groves, Jim{\'{e}}nez-Andrade, Harrington, Karim, Oesch, Sargent,
  Vardoulaki, Bǎdescu, Moser, Bertoldi, Battisti, da~Cunha, Zavala, Vaccari,
  Davidzon, Riechers, \& Aravena}]{Liu2019}
Liu, D., Lang, P., Magnelli, B., {et~al.} 2019, ApJS, 244, 40

\bibitem[{{Lo Faro} {et~al.}(2017){Lo Faro}, {Buat}, {Roehlly},
  {Alvarez-Marquez}, {Burgarella}, {Silva}, \& {Efstathiou}}]{LoFaro2017}
{Lo Faro}, B., {Buat}, V., {Roehlly}, Y., {et~al.} 2017, \mnras, 472, 1372

\bibitem[{{Lo Faro} {et~al.}(2015){Lo Faro}, {Silva}, {Franceschini}, {Miller},
  \& {Efstathiou}}]{Lofaro15}
{Lo Faro}, B., {Silva}, L., {Franceschini}, A., {Miller}, N., \& {Efstathiou},
  A. 2015, \mnras, 447, 3442

\bibitem[{{Lower} {et~al.}(2022){Lower}, {Narayanan}, {Leja}, {Johnson},
  {Conroy}, \& {Dav{\'e}}}]{Lower2021}
{Lower}, S., {Narayanan}, D., {Leja}, J., {et~al.} 2022, \apj, 931, 14

\bibitem[{{Madau} \& {Dickinson}(2014)}]{Madau14}
{Madau}, P. \& {Dickinson}, M. 2014, \araa, 52, 415

\bibitem[{{Magnelli} {et~al.}(2014){Magnelli}, {Lutz}, {Saintonge}, {Berta},
  {Santini}, {Symeonidis}, {Altieri}, {Andreani}, {Aussel}, {B{\'e}thermin},
  {Bock}, {Bongiovanni}, {Cepa}, {Cimatti}, {Conley}, {Daddi}, {Elbaz},
  {F{\"o}rster Schreiber}, {Genzel}, {Ivison}, {Le Floc'h}, {Magdis},
  {Maiolino}, {Nordon}, {Oliver}, {Page}, {P{\'e}rez Garc{\'\i}a}, {Poglitsch},
  {Popesso}, {Pozzi}, {Riguccini}, {Rodighiero}, {Rosario}, {Roseboom},
  {Sanchez-Portal}, {Scott}, {Sturm}, {Tacconi}, {Valtchanov}, {Wang}, \&
  {Wuyts}}]{Magnelli2014}
{Magnelli}, B., {Lutz}, D., {Saintonge}, A., {et~al.} 2014, \aap, 561, A86

\bibitem[{{Ma{\l}ek} {et~al.}(2017){Ma{\l}ek}, {Bankowicz}, {Pollo}, {Buat},
  {Takeuchi}, {Burgarella}, {Goto}, {Malkan}, \& {Matsuhara}}]{Malek2017}
{Ma{\l}ek}, K., {Bankowicz}, M., {Pollo}, A., {et~al.} 2017, \aap, 598, A1

\bibitem[{{Ma{\l}ek} {et~al.}(2018){Ma{\l}ek}, {Buat}, {Roehlly}, {Burgarella},
  {Hurley}, {Shirley}, {Duncan}, {Efstathiou}, {Papadopoulos}, {Vaccari},
  {Farrah}, {Marchetti}, \& {Oliver}}]{Malek2018}
{Ma{\l}ek}, K., {Buat}, V., {Roehlly}, Y., {et~al.} 2018, \aap, 620, A50

\bibitem[{{Ma{\l}ek} {et~al.}(2014){Ma{\l}ek}, {Pollo}, {Takeuchi}, {Buat},
  {Burgarella}, {Malkan}, {Giovannoli}, {Kurek}, \& {Matsuura}}]{Malek2014}
{Ma{\l}ek}, K., {Pollo}, A., {Takeuchi}, T.~T., {et~al.} 2014, \aap, 562, A15

\bibitem[{{Mart{\'\i}-Vidal} {et~al.}(2012){Mart{\'\i}-Vidal},
  {P{\'e}rez-Torres}, \& {Lobanov}}]{visal12}
{Mart{\'\i}-Vidal}, I., {P{\'e}rez-Torres}, M.~A., \& {Lobanov}, A.~P. 2012,
  \aap, 541, A135

\bibitem[{{McLure} {et~al.}(2018){McLure}, {Dunlop}, {Cullen}, {Bourne},
  {Best}, {Khochfar}, {Bowler}, {Biggs}, {Geach}, {Scott}, {Micha{\l}owski},
  {Rujopakarn}, {van Kampen}, {Kirkpatrick}, \& {Pope}}]{McLure2018}
{McLure}, R.~J., {Dunlop}, J.~S., {Cullen}, F., {et~al.} 2018, \mnras, 476,
  3991

\bibitem[{{Micha{\l}owski} {et~al.}(2012){Micha{\l}owski}, {Dunlop},
  {Cirasuolo}, {Hjorth}, {Hayward}, \& {Watson}}]{Michalowski2012}
{Micha{\l}owski}, M.~J., {Dunlop}, J.~S., {Cirasuolo}, M., {et~al.} 2012, \aap,
  541, A85

\bibitem[{{Miyazaki} {et~al.}(2018){Miyazaki}, {Komiyama}, {Kawanomoto}, {Doi},
  {Furusawa}, {Hamana}, {Hayashi}, {Ikeda}, {Kamata}, {Karoji}, {Koike},
  {Kurakami}, {Miyama}, {Morokuma}, {Nakata}, {Namikawa}, {Nakaya}, {Nariai},
  {Obuchi}, {Oishi}, {Okada}, {Okura}, {Tait}, {Takata}, {Tanaka}, {Tanaka},
  {Terai}, {Tomono}, {Uraguchi}, {Usuda}, {Utsumi}, {Yamada}, {Yamanoi},
  {Aihara}, {Fujimori}, {Mineo}, {Miyatake}, {Oguri}, {Uchida}, {Tanaka},
  {Yasuda}, {Takada}, {Murayama}, {Nishizawa}, {Sugiyama}, {Chiba}, {Futamase},
  {Wang}, {Chen}, {Ho}, {Liaw}, {Chiu}, {Ho}, {Lai}, {Lee}, {Jeng}, {Iwamura},
  {Armstrong}, {Bickerton}, {Bosch}, {Gunn}, {Lupton}, {Loomis}, {Price},
  {Smith}, {Strauss}, {Turner}, {Suzuki}, {Miyazaki}, {Muramatsu}, {Yamamoto},
  {Endo}, {Ezaki}, {Ito}, {Kawaguchi}, {Sofuku}, {Taniike}, {Akutsu}, {Dojo},
  {Kasumi}, {Matsuda}, {Imoto}, {Miwa}, {Suzuki}, {Takeshi}, \&
  {Yokota}}]{Miyazaki2018}
{Miyazaki}, S., {Komiyama}, Y., {Kawanomoto}, S., {et~al.} 2018, \pasj, 70, S1

\bibitem[{{Nayyeri} {et~al.}(2014){Nayyeri}, {Mobasher}, {Hemmati}, {De
  Barros}, {Ferguson}, {Wiklind}, {Dahlen}, {Dickinson}, {Giavalisco},
  {Fontana}, {Ashby}, {Barro}, {Guo}, {Hathi}, {Kassin}, {Koekemoer},
  {Willner}, {Dunlop}, {Paris}, \& {Targett}}]{Nayyeri2014}
{Nayyeri}, H., {Mobasher}, B., {Hemmati}, S., {et~al.} 2014, \apj, 794, 68

\bibitem[{{Noeske} {et~al.}(2007){Noeske}, {Weiner}, {Faber}, {Papovich},
  {Koo}, {Somerville}, {Bundy}, {Conselice}, {Newman}, {Schiminovich}, {Le
  Floc'h}, {Coil}, {Rieke}, {Lotz}, {Primack}, {Barmby}, {Cooper}, {Davis},
  {Ellis}, {Fazio}, {Guhathakurta}, {Huang}, {Kassin}, {Martin}, {Phillips},
  {Rich}, {Small}, {Willmer}, \& {Wilson}}]{Noeske2007}
{Noeske}, K.~G., {Weiner}, B.~J., {Faber}, S.~M., {et~al.} 2007, \apjl, 660,
  L43

\bibitem[{{Noll} {et~al.}(2009){Noll}, {Burgarella}, {Giovannoli}, {Buat},
  {Marcillac}, \& {Mu{\~n}oz-Mateos}}]{Noll2009}
{Noll}, S., {Burgarella}, D., {Giovannoli}, E., {et~al.} 2009, \aap, 507, 1793

\bibitem[{{Pantoni} {et~al.}(2021{\natexlab{a}}){Pantoni}, {Lapi}, {Massardi},
  {Donevski}, {Bressan}, {Silva}, {Pozzi}, {Vignali}, {Talia}, {Cimatti},
  {Ronconi}, \& {Danese}}]{Pantoni20212}
{Pantoni}, L., {Lapi}, A., {Massardi}, M., {et~al.} 2021{\natexlab{a}}, \mnras,
  504, 928

\bibitem[{{Pantoni} {et~al.}(2019){Pantoni}, {Lapi}, {Massardi}, {Goswami}, \&
  {Danese}}]{Pantoni2019}
{Pantoni}, L., {Lapi}, A., {Massardi}, M., {Goswami}, S., \& {Danese}, L. 2019,
  \apj, 880, 129

\bibitem[{{Pantoni} {et~al.}(2021{\natexlab{b}}){Pantoni}, {Massardi}, {Lapi},
  {Donevski}, {D'Amato}, {Giulietti}, {Pozzi}, {Talia}, {Vignali}, {Cimatti},
  {Silva}, {Bressan}, \& {Ronconi}}]{Pantoni2021}
{Pantoni}, L., {Massardi}, M., {Lapi}, A., {et~al.} 2021{\natexlab{b}}, \mnras,
  507, 3998

\bibitem[{{Pearson} {et~al.}(2019){Pearson}, {Wang}, {Alpaslan}, {Baldry},
  {Bilicki}, {Brown}, {Grootes}, {Holwerda}, {Kitching}, {Kruk}, \& {van der
  Tak}}]{Pearson2019}
{Pearson}, W.~J., {Wang}, L., {Alpaslan}, M., {et~al.} 2019, \aap, 631, A51

\bibitem[{{Pearson} {et~al.}(2018){Pearson}, {Wang}, {Hurley}, {Ma{\l}ek},
  {Buat}, {Burgarella}, {Farrah}, {Oliver}, {Smith}, \& {van der
  Tak}}]{Pearson2018}
{Pearson}, W.~J., {Wang}, L., {Hurley}, P.~D., {et~al.} 2018, \aap, 615, A146

\bibitem[{{Pearson} {et~al.}(2017){Pearson}, {Wang}, {van der Tak}, {Hurley},
  {Burgarella}, \& {Oliver}}]{Pearson2017}
{Pearson}, W.~J., {Wang}, L., {van der Tak}, F.~F.~S., {et~al.} 2017, \aap,
  603, A102

\bibitem[{{Peng} {et~al.}(2002){Peng}, {Ho}, {Impey}, \& {Rix}}]{Peng2002}
{Peng}, C.~Y., {Ho}, L.~C., {Impey}, C.~D., \& {Rix}, H.-W. 2002, \aj, 124, 266

\bibitem[{{Popping} {et~al.}(2017){Popping}, {Somerville}, \&
  {Galametz}}]{Popping2017}
{Popping}, G., {Somerville}, R.~S., \& {Galametz}, M. 2017, \mnras, 471, 3152

\bibitem[{{Reuter} {et~al.}(2020){Reuter}, {Vieira}, {Spilker}, {Weiss},
  {Aravena}, {Archipley}, {B{\'e}thermin}, {Chapman}, {De Breuck}, {Dong},
  {Everett}, {Fu}, {Greve}, {Hayward}, {Hill}, {Hezaveh}, {Jarugula}, {Litke},
  {Malkan}, {Marrone}, {Narayanan}, {Phadke}, {Stark}, \&
  {Strandet}}]{Reuter2020}
{Reuter}, C., {Vieira}, J.~D., {Spilker}, J.~S., {et~al.} 2020, \apj, 902, 78

\bibitem[{{Rodighiero} {et~al.}(2011){Rodighiero}, {Daddi}, {Baronchelli},
  {Cimatti}, {Renzini}, {Aussel}, {Popesso}, {Lutz}, {Andreani}, {Berta},
  {Cava}, {Elbaz}, {Feltre}, {Fontana}, {F{\"o}rster Schreiber},
  {Franceschini}, {Genzel}, {Grazian}, {Gruppioni}, {Ilbert}, {Le Floch},
  {Magdis}, {Magliocchetti}, {Magnelli}, {Maiolino}, {McCracken}, {Nordon},
  {Poglitsch}, {Santini}, {Pozzi}, {Riguccini}, {Tacconi}, {Wuyts}, \&
  {Zamorani}}]{Rodighiero2011}
{Rodighiero}, G., {Daddi}, E., {Baronchelli}, I., {et~al.} 2011, \apjl, 739,
  L40

\bibitem[{{Rujopakarn} {et~al.}(2016){Rujopakarn}, {Dunlop}, {Rieke}, {Ivison},
  {Cibinel}, {Nyland}, {Jagannathan}, {Silverman}, {Alexander}, {Biggs},
  {Bhatnagar}, {Ballantyne}, {Dickinson}, {Elbaz}, {Geach}, {Hayward},
  {Kirkpatrick}, {McLure}, {Micha{\l}owski}, {Miller}, {Narayanan}, {Owen},
  {Pannella}, {Papovich}, {Pope}, {Rau}, {Robertson}, {Scott}, {Swinbank}, {van
  der Werf}, {van Kampen}, {Weiner}, \& {Windhorst}}]{Rujopakarn16}
{Rujopakarn}, W., {Dunlop}, J.~S., {Rieke}, G.~H., {et~al.} 2016, \apj, 833, 12

\bibitem[{{Salim} \& {Boquien}(2019)}]{Salim2019}
{Salim}, S. \& {Boquien}, M. 2019, \apj, 872, 23

\bibitem[{{Salim} {et~al.}(2018){Salim}, {Boquien}, \& {Lee}}]{salim2018}
{Salim}, S., {Boquien}, M., \& {Lee}, J.~C. 2018, \apj, 859, 11

\bibitem[{{Salim} \& {Narayanan}(2020)}]{Salim2020}
{Salim}, S. \& {Narayanan}, D. 2020, arXiv e-prints, arXiv:2001.03181

\bibitem[{{Schreiber} {et~al.}(2018){Schreiber}, {Glazebrook}, {Nanayakkara},
  {Kacprzak}, {Labb{\'e}}, {Oesch}, {Yuan}, {Tran}, {Papovich}, {Spitler}, \&
  {Straatman}}]{Schreiber2018}
{Schreiber}, C., {Glazebrook}, K., {Nanayakkara}, T., {et~al.} 2018, \aap, 618,
  A85

\bibitem[{{Schreiber} {et~al.}(2015){Schreiber}, {Pannella}, {Elbaz},
  {B{\'e}thermin}, {Inami}, {Dickinson}, {Magnelli}, {Wang}, {Aussel}, {Daddi},
  {Juneau}, {Shu}, {Sargent}, {Buat}, {Faber}, {Ferguson}, {Giavalisco},
  {Koekemoer}, {Magdis}, {Morrison}, {Papovich}, {Santini}, \&
  {Scott}}]{Schreiber15}
{Schreiber}, C., {Pannella}, M., {Elbaz}, D., {et~al.} 2015, \aap, 575, A74

\bibitem[{{Schulz} {et~al.}(2020){Schulz}, {Popping}, {Pillepich}, {Nelson},
  {Vogelsberger}, {Marinacci}, \& {Hernquist}}]{schulz2020}
{Schulz}, S., {Popping}, G., {Pillepich}, A., {et~al.} 2020, \mnras, 497, 4773

\bibitem[{{Scoville} {et~al.}(2007){Scoville}, {Aussel}, {Brusa}, {Capak},
  {Carollo}, {Elvis}, {Giavalisco}, {Guzzo}, {Hasinger}, {Impey}, {Kneib},
  {LeFevre}, {Lilly}, {Mobasher}, {Renzini}, {Rich}, {Sanders}, {Schinnerer},
  {Schminovich}, {Shopbell}, {Taniguchi}, \& {Tyson}}]{Scoville2007}
{Scoville}, N., {Aussel}, H., {Brusa}, M., {et~al.} 2007, \apjs, 172, 1

\bibitem[{{Shirley} {et~al.}(2019){Shirley}, {Roehlly}, {Hurley}, {Buat},
  {Campos Varillas}, {Duivenvoorden}, {Duncan}, {Efstathiou}, {Farrah},
  {Gonz{\'a}lez Solares}, {Malek}, {Marchetti}, {McCheyne}, {Papadopoulos},
  {Pons}, {Scipioni}, {Vaccari}, \& {Oliver}}]{Shirley2019}
{Shirley}, R., {Roehlly}, Y., {Hurley}, P.~D., {et~al.} 2019, \mnras, 490, 634

\bibitem[{{Smol{\v{c}}i{\'c}} {et~al.}(2017){Smol{\v{c}}i{\'c}}, {Novak},
  {Delvecchio}, {Ceraj}, {Bondi}, {Delhaize}, {Marchesi}, {Murphy},
  {Schinnerer}, {Vardoulaki}, \& {Zamorani}}]{Smolcic2017}
{Smol{\v{c}}i{\'c}}, V., {Novak}, M., {Delvecchio}, I., {et~al.} 2017, \aap,
  602, A6

\bibitem[{{Speagle} {et~al.}(2014){Speagle}, {Steinhardt}, {Capak}, \&
  {Silverman}}]{speagle2014}
{Speagle}, J.~S., {Steinhardt}, C.~L., {Capak}, P.~L., \& {Silverman}, J.~D.
  2014, \apjs, 214, 15

\bibitem[{{Strandet} {et~al.}(2016){Strandet}, {Weiss}, {Vieira}, {de Breuck},
  {Aguirre}, {Aravena}, {Ashby}, {B{\'e}thermin}, {Bradford}, {Carlstrom},
  {Chapman}, {Crawford}, {Everett}, {Fassnacht}, {Furstenau}, {Gonzalez},
  {Greve}, {Gullberg}, {Hezaveh}, {Kamenetzky}, {Litke}, {Ma}, {Malkan},
  {Marrone}, {Menten}, {Murphy}, {Nadolski}, {Rotermund}, {Spilker}, {Stark},
  \& {Welikala}}]{Strandet2016}
{Strandet}, M.~L., {Weiss}, A., {Vieira}, J.~D., {et~al.} 2016, \apj, 822, 80

\bibitem[{{Takeuchi} {et~al.}(2005){Takeuchi}, {Buat}, \&
  {Burgarella}}]{Takeuchi2005}
{Takeuchi}, T.~T., {Buat}, V., \& {Burgarella}, D. 2005, \aap, 440, L17

\bibitem[{{Toft} {et~al.}(2014){Toft}, {Smol{\v{c}}i{\'c}}, {Magnelli},
  {Karim}, {Zirm}, {Michalowski}, {Capak}, {Sheth}, {Schawinski}, {Krogager},
  {Wuyts}, {Sanders}, {Man}, {Lutz}, {Staguhn}, {Berta}, {Mccracken}, {Krpan},
  \& {Riechers}}]{Toft2014}
{Toft}, S., {Smol{\v{c}}i{\'c}}, V., {Magnelli}, B., {et~al.} 2014, \apj, 782,
  68

\bibitem[{{Viero} {et~al.}(2013){Viero}, {Wang}, {Zemcov}, {Addison},
  {Amblard}, {Arumugam}, {Aussel}, {B{\'e}thermin}, {Bock}, {Boselli}, {Buat},
  {Burgarella}, {Casey}, {Clements}, {Conley}, {Conversi}, {Cooray}, {De
  Zotti}, {Dowell}, {Farrah}, {Franceschini}, {Glenn}, {Griffin},
  {Hatziminaoglou}, {Heinis}, {Ibar}, {Ivison}, {Lagache}, {Levenson},
  {Marchetti}, {Marsden}, {Nguyen}, {O'Halloran}, {Oliver}, {Omont}, {Page},
  {Papageorgiou}, {Pearson}, {P{\'e}rez-Fournon}, {Pohlen}, {Rigopoulou},
  {Roseboom}, {Rowan-Robinson}, {Schulz}, {Scott}, {Seymour}, {Shupe}, {Smith},
  {Symeonidis}, {Vaccari}, {Valtchanov}, {Vieira}, {Wardlow}, \&
  {Xu}}]{Viero2013}
{Viero}, M.~P., {Wang}, L., {Zemcov}, M., {et~al.} 2013, \apj, 772, 77

\bibitem[{{Wang} {et~al.}(2017){Wang}, {Hirashita}, \& {Hou}}]{Wang2017}
{Wang}, W.-C., {Hirashita}, H., \& {Hou}, K.-C. 2017, \mnras, 465, 3475

\bibitem[{{Wei{\ss}} {et~al.}(2013){Wei{\ss}}, {De Breuck}, {Marrone},
  {Vieira}, {Aguirre}, {Aird}, {Aravena}, {Ashby}, {Bayliss}, {Benson},
  {B{\'e}thermin}, {Biggs}, {Bleem}, {Bock}, {Bothwell}, {Bradford}, {Brodwin},
  {Carlstrom}, {Chang}, {Chapman}, {Crawford}, {Crites}, {de Haan}, {Dobbs},
  {Downes}, {Fassnacht}, {George}, {Gladders}, {Gonzalez}, {Greve},
  {Halverson}, {Hezaveh}, {High}, {Holder}, {Holzapfel}, {Hoover}, {Hrubes},
  {Husband}, {Keisler}, {Lee}, {Leitch}, {Lueker}, {Luong-Van}, {Malkan},
  {McIntyre}, {McMahon}, {Mehl}, {Menten}, {Meyer}, {Murphy}, {Padin},
  {Plagge}, {Reichardt}, {Rest}, {Rosenman}, {Ruel}, {Ruhl}, {Schaffer},
  {Shirokoff}, {Spilker}, {Stalder}, {Staniszewski}, {Stark}, {Story},
  {Vanderlinde}, {Welikala}, \& {Williamson}}]{Weiss2013}
{Wei{\ss}}, A., {De Breuck}, C., {Marrone}, D.~P., {et~al.} 2013, \apj, 767, 88

\bibitem[{{Whitaker} {et~al.}(2017){Whitaker}, {Pope}, {Cybulski}, {Casey},
  {Popping}, \& {Yun}}]{Whitaker2017}
{Whitaker}, K.~E., {Pope}, A., {Cybulski}, R., {et~al.} 2017, \apj, 850, 208

\bibitem[{{Whitaker} {et~al.}(2013){Whitaker}, {van Dokkum}, {Brammer},
  {Momcheva}, {Skelton}, {Franx}, {Kriek}, {Labb{\'e}}, {Fumagalli},
  {Lundgren}, {Nelson}, {Patel}, \& {Rix}}]{Whitaker2013}
{Whitaker}, K.~E., {van Dokkum}, P.~G., {Brammer}, G., {et~al.} 2013, \apjl,
  770, L39

\bibitem[{{Wild} {et~al.}(2011){Wild}, {Charlot}, {Brinchmann}, {Heckman},
  {Vince}, {Pacifici}, \& {Chevallard}}]{Wild2011}
{Wild}, V., {Charlot}, S., {Brinchmann}, J., {et~al.} 2011, \mnras, 417, 1760

\bibitem[{{Zeimann} {et~al.}(2015){Zeimann}, {Ciardullo}, {Gronwall}, {Bridge},
  {Brooks}, {Fox}, {Gawiser}, {Gebhardt}, {Hagen}, {Schneider}, \&
  {Trump}}]{Zeimann2015}
{Zeimann}, G.~R., {Ciardullo}, R., {Gronwall}, C., {et~al.} 2015, \apj, 814,
  162

\end{thebibliography}

\begin{appendix}

\section{Choice of SFH and its parameters}\label{appendix:b}
We tested several SFH models in order to find a proper set of templates which can not only fit the data, but also are physically motivated. Among those models were the simple delayed SFH, which did not succeed in fitting the photometry of the majority of our sample. We therefore added an exponential recent burst. This delayed SFH with an exponential burst fitted the UV part of our SEDs the best (provided significantly better reduced $\chi^2$). We also tested the need to include a recent quench episode in our models, by using a truncated SFH \citep{Ciesla2017} that allows a recent burst or a recent quench episode. This was motivated by previous studies that suggested that SFGs can have significantly low molecular gas masses and therefore might be experiencing a drop in their star formation activity \citep[e.g.,][]{Elbaz2018, Falkendal2019, Hamed21}, and rapidly transitioning to the red sequence.

The truncated SFH did not provide any improvement of the SEDs nor did it change the estimated stellar masses and the SFRs, and there was no evidence of recent quenching episode for the whole sample.

Our analysis suggested a strong dependence of the instantaneous SFR on the choice of the age of recent burst episodes. This was characterized in extremely huge values of derived SFRs (>10000 M$_{\odot}yr^{-1}$) for some of our galaxies when a lower value of burst age was preferred (<5 Myr). Hence we chose a lower limit of the recent burst of 5$\pm$1 Myr.

\section{Redshift distribution of attenuation preference}\label{appendix:a2}

\begin{figure}
\centering
\minipage{0.4\textwidth}
  \includegraphics[width=\linewidth]{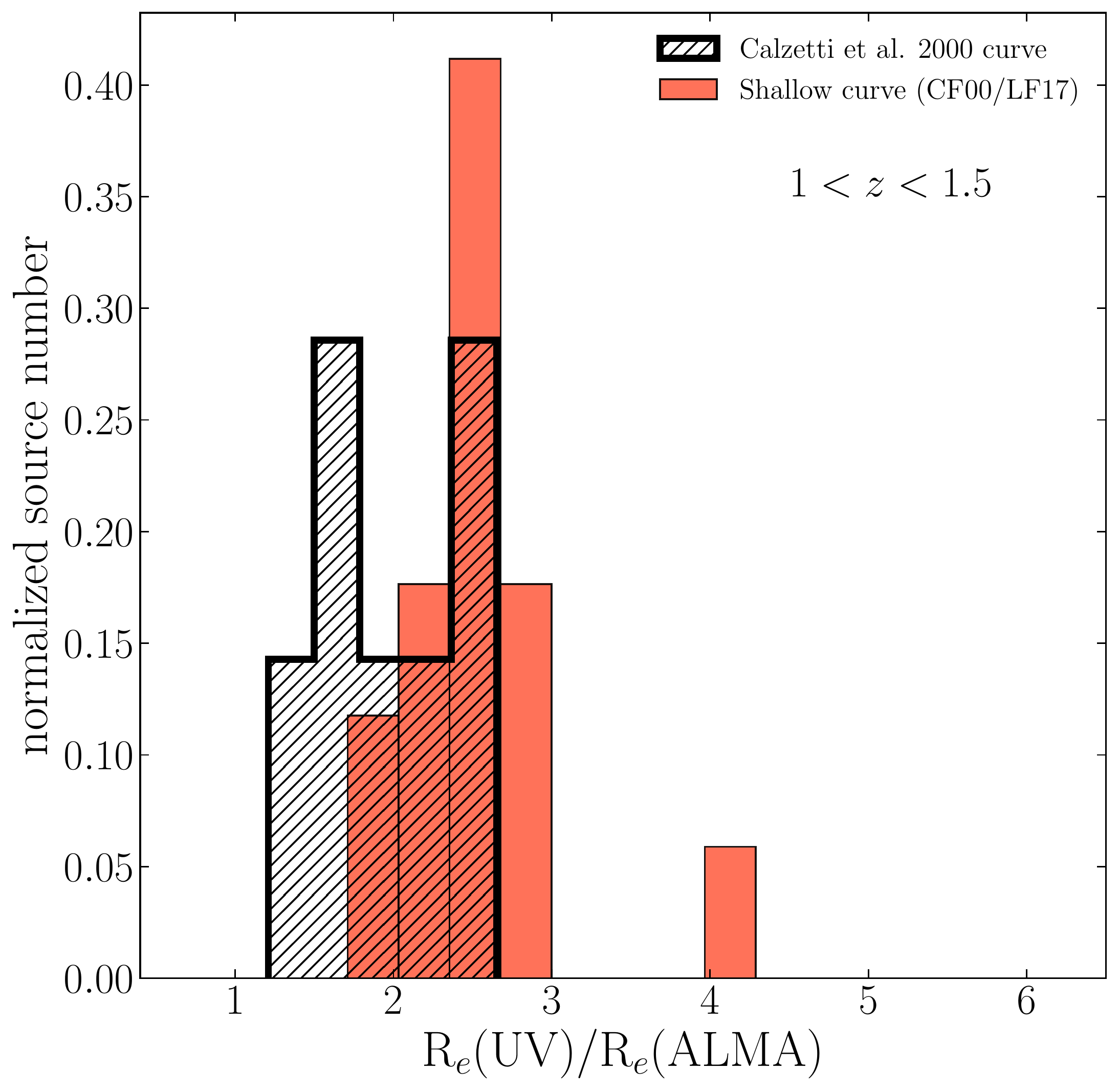}
\endminipage\\
\minipage{0.4\textwidth}
  \includegraphics[width=\linewidth]{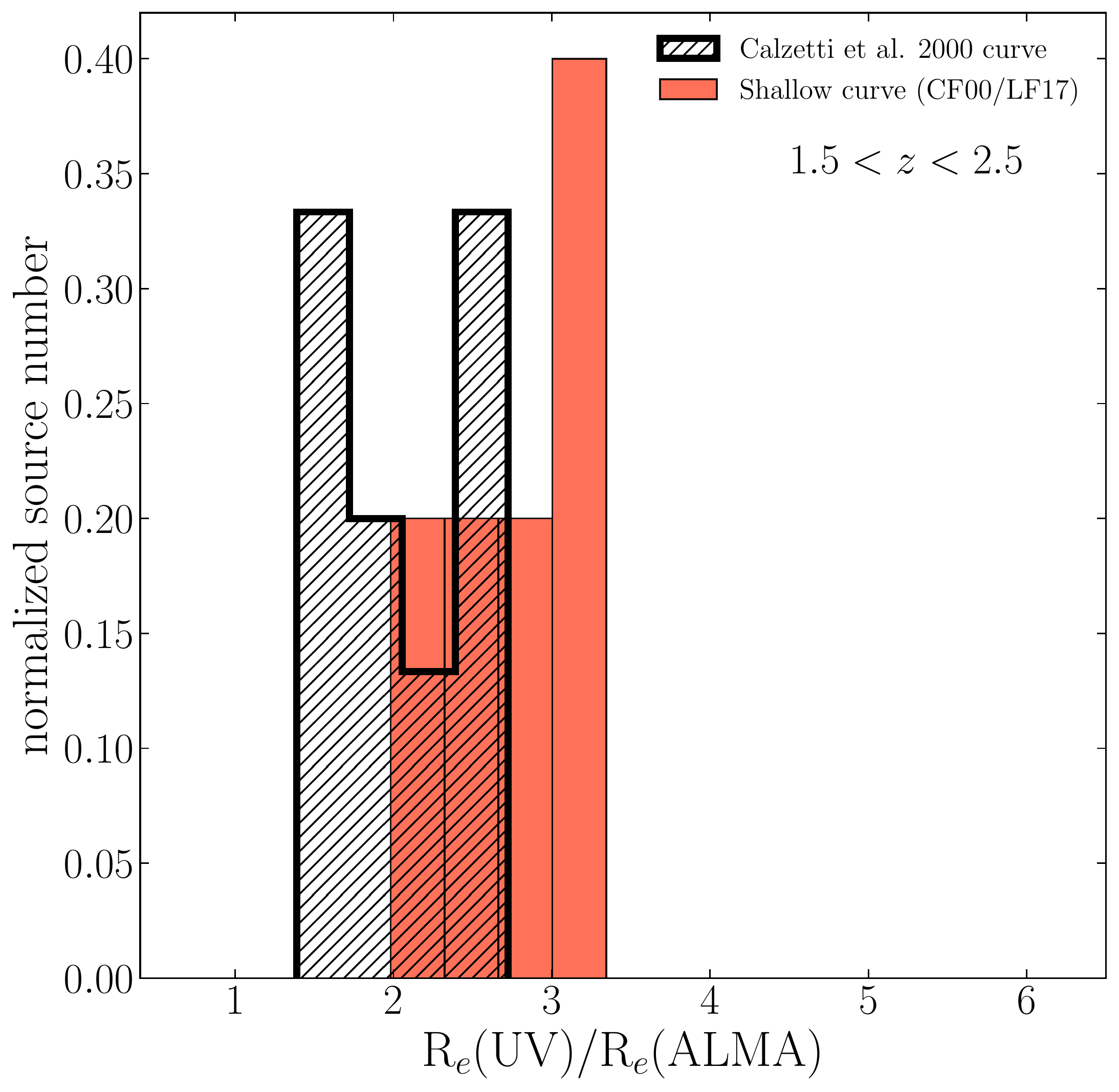}
\endminipage\\
\minipage{0.4\textwidth}%
  \includegraphics[width=\linewidth]{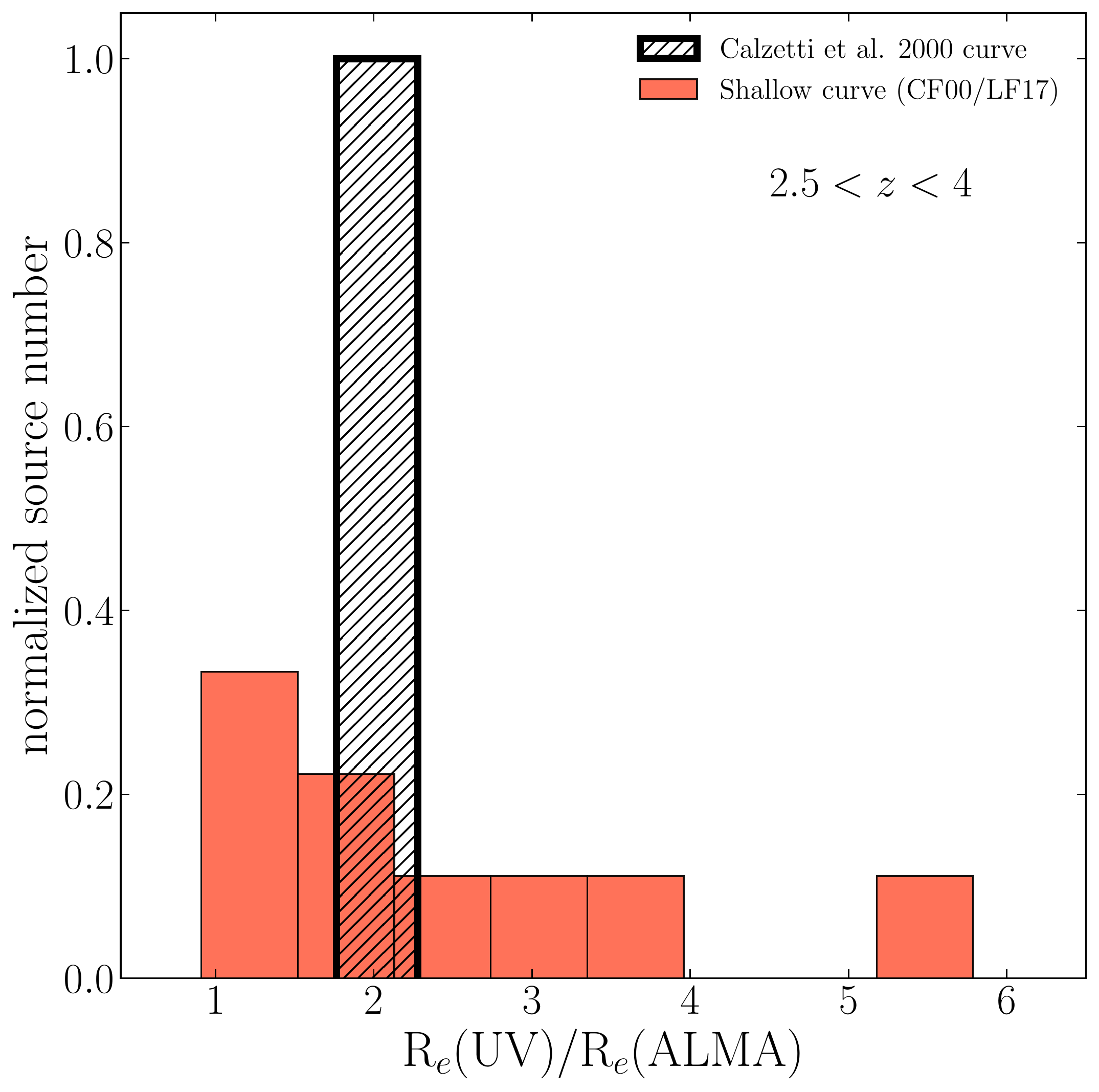}
\endminipage\\
\caption{Histogram of the ratio of HSC's $Y$ band radius to the ALMA radius. The red-filled histograms at different redshift ranges represent galaxies for which a shallow attenuation curve of CF00 or LF17 was critical in order to reproduce the observed stellar light and resulted in better fits. The dashed histogram shows galaxies for which the C00 attenuation curve gave a satisfying fit.}
\label{fig:uv_alma}
\end{figure}

In Fig. \ref{fig:uv_alma} We show the distribution of the relative compactness distribution of our sample based on the preferred attenuation law for each galaxy. Even though the C00 attenuation curve seems to fit the sample better when the R$_{e}$(Y)/R$_{e}$(ALMA) $\sim2$. However, shallow attenuation curves do not seem to have a redshift preference. Despite the fact that we have a large sample that covers a wide redshift range, our sample did not cover a large range statistically in terms of star-to-dust compactness.
\section{Size evolution with physical properties}\label{appendix:a}
\begin{figure}[h!]
    \centering
        \includegraphics[width=0.5\textwidth]{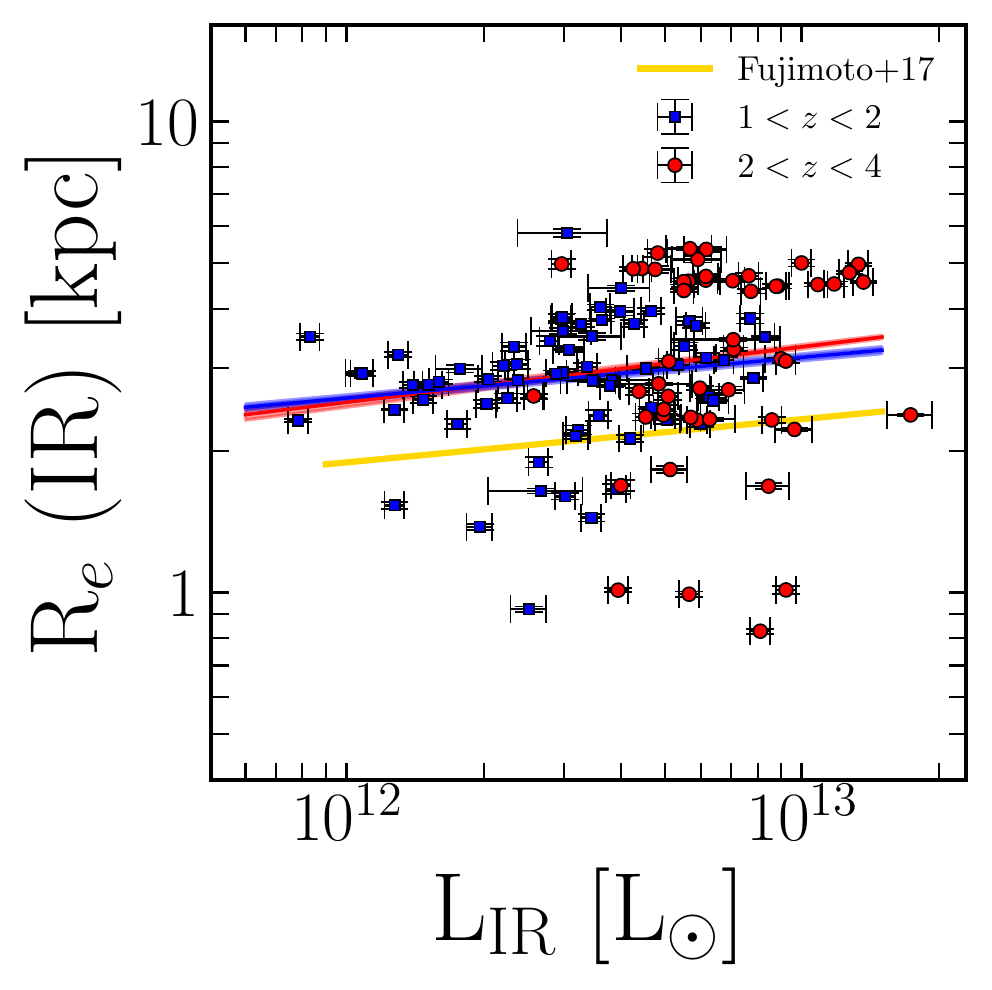}
        \caption{\footnotesize Effective radii of the UV and ALMA detections of our sources, with respect to the IR luminosities obtained from the best fits using \texttt{CIGALE}. The yellow line shows the trend found in \citet{Fujimoto2017}.}
        \label{fig:radius_LIR}
\end{figure}

\begin{figure}[h!]
    \centering
        \includegraphics[width=0.5\textwidth]{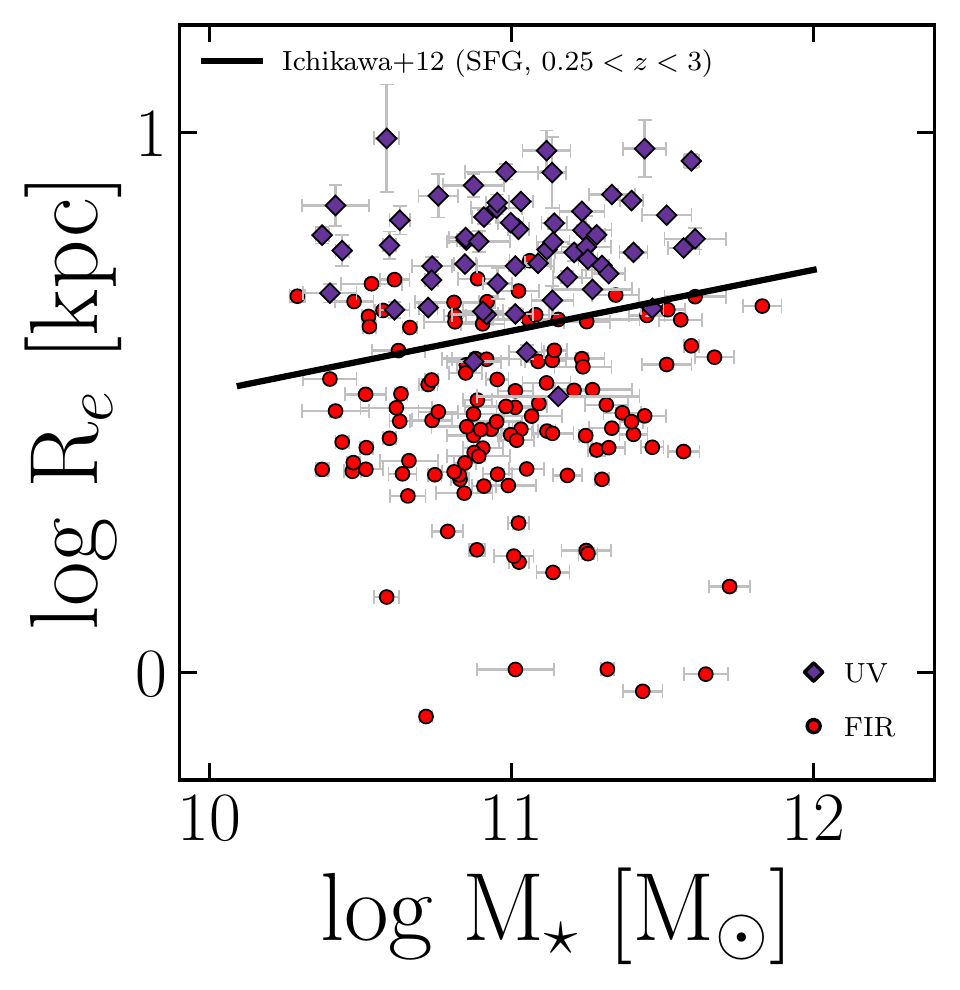}
        \caption{\footnotesize Effective radii of the UV and ALMA detections of our sources, with respect to the stellar masses obtained from the best fits using \texttt{CIGALE}.}
        \label{fig:radius_mstar}
\end{figure}

Here we show the evolution of the derived effective radii of DSFGs with the physical properties of DSFGs, in particular their dust luminosities and the stellar masses. The calculated radii using the two measured bands when available, seem to be constant at different L$_{IR}$ (Fig.~\ref{fig:radius_LIR}) and stellar masses (Fig.~\ref{fig:radius_mstar}) These results agree with \citet{Fujimoto2017}, although the mean value of FIR effective radii of our sample are $\sim$16$\%$ higher than that of \citet{Fujimoto2017}. This difference is due to the different method in calculating these radii in the $uv$ planes, the angle of their fit is similar to the literature at different redshift ranges.  Our results also show that $R_{e_{\mathrm{(FIR)}}}$ are smaller than $R_{e_{\mathrm{(UV,opt)}}}$ for SFGs \citep{Ichikawa12,Fujimoto2017}.

\newpage
\section{CIGALE mock analysis}\label{appendix:c}
\begin{figure*}[ht!]
    \centering
    \minipage{0.32\textwidth}
    \includegraphics[width=\linewidth]{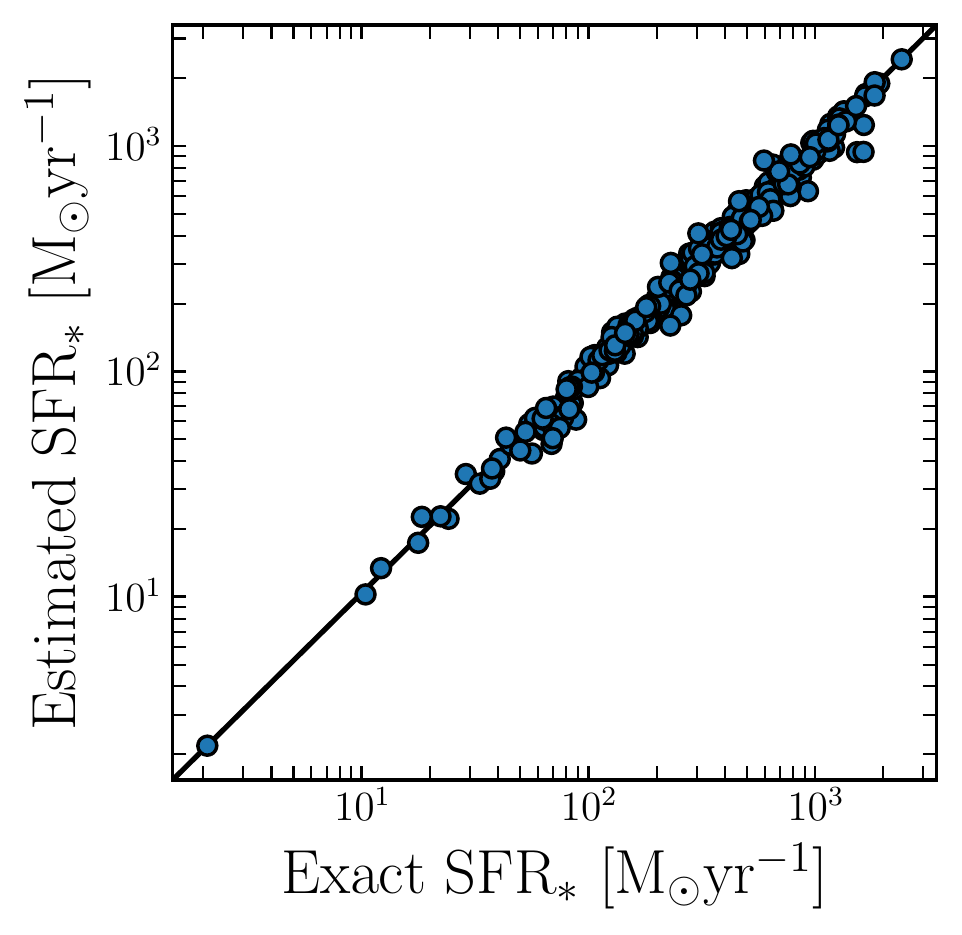}
    \endminipage
    \minipage{0.32\textwidth}
    \includegraphics[width=\linewidth]{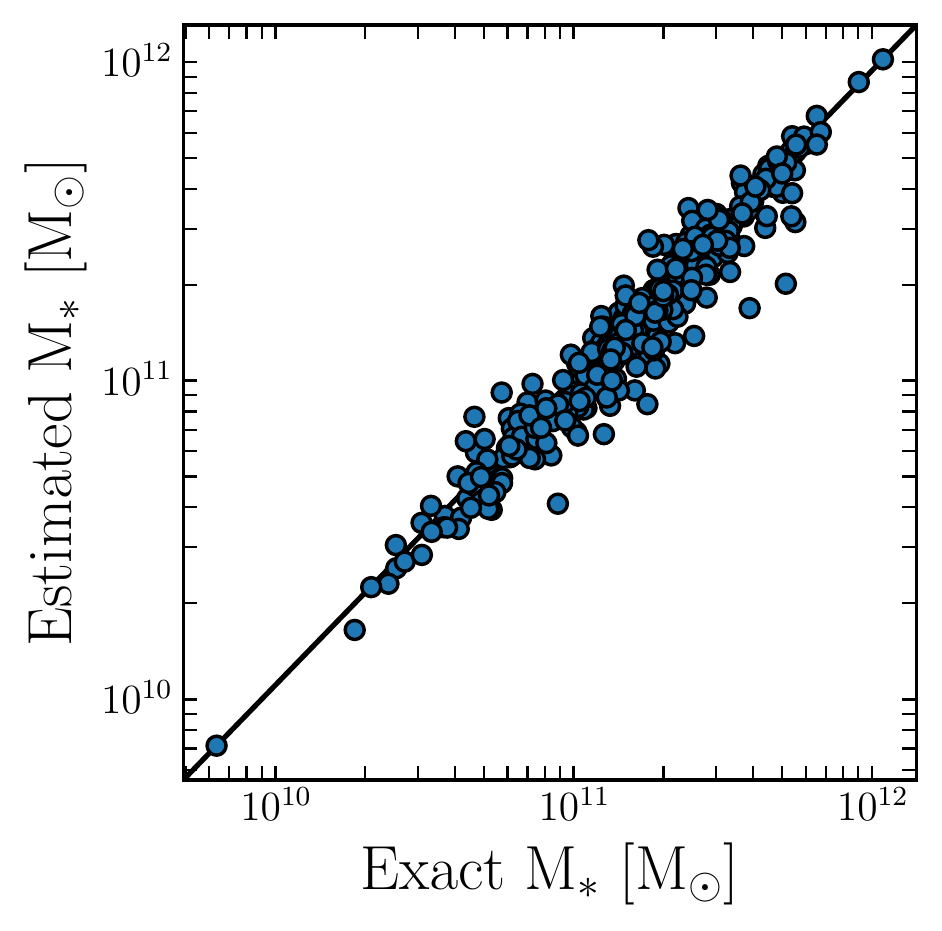}
    \endminipage
    \minipage{0.32\textwidth}
    \includegraphics[width=\linewidth]{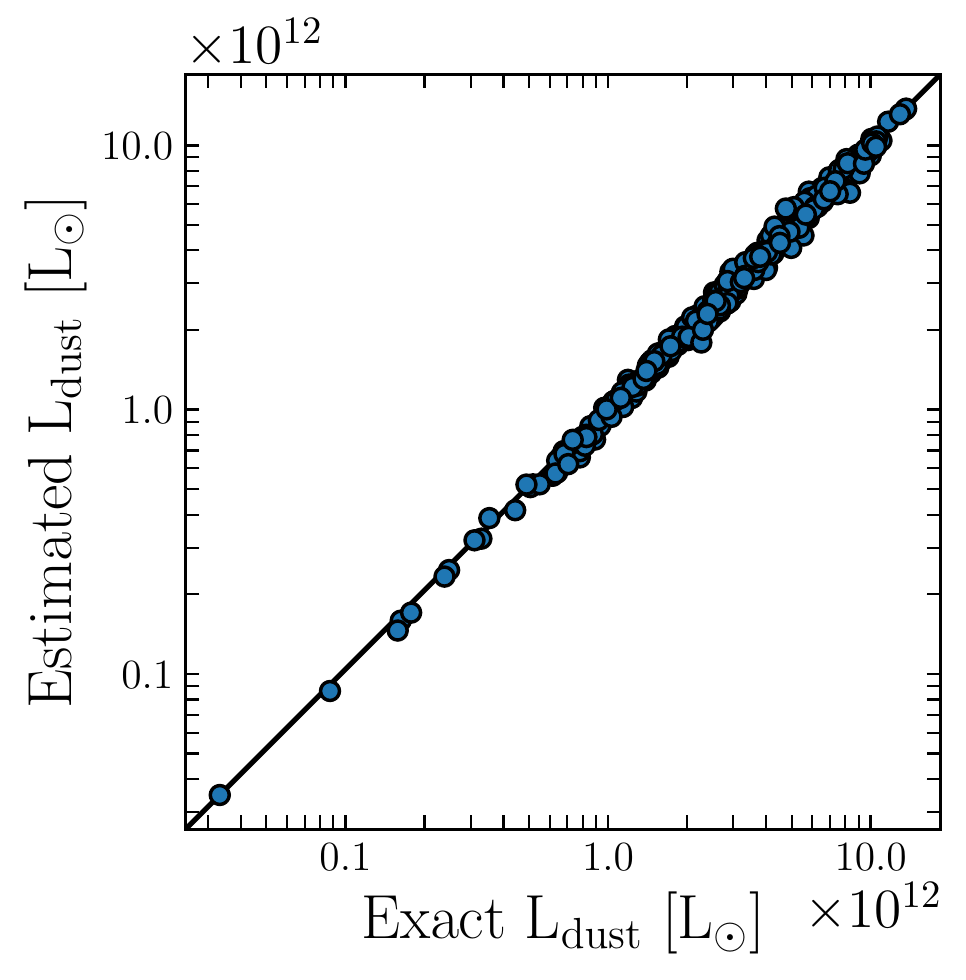}
    \endminipage\\
    \centering
    \minipage{0.332\textwidth}
    \includegraphics[width=\linewidth]{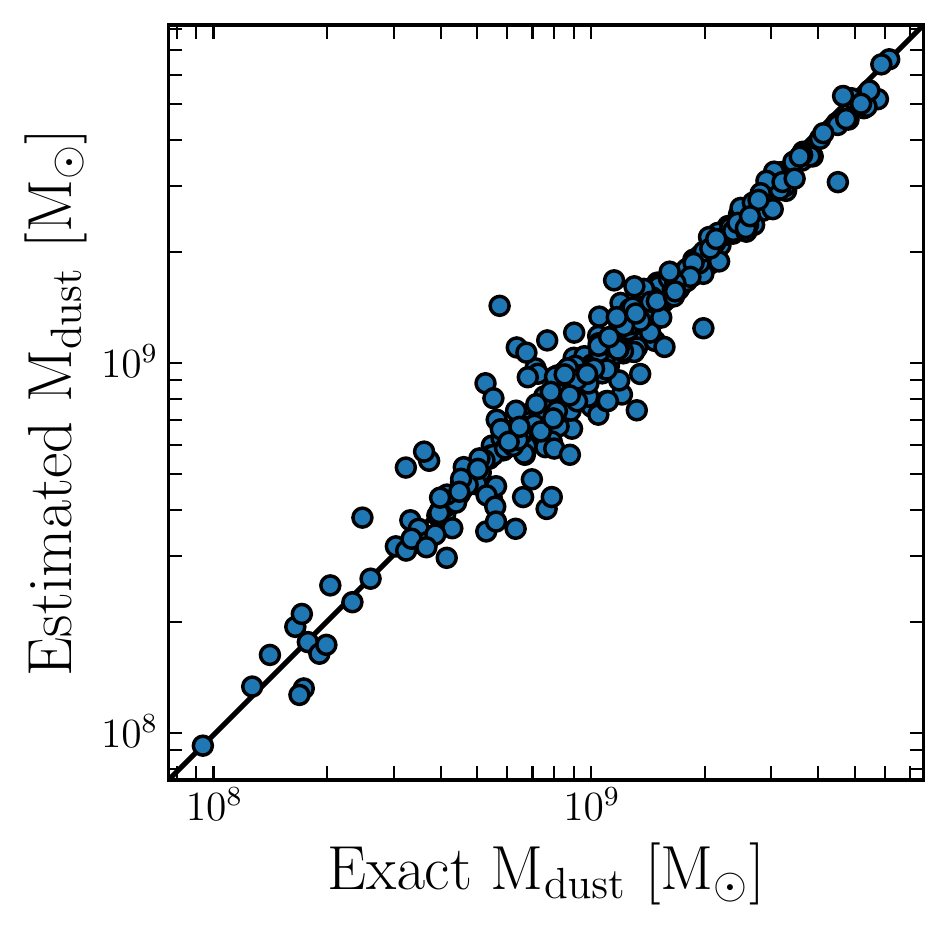}
    \endminipage
    \minipage{0.32\textwidth}
    \includegraphics[width=\linewidth]{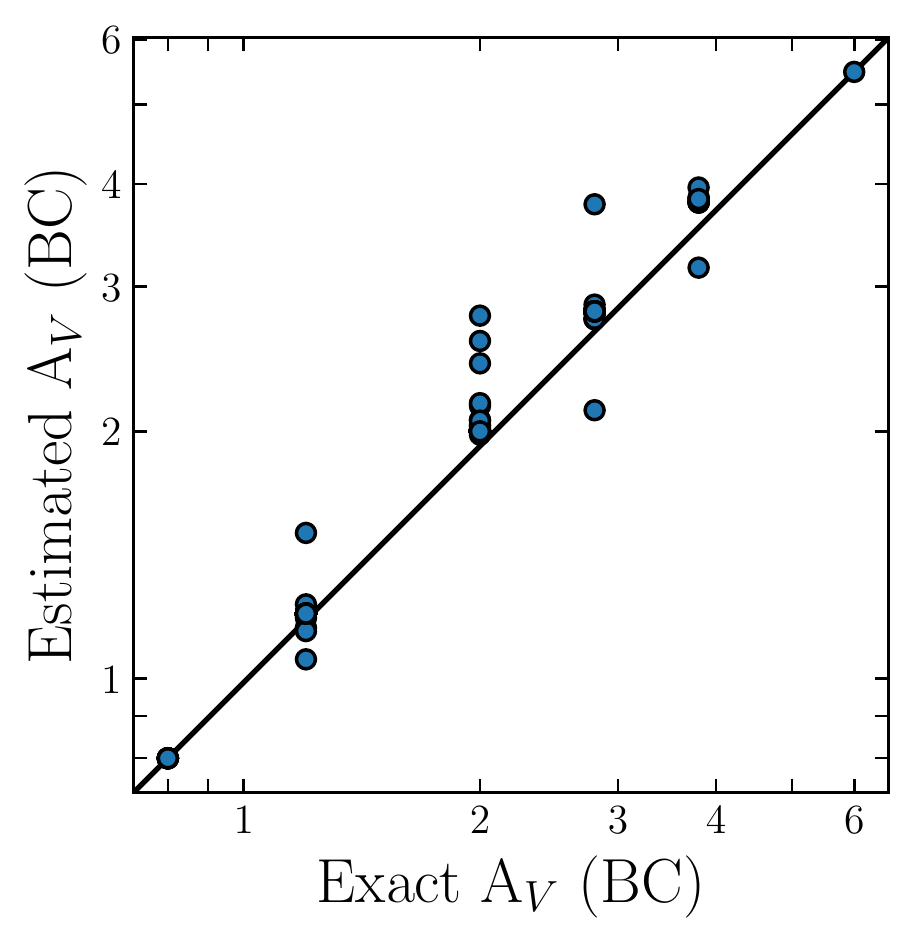}
    \endminipage
    \minipage{0.32\textwidth}
    \includegraphics[width=\linewidth]{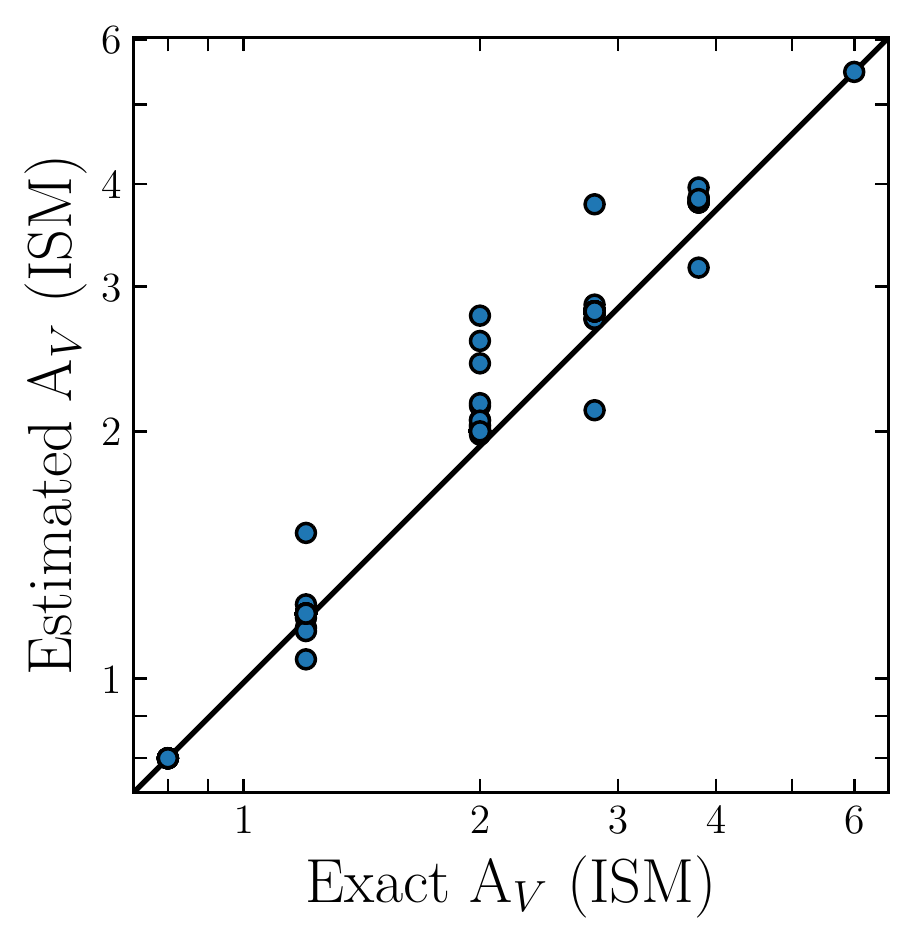}
    \endminipage\\
    \centering
    \minipage{0.32\textwidth}
    \includegraphics[width=\linewidth]{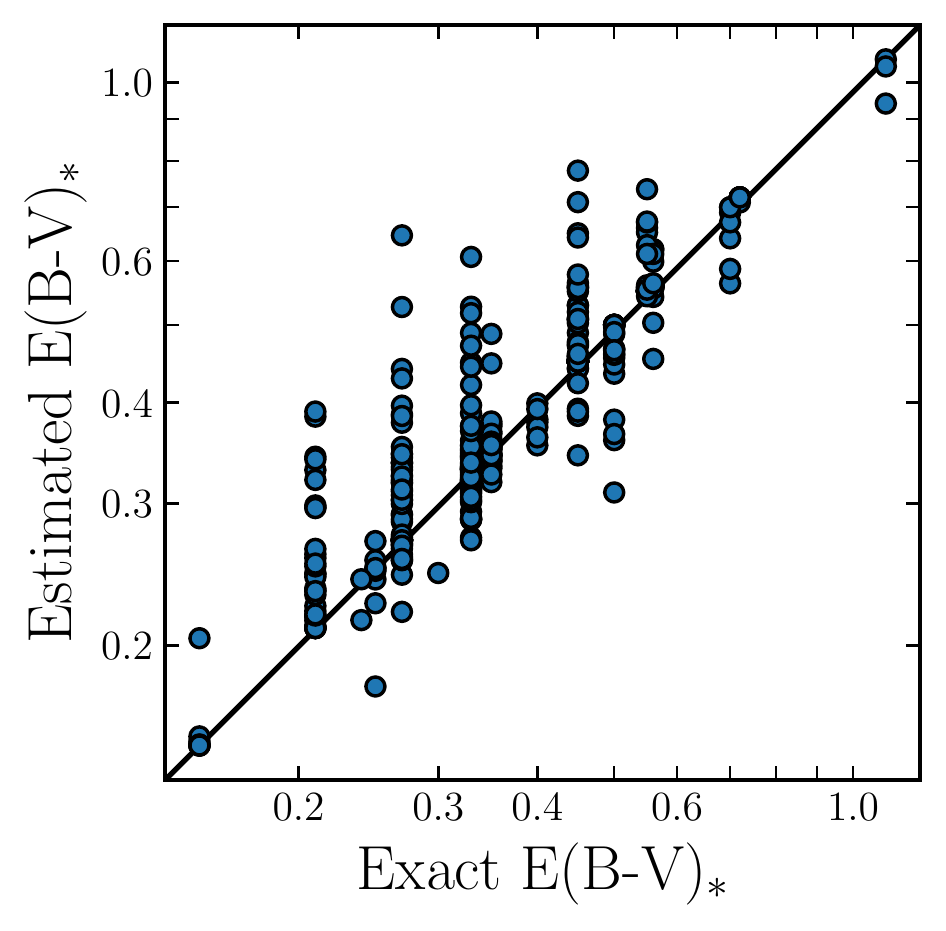}
    \endminipage
    \minipage{0.32\textwidth}
    \includegraphics[width=\linewidth]{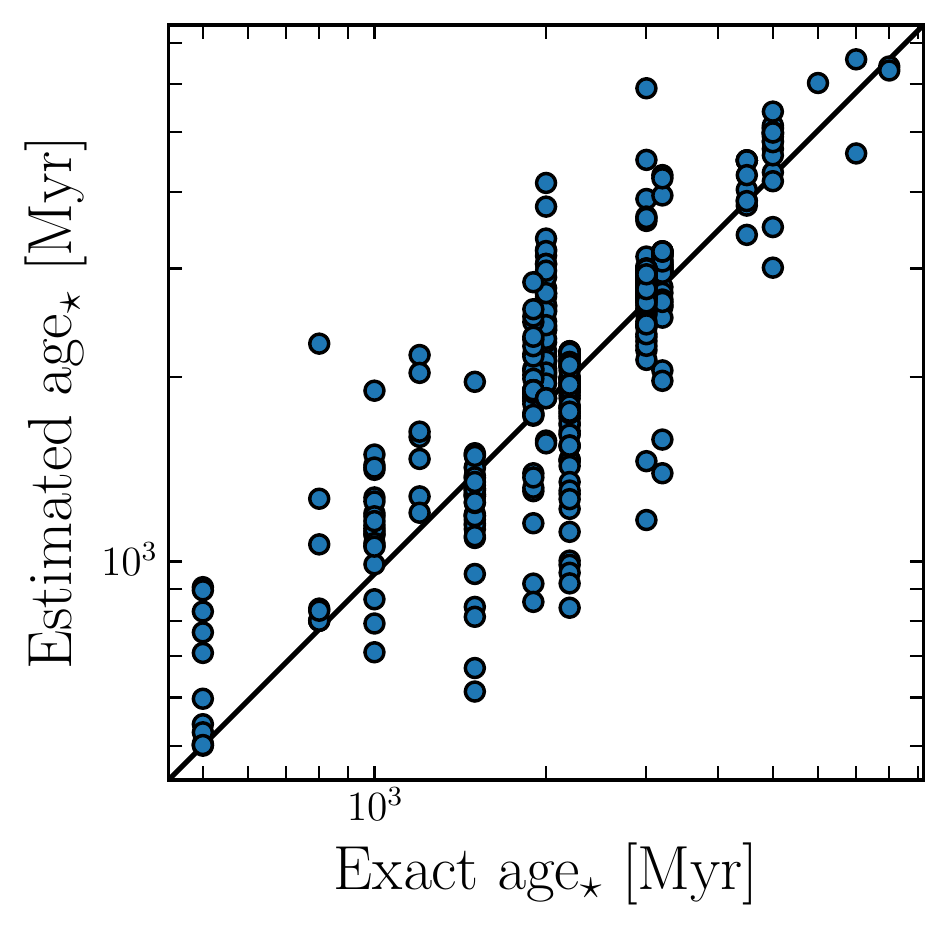}
    \endminipage
    \minipage{0.32\textwidth}
    \includegraphics[width=\linewidth]{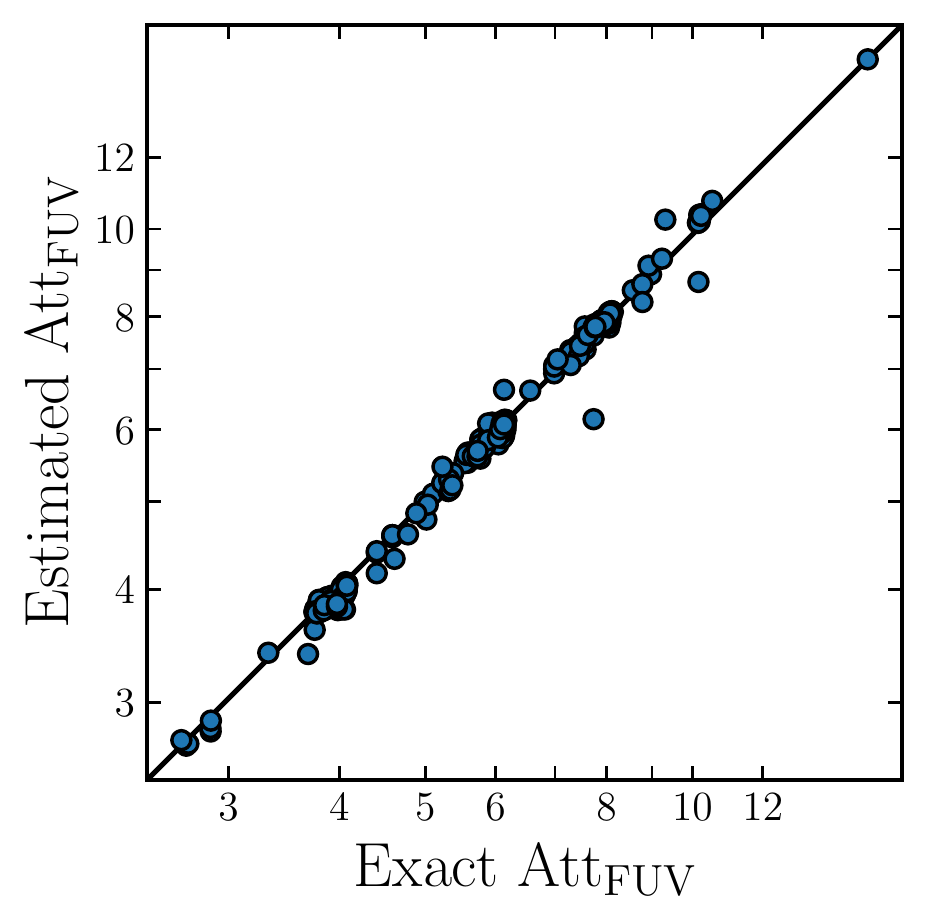}
    \endminipage
    \caption{\footnotesize Comparison between the true parameters of the mock SEDs and the results from the SED modeling with CIGALE (from top left panel) for the SFR, stellar mass (M$_*$), dust luminosity (L$_{dust}$), dust mass (M$_{dust}$), attenuation in the V band (Att$_{V}$) of the birth clouds and the ISM, color excess (B-V), age of the main stellar population (age$_{\star}$) and the attenuation in the FUV (Att$_{FUV}$). A parameter is well estimated and constrained when there is a one-to-one relation between the two.}
       \label{fig:mock}
\end{figure*}

We show the mock analysis of the physical parameters generated by \texttt{CIGALE}, as well as the input parameters that we used in the SED fitting technique, in Fig. \ref{fig:mock}.
\end{appendix}
\end{document}